# An extensive empirical study of inconsistent labels in multi-version-project defect data sets

Shiran Liu[1,2]    Zhaoqiang Guo[1,2]    Yanhui Li[1,2]*    Chuanqi Wang[1,2]
Lin Chen[1,2]    Zhongbin Sun[3]    Yuming Zhou[1,2]*

[1]State Key Laboratory for Novel Software Technology, Nanjing University
[2]Department of Computer Science and Technology, Nanjing University
[3]School of Computer Science & Technology, China University of Mining and Technology
*Corresponding author

*Abstract*—The label quality of defect data sets has a direct influence on the reliability of defect prediction models. In this study, for multi-version-project defect data sets, we propose an approach to automatically detecting instances with "inconsistent labels" (i.e. the phenomena of instances having the same source code but different labels over multiple versions of a software project) and understand their influence on the evaluation and interpretation of defect prediction models. Based on five multi-version-project defect data sets (either widely used or the most up-to-date in the literature) collected by diverse approaches, we find that: (1) most versions (>90%) in the investigated defect data sets contain inconsistent labels with varying degrees; (2) the existence of inconsistent labels in a training data set may considerably change the prediction performance of a defect prediction model as well as can lead to the identification of substantially different true defective modules; and (3) the importance ranking of independent variables in a defect prediction model can be substantially shifted due to the existence of inconsistent labels. The above findings reveal that inconsistent labels in defect data sets can profoundly change the prediction ability and interpretation of a defect prediction model. Therefore, we strongly suggest that practitioners should detect and exclude inconsistent labels in defect data sets to avoid their potential negative influence on defect prediction models. What is more, it is necessary for researchers to improve existing defect label collection approaches to reduce inconsistent labels. Furthermore, there is a need to re-examine the experimental conclusions of previous studies using multi-version-project defect data sets with a high ratio of inconsistent labels.

# 1 Introduction

A multi-version-project defect data set refers to a kind of defect data set that contains defect data of multiple historical versions for each project in the data set. For example, in a multi-version-project defect data set Metrics-Repo-2010 [27], the defect data of five historical versions for the "Ant" project are provided: Ant-1.3, Ant-1.4, Ant-1.5, Ant-1.6, and Ant-1.7. Multi-version-project defect data sets play an important role in the field of defect prediction. It is the basis for establishing cross-version defect prediction (CVDP), a scenario closely associated with real software development. In CVDP, a defect prediction model is first built with the defect data of historical version(s) of a project and then is applied to predict the defects in a target version of the same project. Currently, a few publically available multi-version-project defect data sets such as ECLIPSE-2007 [29] and Metrics-Repo-2010 [27] have been widely used. In the literature [1-10], they are used to evaluate the effectiveness of a variety of defect prediction models in various scenarios, not only in CVDP but also in CPDP (cross-project defect prediction). Clearly, the quality of labels (a label indicates if an instance is defective or not) in multi-version-project defect data sets is essential for the reliability of the experimental results or conclusions in defect prediction studies.

Recently, we are surprised to notice that there is an "inconsistent label" phenomenon in existing multi-version-project defect data sets (either widely used or the most up-to-date in the literature). Specifically, *the "inconsistent label" phenomenon denotes that a module in multiple versions has the same source code (non-blank, non-comment) but the corresponding instances in these versions have different labels* (it should be pointed out that this concept is different from the concept of "inconsistent instances" [110, 111, 124], which denotes two instances having identical values for all features, rather than the same source code, but different labels). To the best of our knowledge, there is no previous work that reports the "inconsistent label" phenomenon, not to mention the understanding on what they really mean, how inconsistent labels are generated, to what extent they exist, and how they influence defect prediction. In order to fill in this gap, we carry out a detailed empirical investigation on inconsistent labels. We analyze and elaborate the underlying rationale behind the generation of inconsistent labels and the implications of inconsistent labels from multiple perspectives. We find that inconsistent labels are generally caused by mislabels and extrinsic bugs [39]. In either case, the existence of inconsistent labels is not conducive to defect prediction. Therefore, in the sense of prediction model, inconsistent labels mean "suspected" noise. The reason why the inconsistent labels are called "suspected" noise is that in the absence of sufficient background information about the target data set (such as the issue reports used when collecting defect labels), it is not possible to manually distinguish between the correct and wrong labels. Neglecting such inconsistent labels in defect data sets may lead to misleading results when studying the validity of prediction models. Therefore, we further propose an approach to automatically detect inconsistent labels in multi-version-project defect data sets and analyze their impact on defect prediction. Based on five multi-version-project defect data sets collected by diverse approaches, we attempt to investigate the degree of existence of inconsistent labels as well as their influences on the prediction performance, detected actual defects, and model interpretability of a defect prediction model.

As far as we are aware, this is the first study to systematically investigate the existence and influence of inconsistent labels in multi-version-project defect data sets. The main contributions of our study are summarized as follows:

(1) By combining the latest idea of intrinsic and extrinsic bugs with the mechanism of diverse defect data collection approaches, we uncover the underlying rationale behind the generation of inconsistent labels and their real meaning. Through a detailed analysis, we show that in the sense of defect prediction models, inconsistent labels generally indicate noise.

(2) We propose an approach to automatically detecting inconsistent labels and analyze their influence on the evaluation and interpretation of defect prediction models. We find that most versions (>90%) in the investigated five multi-version-project defect data sets collected by current common defect data collection approaches contain inconsistent labels with varying degrees. In particular, inconsistent labels can profoundly change the prediction ability and interpretation of a defect prediction model.

(3) Our results uncover an important fact: even the latest proposed defect label collection approaches can lead to inconsistent labels, including the affected version approach (believed as a realistic approach in the literature) [23] and Herbold et al.'s semi-automatic SZZ-based defect label collection approach [90]. This means that, on the one hand, there is still a strong need to further improve the reliability of defect label collection, and on the other hand, our approach is effective in detecting inconsistent labels.

(4) We share code and data sets used in this work, which allows for replication and extension of our study and can also be used for other studies on defect prediction[1].

Our work has important implications for defect prediction community. On the one hand, our work provides practitioners an effective way to examine the quality of labels in multi-version-project defect data sets before building defect prediction models. On the other hand, our work provides researchers an in-depth understanding on the reason why inconsistent labels are generated, thus facilitating them to develop better defect label collection approaches. In particular, we find that Metrics-Repo-2010 [27], the most widely used multi-version-project defect data set in the literature, contains a high proportion of inconsistent labels. Therefore, there is a need to re-examine the experimental conclusions of previous studies using this data set.

Section 2 introduces the background knowledge. Section 3 shows examples to motivate our study. Section 4 analyzes the rationale behind the generation of inconsistent labels and their meanings. Section 5 presents our three-stage approach to identifying inconsistent labels in a multi-version-project defect data set. Section 6 describes the experimental design used in our study. Section 7 reports the experimental results in detail. Section 8 discusses our findings. Section 9 gives the implications of our study. Section 10 summarizes related studies and compares them with our study. Section 11 analyzes the threats to the validity of our study. Section 12 summarizes the paper and outlines the directions for the future work.

## 2 Background

In this section, we introduce the background knowledge serving to better elaborate this paper, including defect label collection and intrinsic verse extrinsic bugs.

**2.1 Defect label collection**

For a specific version of a project, the corresponding defect data set is composed of instances, each corresponding to a module (e.g. class) in the version. Furthermore, each instance consists of the following two types of information: code related features (such as size, complexity, historical change characteristics, historical defect information, and development process characteristics) and defect label (a binary variable indicating whether the corresponding module is defective or clean)[2]. In defect prediction community, it is common to: (1) collect code related features by analyzing source code and historical commits / issue reports recorded in VCS (version control system, e.g., CVS[3], Git[4]) / ITS (issue tracking system, e.g., Bugzilla[5], JIRA[6]); and (2) collect the defect label by analyzing bug-fixing commits in VCS and issue reports in ITS.

---

1 http://github.com/sticeran/InconsistentLabels
2 In the literature, these features are called the independent variables, while the label information is called the dependent variable.

In order to save the time and cost required by a manual method to collect defect labels, many automatic defect label collection approaches have been developed by mining the data in VCS and ITS [11-30]. According to different implementation mechanisms, these approaches can be roughly divided into three categories: SZZ-based approach [11-22, 90-93], time-window approach [23-28], and affected version approach [13, 23, 29, 30]. At a high level, for a given project, these defect label collection approaches consist of the following two steps:

- At the first step, identify BFCs (*bug-fixing commits*[7], the code changes that occur when developers fix bugs). It is common to use regular expression (or similar approaches) to match specific keywords (such as "bug" and "fix") and bug identifiers (recorded in ITS, e.g. "AMQ-7178") in the logs of historical commits in VCS to identify BFCs.
- At the second step, analyze BFCs to determine which modules in a version are defective.

The three categories of approaches are similar at the first step but are different at the second step. In the following, we describe the second step in detail for these approaches.

**SZZ-based approach.** The idea of the SZZ-based approach is to leverage BFCs and their corresponding BICs (bug-introducing commits) to link defective modules to versions [21, 90, 93]. SZZ [11] is a popular algorithm originally for collecting defect label data at the commit granularity [11-20, 31-38, 91, 92]. As shown in Fig. 1, a SZZ-based approach uses SZZ as the basis to identify the defect labels at the module granularity. Specifically, given the identified BFCs, a SZZ-based approach proceeds as follows. First, for each BFC, backtrack the corresponding modified lines to obtain BIC candidates. These modified lines can be identified either by *diff* command in CVS or by *git show* command in GIT[8], depending on the type of VCS. The assumption of SZZ [11] and its variants [12-16] are that the modified lines in a BFC are the lines that introduce bugs. By a code backtracking command provided by VCS (e.g., *annotate* in CVS or *git blame* in GIT), SZZ tracks back to identify the commits that introduce these buggy lines. These identified commits are called BIC candidates. Second, apply heuristic rules to rule out illogical BIC candidates to obtain correct BIC candidates. For example, a BIC candidate should be excluded if its commit time is later than the time that the corresponding bug report was submitted. Third, link defective modules to versions to determine which modules in a version are defective. To this end, it is common to leverage the period between the (starting) time a bug was introduced and the time it was fixed (BFC) to assign defective modules to versions [21]. In the literature, for a given BFC, different criteria were used to determine the starting time that there was a (potential) bug in the software [90, 93]. For example, Liu et al. considered the submit time of each BIC (as each BIC contributed to the bug) [93], while Herbold et al. only considered the submit time of the last BIC (as this was time that the bug took place) [90]. For each of the corresponding BIC(s), identify the modules containing the corresponding bug-introducing lines as defective modules. Given each identified defective module, its association versions can be determined by the following rule: the versions released between the submission times of the corresponding BIC and BFC. In other words, in these versions, the corresponding module will be labeled as "defective". After analyzing all BFCs, all the modules in each version not labeled as "defective" will be labeled as "clean".

---

[3] http://www.nongnu.org/cvs/
[4] https://git-scm.com
[5] http://www.bugzilla.org/
[6] http://www.atlassian.com/JIRA
[7] Some existing literatures use "bug-fixing change" to refer to BFC and some existing literatures use "bug-fixing commit" to refer to BFC. We use "bug-fixing commit" to refer to BFC because a "change" is recorded in the VCS in the form of a "commit".
[8] https://git-scm.com

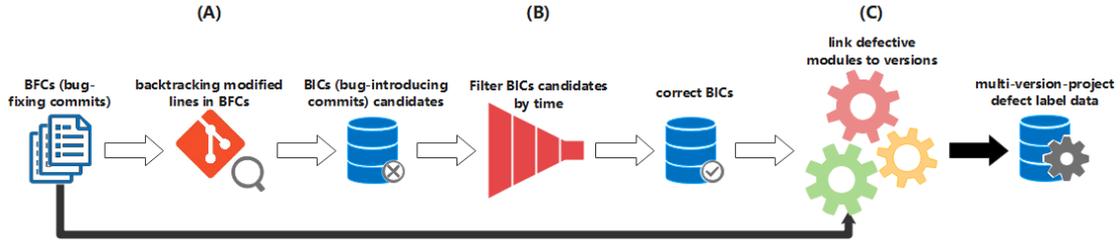

Fig. 1. Schematic diagram of defect label data collection by SZZ-based approach

**Time-window approach.** The idea of a time-window approach is to leverage a post-release time window to link defective modules to versions [23-28]. For a target version (i.e. a version of interest), all the BFCs submitted within its post-release time window are assumed as the bug fixes for the target version. Based on this assumption, for the target version, all modules modified in these BFCs will be labeled as "defective" and the remaining modules will be labeled as "clean". We next use an example shown in Fig. 2 (adapted from [23]) to explain the time-window approach. For the simplicity of presentation, we assume that the target version is 1.0. If the post-release window is set to six months [23], $C_2$ and $C_4$ will be considered as BFCs for version 1.0. Since module A is modified in $C_2$ and module B is modified in $C_4$, A and B in version 1.0 will be labeled as "defective" and the remaining modules will be labeled as "clean". If the post-release window is set to the period between the target version and its next version [27, 28], $C_2$, $C_4$, and $C_5$ will be considered as BFCs for version 1.0. Consequently, in version 1.0, A, B, and C will be labeled as "defective" and the remaining modules will be labeled as "clean". According to [27, 28], the widely used "Metrics-Repo-2010" data set was generated using such a time window approach.

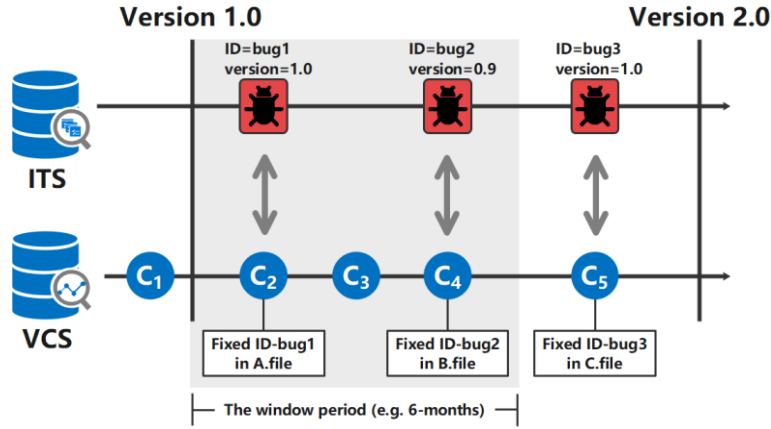

Fig. 2. Schematic diagram of defect label data collection by a time-window approach

**Affected version approach.** The idea of the affected version approach[9] is to link defective modules to versions using the version information recorded in issue report in ITS [13, 23, 29, 30]. Given a BFC, the corresponding issue report recorded in the ITS has a field (called "version" in Bugzilla and "affected version" in JIRA) listing the versions affected by the bug (provided by a software development team). This approach uses the following steps when mapping defective modules to versions. First, retrieve all BFCs satisfying the condition that the target version is the earliest affected version listed in their corresponding bug reports. Second, analyze these BFCs to determine the modified modules. Third, in the target version, these modules modified in BFCs will be labeled as "defective" and the remaining modules will be labeled as "clean". For example, for the target version 1.0 shown in Fig. 2, the following two BFCs will be retrieved: $C_2$ and $C_5$. Consequently, in version 1.0, module A and C will be labeled as "defective" and the remaining modules will be labeled as "clean". In [23], for a target version, all the BFCs whose earliest affected version is the target version are used to map defective modules to versions. However, in practice, it is not uncommon to see that a time window is set to limit which BFCs should be considered. For example, for the widely used Eclipse post-release defect data set [29], Zimmermann et al. only considered BFCs whose corresponding bugs were reported in the first six months after the release. For the example shown in Fig. 2, we assume that $C_2$ and $C_4$ are the BFCs whose bug reports are within the six-month time window. In this case, for the target version 1.0, the affected version approach will lead to that only $C_2$ is considered. As a result, in version 1.0, module A will be labeled as "defective" and all the other modules will be labeled as "clean".

---

[9] Note: the affected version approach is called "realistic approach" in the literature [23].

In light of the importance of defect label quality, recent studies pay attention to investigate which defect label collection approach is more accurate. In [23], Yatish et al. reported that the affected version approach was more accurate than a time window approach. However, as observed by Herbold et al. [90], it is not uncommon to see that the "affected version" field of a bug report is not updated, incorrectly typed, or even not used at all. Therefore, in practice, the affected version approach may also lead to inaccurate defect labels. Furthermore, Herbold et al. proposed a semi-automated SZZ-based defect label collection approach IND-JLMIV+R, which combined manual validation and improved heuristics for finding BFCs and improved the heuristic for identifying BICs. In nature, in their work, the focus is whether a defect label collection approach will lead to incorrect defect labels in a version. Compared with their work, our study investigates the quality of defect labels from a different dimension, i.e. whether a defect label collection approach will lead to inconsistent defect labels over multiple versions of a project. In the following sections, we will show that all the SZZ-based approach, time window approach, and affected version approach may lead to inconsistent defect labels with varying degrees.

**2.2 Intrinsic verse extrinsic bugs**

Given a BFC, if the origin of the corresponding bug (i.e. bug-inducing changes) can be identified, we will know the time point when the bug is introduced. This information is very helpful for accurately inferring which versions of a project will be affected by the bug. In an SZZ-based approach, the BICs traced back by the SZZ algorithm are believed as the origins of the corresponding bug. As a result, all the versions between the BICs and BFC are linked to the bug. In both the time window and affected version approaches, the origin of the bug is not explicitly traced. For the former, it is simply believed that all the versions, whose post-release time windows cover the BFC, have the corresponding bug. For the latter, it is simply believed that the bug occurs in the earliest version listed in the "affected version" field of the corresponding bug report. As can be seen, intuitively, the SZZ-based approach should generate more accurate defect labels, if the issue type and the identification of BFC and BICs are accurate.

The SZZ algorithm [11-16] has a basic assumption when collecting defect data: "a bug was introduced by the lines of code that were modified to fix it". The implication of this assumption is that bugs are caused by defective code. However, as pointed out by Rodríguez-Pérez et al. [39], this assumption is inaccurate. The reason is that, for some bugs, an explicit code change introducing them does not exist, for example, "software may work properly until the system where it runs on upgrades a library it depends on" (i.e. "the same snapshot does not exhibit the bug before the library upgrade, but exhibits the bug after") [39, 87, 88]. According to the causes of bugs, Rodríguez-Pérez et al. [39] classified bugs[10] into two categories: intrinsic and extrinsic bugs.

- **Intrinsic bugs**: "bugs that were introduced by one or more specific changes to the source code" [39]. That is to say, intrinsic bugs are caused by code itself. The introduction of defective codes occurs in the process of code change (adding, modifying, and/or deleting codes) of a module. In general, VCS can record BICs corresponding to intrinsic bugs (in practice, BICs may also not be recorded or not recorded completely for various reasons). Therefore, if a module has an intrinsic bug, it is possible to mine the corresponding BIC(s) from VCS.

---

[10] Note: the standard definition of the term "bug" is the same as the term "defect" [40, 41]. Unless otherwise stated, "bug" and "defect" have the same meaning in this paper.

- **Extrinsic bugs**: bugs are caused by the external factors outside the code, including changes in requirements, dependencies on the run-time environment, changes to the environment, and bugs in external APIs. In general, the cause of an extrinsic bug is not recorded in VCS. In other words, an extrinsic bug is not the result of an explicit change recorded in VCS. As a result, if a module has an extrinsic bug, it is not possible to mine the corresponding BIC(s) from VCS. At present, the identification of the cause of extrinsic bugs needs manual analysis.

According to this classification, a bug is caused either by code changes (for intrinsic bugs) or by context changes (for extrinsic bugs).

In light of the existence of intrinsic and extrinsic bugs, Rodríguez-Pérez et al. distinguished a commit first exhibiting a bug (i.e. before the commit the system does not show a failure but, after it, a failure appears) from a commit introducing a bug. The reason is that they have observed that: "the fact that before a given change the system does not exhibit a bug, but after it, the bug appears, is not enough to consider that the change introduced the bug" [39]. In order to help determine the origin of a bug, Rodríguez-Pérez et al. introduced the following two concepts: FFM (first-failing moment) and FFC (first-failing commit). The former denotes the moment that the bug manifests itself for the first time (which is not recorded in the VCS), while the latter denotes that the commit corresponds to the first time that the system manifested the bug.

Based on the concepts of BIC, FFM, and FFC, Rodríguez-Pérez et al. explained the differences between intrinsic and extrinsic bugs as follows. For an intrinsic bug, there is a BIC in the VCS and the BIC coincides with FFM and FFC (as shown in Fig. 3), i.e. BIC, FFM, and FFC occur at the same time point. However, for an extrinsic bug, there is no BIC recorded in the VCS and the corresponding FFC is the commit in the VCS after the FFM occurs (as shown in Fig. 4). Specifically, since FFC is not the commit causing the failure at FFM moment, it is not the BIC. The commit before FFC is not the BIC either because the external factors outside the code at that time had not changed and there were no bugs (for example, the external API that the module depends on had not been updated and the call to the external API would not cause bugs). Therefore, an extrinsic bug does not have BIC recorded in the VCS, and FFC and FFM are not equal.

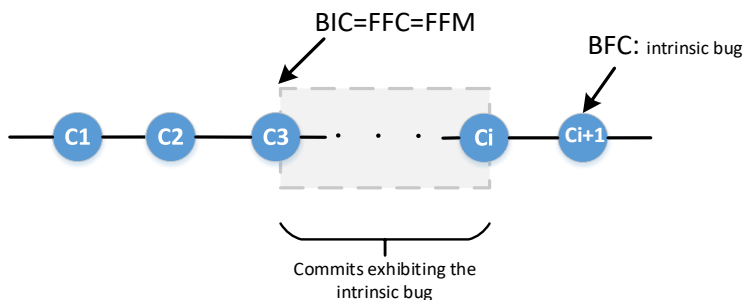
Fig. 3. Intrinsic bug: BIC=FFC=FFM

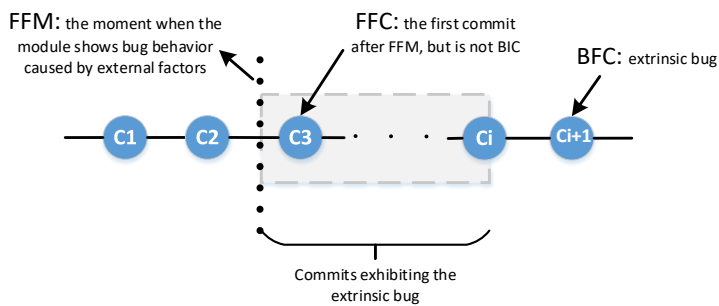

Fig. 4. Extrinsic bug: no BIC and FFC≠FFM

In summary, Rodríguez-Pérez et al.'s work shows that, in theory, in the current version control systems such as CVS and Git, intrinsic bugs have the corresponding BICs, while extrinsic bugs have no corresponding BICs. Therefore, distinguishing bug categories (intrinsic or extrinsic) can help analyze the real time period of bugs in modules:

- If the bug fixed by a BFC is an intrinsic bug, the real time period of the bug in a module should be the time period from the BIC to the BFC;
- If the bug fixed by a BFC is an extrinsic bug, the real time period of bugs in a module should be the time period from the FFM to the BFC.

It should be noted that the current VCS and ITS do not distinguish between intrinsic and extrinsic bugs [39]. In other words, there is no difference between intrinsic and extrinsic bugs in the form of issue reports or bug-fixing commits (BFCs). As a result, all the three categories of defect label collection approaches are not aware of bug categories. Nonetheless, when inconsistent labels are found on multiple versions, such a division will help analyze the real meaning of inconsistent labels. In Section 4, we will show that extrinsic bugs may lead to inconsistent labels.

## 3   Motivational examples

In the following, we use three examples to show inconsistent label phenomena in multi-version-project data sets generated by different approaches. The first example is from IND-JLMIV+R-2020 [90], a recently published data set generated using an enhanced SZZ-based approach (manually validated links between commits and issues, manually validated issues of type BUG, improved heuristics for determining BFCs, and improved heuristics for identifying BICs). The Tika project in IND-JLMIV+R-2020 data set contains 28 versions (tika-0.1~tika-1.17). As shown in Fig. 5, of the 28 versions, 18 versions (tika-1.0~tika-1.17) contain the following Java file with the same code, including blank and comment lines.

   *"tika-core/src/main/java/org/apache/tika/sax/xpath/XPathParser.java"*

However, the corresponding eighteen instances on these versions do not have the same label: the cross-version instance on tika-1.0~tika-1.9 have a label of "1", while the instances on tika-1.10~tika-1.17 have a label of "0".

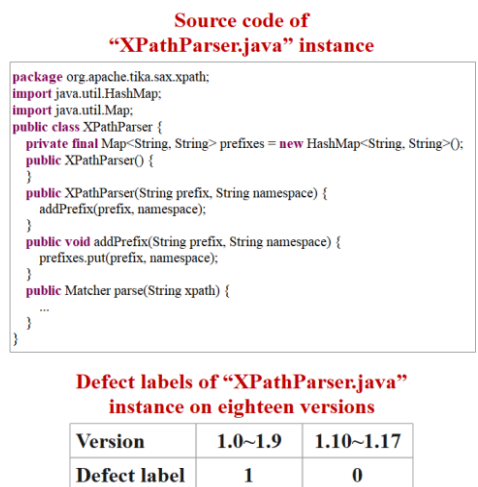

Fig. 5. The source code and defect labels of eighteen instances corresponding to "XPathParser.java"

The second example is from Metrics-Repo-2010 [27], a widely used data set generated using a time window approach. The Xalan project in Metrics-Repo-2010 contains four versions (Xalan-2.4, 2.5, 2.6, and 2.7). As shown in Fig. 6, all the four versions contain the following Java file with the same (non-blank, non-comment) source code:

"org/apache/xpath/SourceTree.java"

However, the corresponding four instances on these versions do not have the same label: the instances on Xalan-2.4 and Xalan-2.6 have a label of "0", while the instances on Xalan-2.5 and Xalan-2.7 have a label of "1".

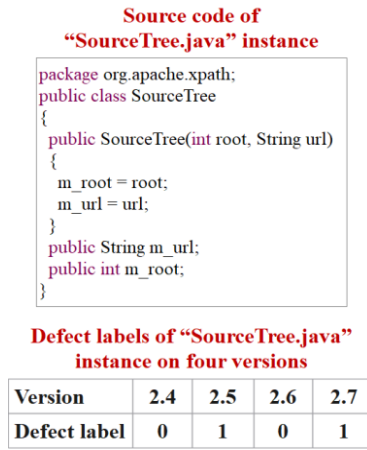

Fig. 6. The source code and defect labels of four instances

corresponding to "SourceTree.java"

The third example is from JIRA-RA-2019 [23], a recently published data set generated using an affected version approach (Yatish et al. regarded it as a realistic approach and believed that it can generate accurate defect label data). The Activemq project in JIRA-RA-2019 contains five versions (Activemq-5.0.0, 5.1.0, 5.2.0, 5.3.0, and 5.8.0). As shown in Fig. 7, of the 5 versions, 4 versions (Activemq-5.0.0~Activemq-5.3.0) contain the following Java file with the same code, including blank and non-comment lines.

"activemq-core/src/main/java/org/apache/activemq/filter/DestinationMap.java"

However, the corresponding four instances do not have the same label: the instances on Activemq-5.0.0, Activemq-5.1.0, and Activemq-5.3.0 have a label of "0", while the instance on Activemq-5.2.0 has a label of "1".

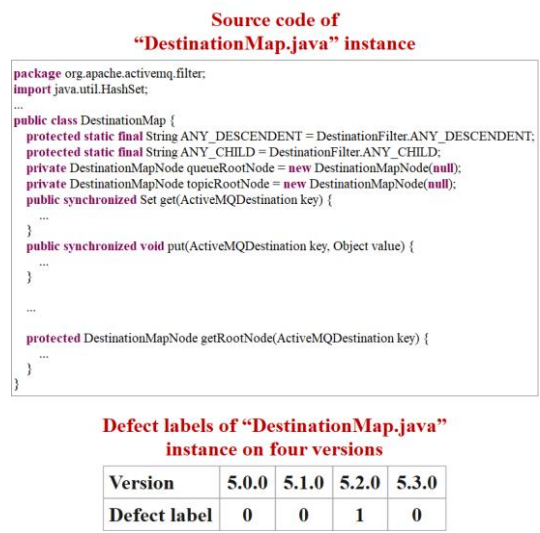

Fig. 7. The source code and defect labels of four "DestinationMap.java"

corresponding to "DestinationMap.java"

For each of the above-mentioned examples, the instances on different versions have the same source code. To the best of our knowledge, such an inconsistent label phenomenon is not reported by prior studies. What do inconsistent defect labels mean? Why does a defect label collection approach generate inconsistent defect labels? How will inconsistent defect labels affect the performance and interpretability of a defect prediction model? In the following sections, we will answer these questions. Our motivation is to make an in-depth understanding on inconsistent labels by analyzing the rationale behind the generation of inconsistent labels and evaluating their impact on the performance and interpretability of the defect prediction model.

## 4 Inconsistent defect labels: essence, cause, and implication

In this section, we provide an in-depth understanding of inconsistent defect labels. First, we analyze the essence of inconsistent labels. Then, we disclose the causes leading to inconsistent labels. Finally, we examine the implication of inconsistent labels for defect prediction.

**4.1 What is the essence of inconsistent defect labels?**

If we observe that there is a phenomenon of inconsistent defect labels on multiple instances, will this mean that at least one instance is mislabeled? Is it possible that all of their labels are correct? If all the labels are correct, what does this mean? In order to answer these questions, we need to understand the essence of an inconsistent label phenomenon. To this end, we first describe the following preliminary concepts.

- *Module*: a concept in source code. A module represents a piece of packaged source code that provides specific and tightly coupled functionality. For a given project, a version consists of a number of modules.
- *Instance*: a concept in data set. For a given version, the corresponding data set consists of instances, each corresponding to a module in the version. An instance in a data set is a row of data composed of a module name, a number of features indicating module-related characteristics, and a defect label indicating whether the corresponding module is defective or non-defective.
- *Cross-version module*. For a project with multiple versions, multiple versions can have modules with the same name. We refer to such modules as cross-version modules. Cross-version modules can have the same or different source code on different versions.
- *Cross-version instance*. In a multi-version-project data set, we refer to instances corresponding to cross-version modules as cross-version instances. Each cross-version instance has a corresponding cross-version module.

Therefore, an inconsistent label phenomenon will be observed if cross-version instances have different labels, but their corresponding cross-version modules have the same source code.

Without the loss of generalization, we assume that: (1) module A and module B are cross-version modules, i.e. they are from different versions of a project and have the same module name (i.e. A = B); (2) module A and module B have the same (non-blank, non-comment) source code; and (3) instance IA (corresponding to module A) and instance IB (corresponding to module B) have different defect labels, i.e. they are cross-version instances with different defect labels. Given this situation, there is an inconsistent label phenomenon on instances IA and IB. For the simplicity of presentation, we assume that instances IA has a label "buggy" and instances IB has a label "clean"[11]. In the following, we examine under which situation such an inconsistent label phenomenon may occur and what it means when taking into account intrinsic and extrinsic bugs.

**Under no bug scenario.** We use an example shown in Fig. 8 to illustrate the meaning of inconsistent labels under this scenario. For the simplicity of presentation, we assume that: (1) the commits $C_1$ and $C_2$ do not change the code in module X; and (2) module X is clean in $V_{1.0}$ and $V_{1.1}$. As a result, module X in $V_{1.0}$ has the same source code as module X in $V_{1.1}$. In this context, if A denotes X in $V_{1.0}$ and B denotes X in $V_{1.1}$, then an inconsistent label phenomenon will happen. At this time, both A and B are actually clean. As a result, the "buggy" label of instance IA is mislabeled (FP, false positive), while the "clean" label of instance IB is correct. Therefore, under no bug scenario, an inconsistent label phenomenon means that there is a mislabeling problem in the data set, i.e. **inconsistent labels contain incorrect labels**.

---

[11] In this study, "buggy" indicates defective and "clean" indicates non-defective. In the following, they will be used interchangeably.

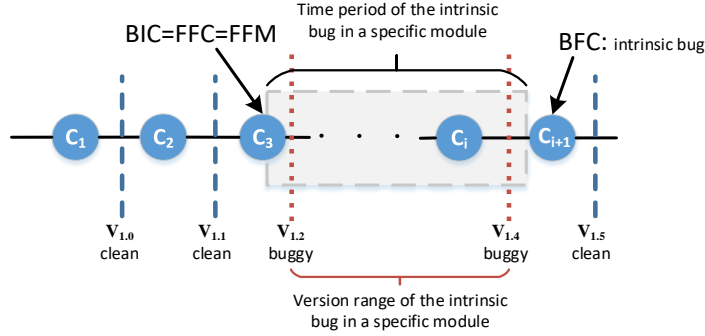

Fig. 8. Inconsistent labels under the intrinsic bug scenario

**Under the intrinsic bug scenario**. We still use the example in Fig. 8 to illustrate the meaning of inconsistent labels under this scenario. We further assume that: (1) $C_{i+1}$ is a BFC for fixing an intrinsic bug in module X and $C_3$ is the corresponding BIC; and (2) the commits from $C_4$ to $C_i$ do not change the code in module X. As a result, module X is defective in $V_{1.2}$, $V_{1.3}$, and $V_{1.4}$ and has the same source code in $V_{1.2}$, $V_{1.3}$, and $V_{1.4}$. In this context, if A denotes X in $V_{1.2}$ and B denotes X in $V_{1.3}$, an inconsistent label phenomenon will happen. At this time, both A and B are actually defective. Consequently, the "buggy" label of instance IA is correct, while the "clean" label of instance IB is mislabeled (FN, false negative). Therefore, under the intrinsic bug scenario, an inconsistent label phenomenon also indicates that there is a mislabeling problem in the data set, i.e. **inconsistent labels contain incorrect labels**.

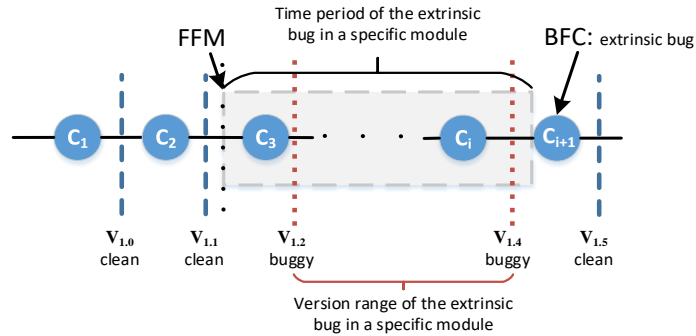

Fig. 9. Inconsistent labels under the extrinsic bug scenario

**Under the extrinsic bug scenario**. We use an example shown in Fig. 9 to illustrate the meaning of inconsistent labels under this scenario. Different from Fig. 8, $C_{i+1}$ is a BFC for fixing an extrinsic bug (rather than an intrinsic bug) in module X and $C_3$ is the FFC corresponding to FFM. In particular, we assume that all the commits except $C_{i+1}$ do not change the code in module X. As a result, module X in $V_{1.0}$ to $V_{1.4}$ has the same source code. In this context, an inconsistent label phenomenon related to the extrinsic bug can happen under the following cases:
(1) Module A is actually clean, but module B is actually defective (e.g. A denotes X in $V_{1.1}$ and B denotes X in $V_{1.2}$). In this case, not only the "buggy" label of instance IA is mislabeled (FP), but also the "clean" label of instance IB is mislabeled (FN). At this time, **inconsistent labels are incorrect labels**.
(2) Module A is actually defective, but module B is actually clean (e.g. A denotes X in $V_{1.2}$ and B denotes X in $V_{1.1}$). In this case, both the "buggy" label of instance IA and the "clean" label of instance IB are correct. At this time, **inconsistent labels are correct labels**.

Therefore, under the extrinsic bug scenario, an inconsistent label phenomenon does not necessarily mean that there is a mislabeling problem in the data set.

The above analyses reveal that, in essence, inconsistent labels are caused by extrinsic bugs and/or mislabeling. For inconsistent labels in a multi-version-project data set, there are three possible cases: (1) all-incorrect: all the involved labels are incorrect labels; (2) partly-correct: the involved labels contain correct labels as well as incorrect labels; and (3) all-correct: all the involved labels are correct labels. In particular, incorrect labels can be false positive or false negative.

**4.2 What factors cause inconsistent labels?**

From Section 4.1, we can see that extrinsic bugs and mislabeling are two sources of inconsistent defect labels. On the one hand, extrinsic bugs are caused by many factors in external factors outside the code of a project, including changes in requirements, dependencies on the run-time environment, changes to the environment, and bugs in external APIs [39, 87, 88]. If there is no mislabeling, extrinsic bugs will lead to inconsistent but correct defect labels. On the other hand, mislabeling can also lead to inconsistent defect labels, regardless of whether intrinsic or extrinsic bugs are involved. The factors causing mislabeling can be classified to two categories: (1) incomplete or incorrect data in VCS and ITS; and (2) an imperfect defect label collection mechanism or implementation. In practice, multiple factors with respect to extrinsic bugs and mislabeling may be tangled together, leading to inconsistent defect labels in a multi-version-project data set.

When collecting defect label, all the SZZ-based, time window, and affected version approaches are not aware of bug categories (intrinsic and extrinsic). As mentioned before, at a high level, they consist of two steps: (1) identify BFCs usually by linking commits for fixing bugs recorded in VCS to issue reports recorded in ITS; and (2) analyze BFCs to determine which modules in a version are defective. At the first step, incorrect or incomplete data in VCS and ITS may result in many missing links (i.e., many BFCs are not found) and many incorrect links (i.e., many identified BFCs are not correct). For the former, the main reason is that developers may forget to write specific keywords in the logs of commits in VCS or leave links for commit log in issue report description in ITS [94-102]. It is also possible that some BFCs are not recorded in VCS or some issues are not recorded in ITS [106, 107]. For the latter, the main reason is that many issues reported as bugs in ITS are actually requests for new features, bad documentation, or refactoring [105]. In particular, BFCs may contain non-fixing changes [103, 104]. At the second step, even if all the BFCs are correctly identified at the first step, an inaccurate defect label collection mechanism can also introduce mislabels. To make it clear, we next analyze why the SZZ-based, time window, and affected version approaches can introduce mislabels that lead to inconsistent labels.

**SZZ-based approach.** The original SZZ algorithm [11] is based on an implicit assumption that the modified lines in a BFC are the lines that introduce bugs. However, this assumption does not always hold. The reason is that a BFC may include non source code modification (such as modifications on comments or formats). As a result, the BICs identified from the modified lines in a BFC (i.e. the commits last touched the modified lines) may include those commits that do not modify source codes [12-14]. This can lead to mislabels, which may further lead to inconsistent defect labels. We next use an example shown in Fig. 10 to explain how this may happen. Assume that: (1) $C_2$ is a commit for adding comments to module A; (2) $C_4$ is a commit for adding buggy source code lines to module B; (3) $C_{i+1}$ is a mixed-purpose commit[12] for fixing bugs in module B as well as modifying the comments in module A; and (4) module A contains no bug in version 1.1 (i.e., $V_{1.1}$). In this context, since module A has the same (non-blank, non-comment) source code in version 1.1 and version 1.2 (i.e., $V_{1.2}$), the corresponding instances in version 1.1 and version 1.2 should have the same actual label "clean". However, according to the original SZZ algorithm, $C_2$ will be incorrectly identified as a BIC. Therefore, the instance corresponding to module A in version 1.2 will be mislabeled as "buggy". Consequently, due to the introduction of such a mislabel, the instance labels of module A on ($V_{1.1}$, $V_{1.2}$) are inconsistent defect labels: (clean, buggy).

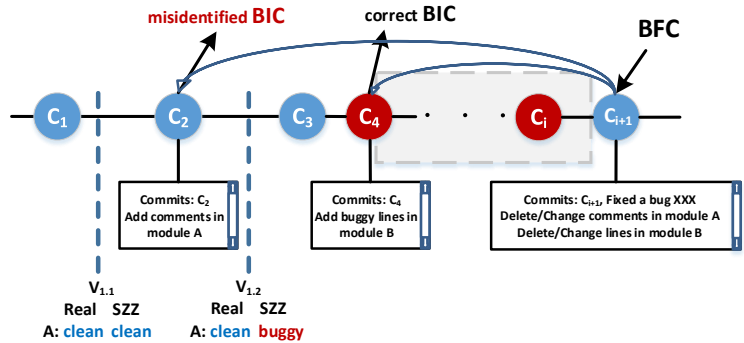

Fig. 10. SZZ-based approach: An example of inconsistent labels caused by non source code modification in a BFC

After the original SZZ algorithm is published, various variants have been proposed to improve its accuracy [12-16, 90-92]. In particular, much effort has been made to accurately exclude non-source-code modifications in a BFC during the identification of buggy lines. However, even if all the identified buggy lines are accurate, the mechanism itself of the SZZ algorithm can still lead to mislabel. We next use a "rollback change" example shown in Fig. 11 to explain how this may happen. Assume that: (1) $C_1$ is a commit adding a buggy line "if (a>1)" in module A; (2) $C_2$ is a commit changing "if (a > 1)" to "if (a ≥ 1)" in module A; (3) $C_4$ is a commit changing "if (a ≥ 1)" to "if (a > 1)" in module A, i.e. it is a rollback commit; and (4) $C_{i+1}$ is a commit for fixing bugs in module A. As can be seen, module A has the same code in version $V_{1.1}$ and $V_{1.2}$ and the corresponding instances in these two versions should have the same actual label "buggy". As can be seen, by backtracking the modified line, the SZZ algorithm will identify $C_4$ as a BIC but will miss $C_1$ that actually is also a BIC. Consequently, the instance corresponding to module A in version 1.1 will be mislabeled as "clean". Due to the introduction of such a mislabel, the instance labels of module A on ($V_{1.1}$, $V_{1.2}$) are inconsistent defect labels: (clean, buggy). To the best of our knowledge, such a problem does exist even if the state-of-the-art SZZ algorithm variants are considered.

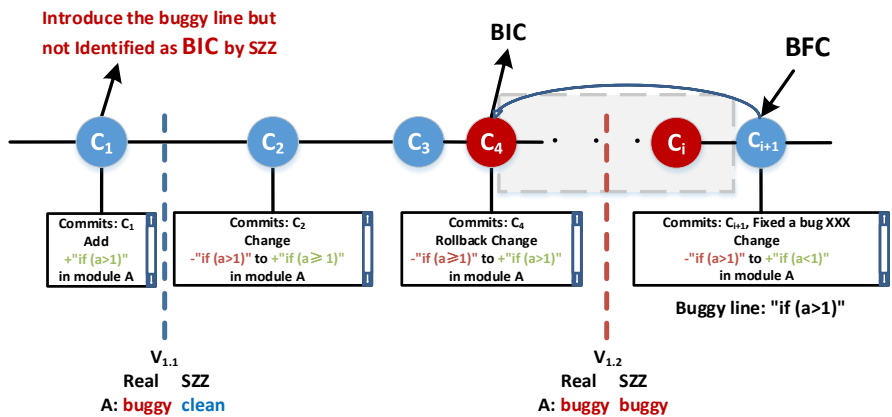

---

[12] A mix-purpose BFC includes changes to fix a bug as well as changes for other purposes such as refactoring and maintenance [103].

Fig. 11. SZZ-based approach: An example of inconsistent labels

caused by source code modification in a BFC

**Time-window approach.** The following two factors cause that a time window approach may introduce mislabels that lead to inconsistent labels. On the one hand, the length of the time window may be not large enough to cover all the BFCs that fix the bug in the target version. Consequently, false negative labels may be introduced. On the other hand, mixed-purpose BFCs [103] are not analyzed to exclude non-fixing changes. As a result, false positive labels may be introduced. In both cases, inconsistent labels can be produced. We next use an example shown in Fig. 12 to explain how this may happen. Assume that: (1) $C_1$ is a commit adding module A and B, in which module A has a bug (ID-bug1); (2) $C_4$ is a mixed-purpose BFC that includes fixing-change for the bug in module A and non-fixing change in module B; and (3) $C_2$ and $C_3$ are commits that are not related to module A and module B. As can be seen, module A has the same code in version $V_{1.1}$ and $V_{1.2}$ and the corresponding instances in these two versions should have the same actual label "buggy". However, for the target version $V_{1.1}$, the length of the time window is not enough to cover $C_4$. Consequently, the instance corresponding to module A in $V_{1.1}$ will be mislabeled as "clean" (false negative). Due to the introduction of such a mislabel, the instance labels of module A on ($V_{1.1}$, $V_{1.2}$) are inconsistent labels: (clean, buggy). In addition, module B has the same (non-blank, non-comment) source code in version $V_{1.1}$ and $V_{1.2}$. Therefore, the corresponding instances in these three versions should have the same actual label "clean". However, for the target version $V_{1.2}$, since the time window approach does not exclude non-fixing change from $C_4$, the instance corresponding to module B in $V_{1.2}$ will be mislabeled as "buggy" (false positive). Consequently, due to the introduction of such a mislabel, the instance labels of module B on ($V_{1.1}$, $V_{1.2}$) are inconsistent defect labels: (clean, buggy).

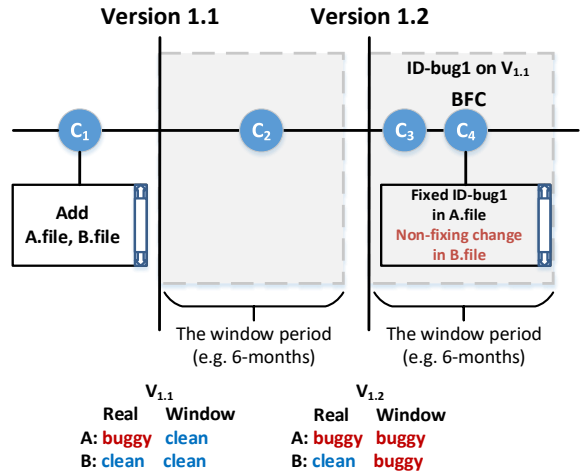

Fig. 12. Time-window approach: Example inconsistent labels

caused by the length of a time window and mixed-purpose BFCs

**Affected version approach.** The affected version approach uses the version numbers recorded in the *affected version* field / *version* field[13] of issue reports to determine defective modules in a version. In the literature [13, 23, 29, 30], it is common to use the recorded earliest affected version to map issue reports to versions (i.e. the earliest affected version approach). As stated in [29], the reason for this is that "the values of the field may change during the life cycle of a bug (e.g. when a bug is carried over to the next release)". The following two factors can cause that the earliest affected version approach may introduce mislabels leading to inconsistent labels. On the one hand, the ignorance of the non-earliest affected versions may introduce false negative labels. On the other hand, not excluding non-fixing changes from mixed-purpose BFCs may introduce false positive labels. In both cases, inconsistent labels can be produced. We next use an example shown in Fig. 13 to explain how this may happen. Assume that: (1) a defective module A has the same (non-blank, non-comment) source code in $V_{1.1}$ and $V_{1.2}$; (2) a non-defective module B has the same (non-blank, non-comment) source code in $V_{1.1}$ and $V_{1.2}$; (3) $C_4$ is a mixed-purpose BFC that includes fixing-change for the bug in module A and non-fixing changes in module B; (4) the issue report corresponding to $C_4$ records $\{V_{1.1}, V_{1.2}\}$ in the *affected version* field; and (5) $C_2$ and $C_3$ are commits that are not related to module A and module B. In this context, the real instance labels of module A on ($V_{1.1}$, $V_{1.2}$) should be (buggy, buggy). However, since the earliest affected version approach only takes into account the earliest version $V_{1.1}$, only the instance of module A on $V_{1.1}$ will be labeled as "buggy". Consequently, the instance of module A on $V_{1.2}$ will be mislabeled as "clean" (false negative). This leads to that the instance labels of module A on ($V_{1.1}$, $V_{1.2}$) are inconsistent defect labels: (buggy, clean). In addition, from module B, since the earliest affected version approach does not exclude non-fixing change from $C_4$, the instance corresponding to module B in $V_{1.1}$ will be mislabeled as "buggy" (false positive). Due to the introduction of such a mislabel, the instance labels of module B on ($V_{1.1}$, $V_{1.2}$) are inconsistent defect labels: (buggy, clean).

---

[13] For brevity, *version* field and *affected version* field are represented by *affected version* field in the following parts of this paper unless otherwise specified.

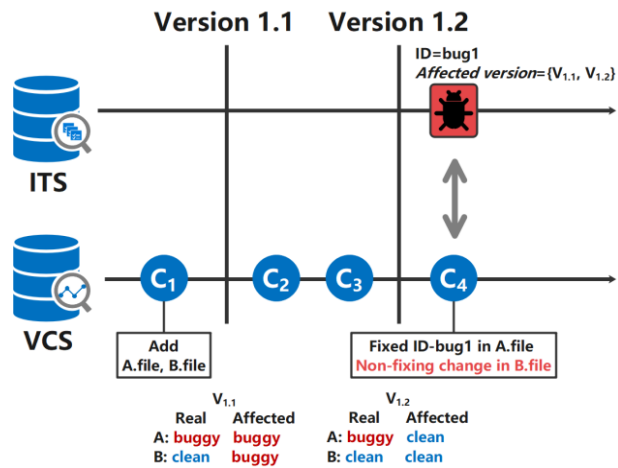

Fig. 13. Affected version approach: Example inconsistent labels
caused by only using the earliest version and mixed-purpose BFCs

If all the recorded affected versions in the *affected version* field / *version* field are used to map issue reports to versions (i.e. all the affected versions approach), can the phenomenon of inconsistent labels be avoided? The answer is "No", even if all the BFCs do not include non-fixing changes. The reason is that some affected versions may not be recorded in the affected version field [90]. For example, according to [90], the affected version field in the issue report CAY-1657[14] only records "3.1M3", but the author writes in the description: "I am sure this affects ALL versions of Cayenne, but my testing is done on 3.1 M3/M4". Another example is the issue AMQ-7346[15], in which the *affected version* field is empty but the *fix version* field records 5.16.0 and 5.15.12. We next use an example shown in Fig. 14 to explain how all the affected versions approach may lead to inconsistent labels. Assume that: (1) a defective module A has the same (non-blank, non-comment) source code in $V_{1.1}$ and $V_{1.2}$; (2) $C_4$ is a BFC that fixed the bug in module A; (3) the issue report corresponding to $C_4$ only recorded $V_{1.2}$ but miss $V_{1.1}$ in the *affected version* field due to incomplete recorded affected versions; and (4) $C_2$ and $C_3$ are commits that are not related to module A and module B. We can see that the real instance labels of module A on ($V_{1.1}$, $V_{1.2}$) should be (buggy, buggy). However, the instance of module A on $V_{1.1}$ will be mislabeled as "clean" (false negative), while the instance of module A on $V_{1.2}$ will be labeled as "buggy". Due to the introduction of such a mislabel, the instance labels of module A on ($V_{1.1}$, $V_{1.2}$) are inconsistent labels: (clean, buggy).

---

[14] https://issues.apache.org/jira/browse/CAY-1657
[15] https://issues.apache.org/jira/projects/AMQ/issues/AMQ-7346

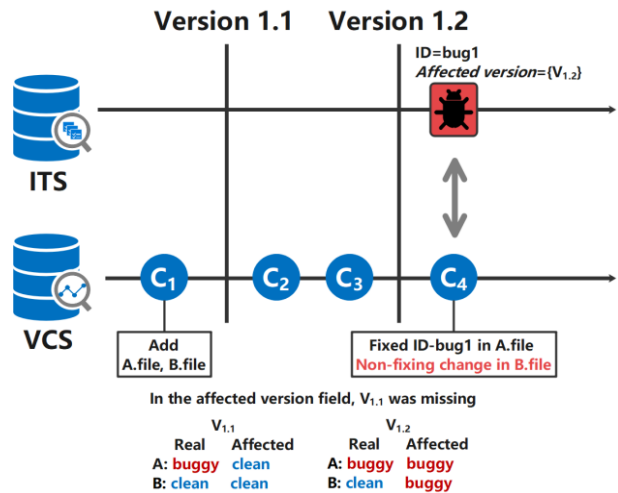

Fig. 14. Affected version approach: Example inconsistent labels caused by missing version(s) and mixed-purpose BFCs

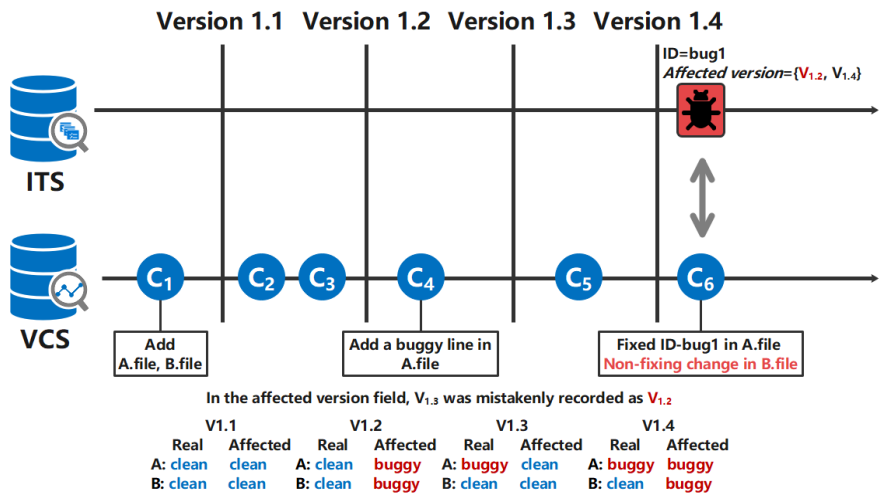

Fig. 15. Affected version approach: Example inconsistent labels caused by error records and mixed-purpose BFCs

More importantly, it is not uncommon to see that incorrect values are filled in the *affected version* field [90]. Indeed, such incorrect values may also introduce mislabels leading to inconsistent labels. We next use an example shown in Fig. 15 to explain, for all the affected versions approach, how this may happen (note that similar problems can be observed for the earliest affected version approach). Assume that: (1) a clean module A has the same (non-blank, non-comment) source code in $V_{1.1}$ and $V_{1.2}$; (2) $C_4$ adds a buggy source code line in module A; (3) module A has the same (non-blank, non-comment) source code in $V_{1.3}$ and $V_{1.4}$, but contains a bug; (4) $C_6$ is a BFC fixing the bug in module A; (5) the issue report corresponding to $C_6$ record $\{V_{1.2}, V_{1.4}\}$ in the *affected version* field but $V_{1.2}$ is an error record caused by developers mistakenly recording $V_{1.3}$ as $V_{1.2}$; that is to say, the actual affected version $V_{1.3}$ is not recorded in this field, while the actual unaffected version $V_{1.2}$ is recorded in this field; and (6) $C_2$, $C_3$, and $C_5$ are commits that are not related to module A and module B. In this context, the real instance labels of module A on ($V_{1.1}$, $V_{1.2}$) should be (clean, clean) and on ($V_{1.3}$, $V_{1.4}$) should be (buggy, buggy). However, when all versions $\{V_{1.2}$, $V_{1.4}$} in the *affected version* field are used, the instance of module A on $V_{1.2}$ will be will be mislabeled as "buggy" (false positive) and on $V_{1.3}$ will be will be mislabeled as mislabeled as "clean" (false negative). Due to the introduction of such mislabels, the instance labels of module A on ($V_{1.1}$, $V_{1.2}$) or on ($V_{1.3}$, $V_{1.4}$) are inconsistent labels: (clean, buggy).

The above analyses reveal that, from the viewpoint of labeling mechanism itself, all the SZZ-based, time window, and affected version approaches can lead to inconsistent defect labels. Since they are based on the data recorded in VCS and ITS, low quality data such as incomplete and inaccurate data would make more inconsistent defect labels. In particular, due to the existence of extrinsic bugs, inconsistent defect labels cannot be avoided. In this sense, it is hard, if not possible, to obtain multi-version-project defect data sets without inconsistent labels.

**4.3 What do inconsistent labels imply for defect prediction?**

In the last decades, a lot of literature pointed out that real-world data sets may contain noise [42-47, 94-107], which is defined in [89] as anything that obscures the relationship between the features of an instance and its class label. Generally speaking, there are two types of noise source in a data set [43, 46]: (a) feature noise; and (b) label noise. According to [46], on the one hand, the possible causes for feature noise include erroneous values, missing (or unknown) values, and incomplete values. On the other hand, there are at least two possible causes for label noise: (1) contradictory instances, i.e., multiple instances have the same feature values but different labels; and (2) mislabeled instances: instances labeled with wrong labels.

In defect prediction context, a module is represented by an instance, i.e. a feature vector with a class label. In nature, for an instance, its label is used to indicate whether the corresponding module is defective or not. According to Section 4.1, in a multi-version-project data set, a pair of instances is regarded as having inconsistent labels if the following conditions are satisfied: (1) they are cross-version instances; (2) the corresponding modules have the same (non-blank, non-comment) source code; and (3) their labels are inconsistent. Furthermore, there are three types of inconsistent labels: both labels are incorrect, only one label is correct, and both labels are correct. In the following, we examine what they mean for defect prediction. Since these two instances are cross-version, for the simplicity of presentation, we assume that one instance (called IA) is in the training set and another instance (called IB) is in the testing set.

- Both labels are incorrect. On the one hand, IA is a mislabeled instance in the training set. If we use it to build a defect prediction model, the resulting model will be biased. On the other hand, IB is a mislabeled instance in the testing data. If we use it to evaluate a model, the resulting performance score will be misleading. As can be seen, IA is a noise instance in the training set, while IB is a noise instance in the testing set.
- Only one label is correct. If IB has an incorrect label, IB is noise in the testing set. At this time, from the viewpoint of the training set self, IA is not noise due to that it has a correct label. If IA has an incorrect label, IA is noise in the training data. From the viewpoint of the testing set itself, IB is not noise due to that it has a correct label.
- Both labels are correct. This case is caused by an extrinsic bug. Traditionally, the goal of building defect prediction models is to explore whether code characteristics and code-related development activities can predict intrinsic bugs in a module. Therefore, in a defect data set, the features of an instance usually only consist of code metrics and process metrics and do not include the external factors causing extrinsic bugs. As a result, bugs caused by external factors should not be used to study the association of intrinsic bugs with code characteristics and code-related development activities. In other words, a defect prediction model built with traditional defect data sets is not applicable to the prediction of extrinsic bugs. In this sense, for a pair of instances with inconsistent labels caused by an extrinsic bug, the instance with a "buggy" label is noise, regardless of whether it is in the training set or the testing set. Indeed, our opinion is in accordance with a recent study by Rodríguez-Pérez et al. [51]. In their study, Rodríguez-Pérez et al. stated that extrinsic bugs should be removed when building and evaluating a defect prediction model.

As can be seen, when a pair of instances with inconsistent labels are found, we can conclude that at least one instance is noise for defect prediction. However, we do not know which one(s) is(are) noise if no additional information is given. When building a defect prediction model, such a pair of instances will lead to a contradictory relationship between the features of an instance and its class label. Therefore, from the viewpoint of practical application, if a pair of instances have inconsistent labels, both can be considered as "noise" data.

As we know, training a model with noise data will result in a biased model, while testing a model with noise data will lead to incorrect performance scores. As a result, not considering inconsistent labels in multi-version-project defect data sets may lead to misleading results or conclusions in the context of defect prediction. This is especially true when taking into account the fact that "class noise is potentially more harmful than feature noise, what highlights the importance of dealing with this type of noise" [46, 47]. Before dealing with inconsistent labels, we should have an approach to accurately detecting inconsistent labels in a multi-version-project defect data set.

## 5 TSILI: A three stage inconsistent label identification approach

In this section, we propose an approach called **T**hree **S**tage **I**nconsistent **L**abel **I**dentification (TSILI) to automatically identify inconsistent labels in a multi-version-project defect data set. For the simplicity of presentation, we assume that a project consists of $n$ versions. For each version $V_i$, let $DV_i$ be the corresponding defect data set and $SV_i$ be the corresponding source code database, $1 \leq i \leq n$. In this context, the task of TSILI is to identify inconsistent labels for the multi-version-project defect data set consisting of $DV_1$, $DV_2$, …, and $DV_n$ (in the following, $DV_i$ will be called a component defect data set). Specifically, for each component defect data set $DV_i$, TSILI aims to add a feature to indicate which instances have inconsistent labels, $i \leq n$. To this end, TSILI proceeds as follows. At the first stage (see Section 5.1), given the inputs of a multi-version-project defect data set and the corresponding source code databases, an information table is generated to record those instances whose source codes can be found. At the second stage, the elements in the information table are analyzed to identify instances with inconsistent labels, i.e. those cross-version instances with the same name, the same source code, but different defect labels (see Section 5.2). At the third stage, for each instance in the multi-version-project defect data set, a feature is added to indicate whether its label is inconsistent or not based on the inconsistent label information recorded in the information table (see Section 5.3). Finally, we analyze the time complexity of TSILI (see Section 5.4).

**5.1 Stage 1: Generate an information table recording instances in all versions**

For the first stage, the inputs are a multi-version-project defect data set (i.e. $DV_1$, $DV_2$, …, and $DV_n$) and the corresponding source code databases (i.e. $SV_1$, $SV_2$, …, and $SV_n$). The output is an information table *moduleInfo* recording the following information for those instances whose source codes can be found in the source code databases: (1) "name": the name of an instance in multi-version-project defect data set (i.e. the name of the corresponding module in the source code databases); (2) "version": the version number of the project that the corresponding module of an instance belongs to; (3) "codePath": the path of the file in which the source code of the corresponding module of an instance is located; (4) "defectLabel": the defect label of an instance in the multi-version-project defect data set ("0" is clean and "1" is buggy); and (5) "isInconsistentLabel": whether the defect label is an inconsistent label ("NO" means non-inconsistent label, while "YES" means an inconsistent label).

As shown in Fig. 16, the first stage proceeds as follows. First, set *moduleInfo* as an empty table (b1). Second, analyze the multi-version-project defect data set and the corresponding source code databases to generate *moduleInfo* (b2~b10). Specifically, for each instance in $DV_i$, examine whether its name appears in $SV_i$ (b7). If the answer is "Yes", obtain the path of the corresponding source code recorded in $SV_i$ and the corresponding defect label recorded in $DV_i$. With such information, a five-tuple (i.e. <name, version, codePath, defectLabel, isInconsistentLabel>) is generated and added to *moduleInfo*, in which "isInconsistentLabel" is set as "NO" (b8). Third, sort the elements (each element is a five-tuple, corresponding to an instance) in *moduleInfo* in ascending order according to the instance name (b11). As a result, the instances with the same name on different versions (i.e. cross-version instances) are grouped together. This will facilitate the identification of cross-version instances with inconsistent labels in the next stage.

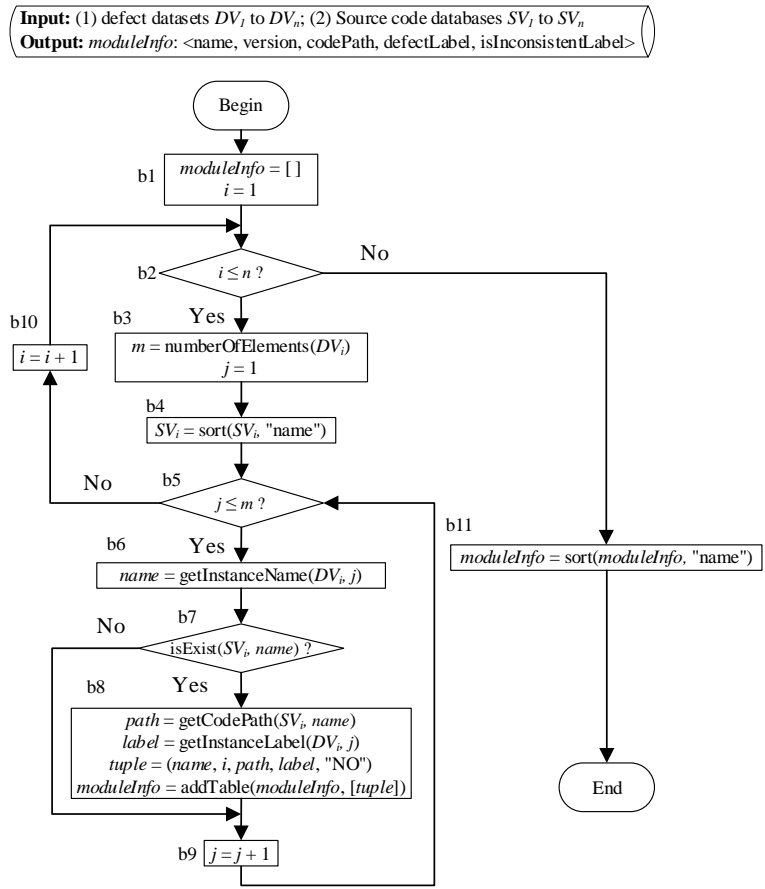

Fig. 16. The flowchart of generating an information table recording all instances in all versions

**5.2 Stage 2: Identify cross-version instances with inconsistent labels**

For the second stage, the input is the information table *moduleInfo* generated in the first stage, in which all the instances have the same "NO" value for the attribute "isInconsitentLabel". The output is an updated *moduleInfo* in which inconsistent labels have been recorded, i.e. instances with inconsistent labels have a value "Yes" for the attribute "isInconsitentLabel".

As shown in Fig. 17, the second stage proceeds as follows. First, divide the elements (i.e. instances) in *moduleInfo* into different groups by their "name" values (b1). Within each group, the instances have the same instance name but are from different versions, i.e. they are cross-version instances. Second, for each group, identify which instances have the same source code but different defect labels (b2~b19). Specifically, for each group, obtain the corresponding set of defect labels *labelSet*. If |*labelSet*| == 1, skip this group as there is no inconsistent label (b19). Otherwise, the instances in the group are partitioned into equivalence classes by comparing their source codes (b6~b18). Within each equivalence class, the instances not only have the same name but also have the same source code. In order to reduce the influence of code format, the following measures are taken to format the code (b9) before the partition (b11): filter out comments and replace consecutive whitespaces and newlines with one whitespace character. For each equivalence class, examine whether all the instances have the same label (b16). If the answer is "No", this means that the instances in this equivalence class have inconsistent labels and update *moduleInfo* to record this information accordingly (b17). When the second stage terminates, it outputs *moduleInfo* with the updated "isInconsitentLabel" attribute.

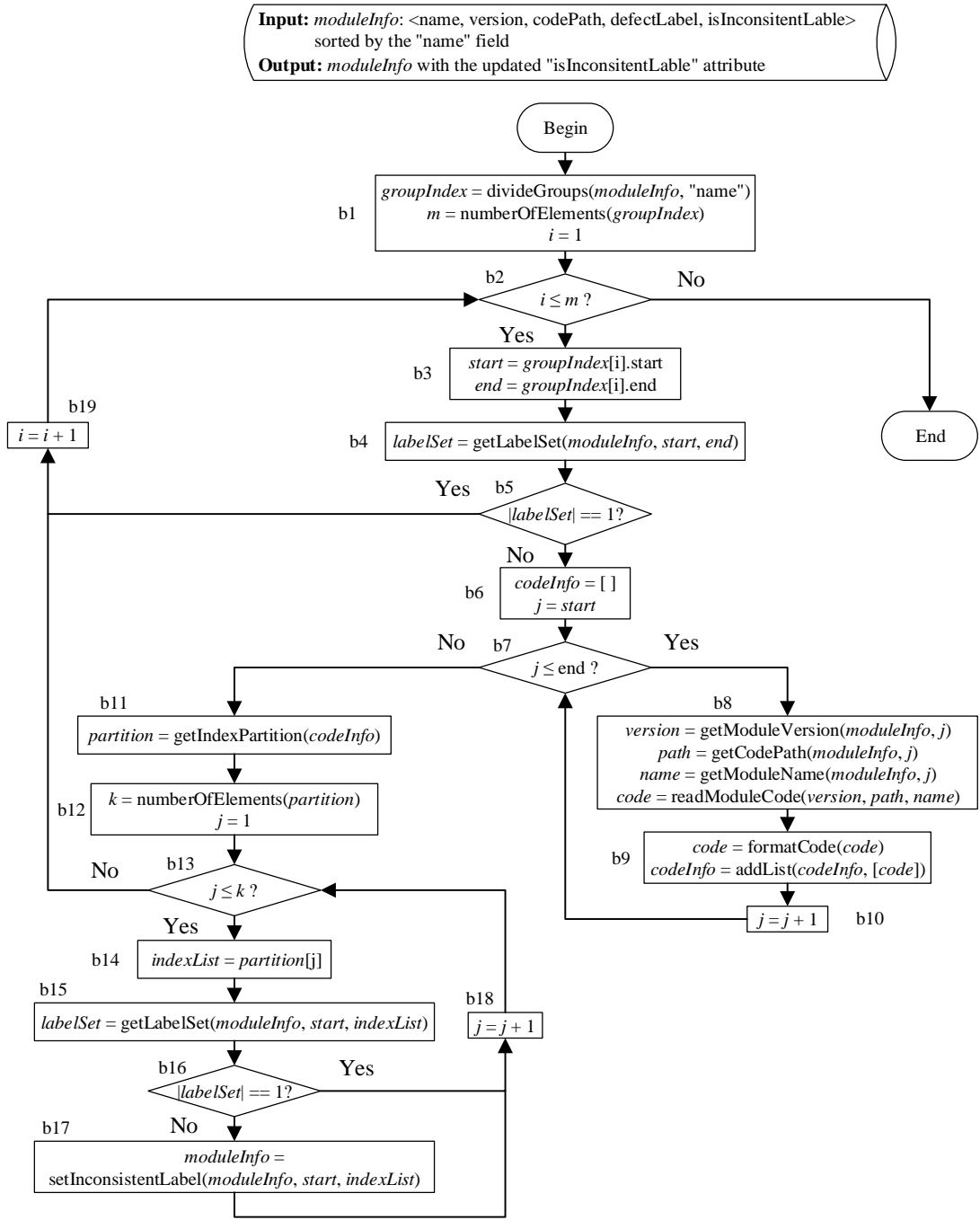

Fig. 17. The flowchart of identifying cross-version instances with inconsistent labels

**5.3 Stage 3: Augmenting the multi-version-project defect data set with inconsistent labels**

For the third stage, the input is the multi-version-project defect data set (i.e. $DV_1$, $DV_2$, …, and $DV_n$) and the information table *moduleInfo* generated at the second stage (in which all the instances have the "Yes" or "NO" value for the feature "isInconsitentLabel"). The output is the multi-version-project defect data set augmented with the feature "isInconsitentLabel" that indicates whether an instance has an inconsistent label: "NO" means non-inconsistent, "YES" means inconsistent, and "NA" means unknown (for an instance, if the corresponding source code cannot be found, its feature "isInconsitentLabel" will be assigned an "NA" value).

As shown in Fig. 18, the third stage proceeds as follows. First, sort the elements in *moduleInfo* according to their "version" values so that the instances whose corresponding modules belong to the same version are grouped together (b1). Second, for each component defect data set, add a feature "isInconsitentLabel" and set its value for each instance based on the inconsistent label information recorded in *moduleInfo* (b2~b13). Specifically, for $DV_i$, take all the elements belonging to the *i*th version from *moduleInfo* to *curModuleInfo* (b3). For each instance in $DV_i$, examine whether its name appears in *curModuleInfo* (b8). If the answer is "Yes", set its "isInconsitentLabel" as the corresponding value recorded in *curModuleInfo*; otherwise, set its "isInconsitentLabel" as a value of "NA" (b9~b11). When the third stage terminates, it outputs a multi-version-project defect data set augmented with inconsistent label information.

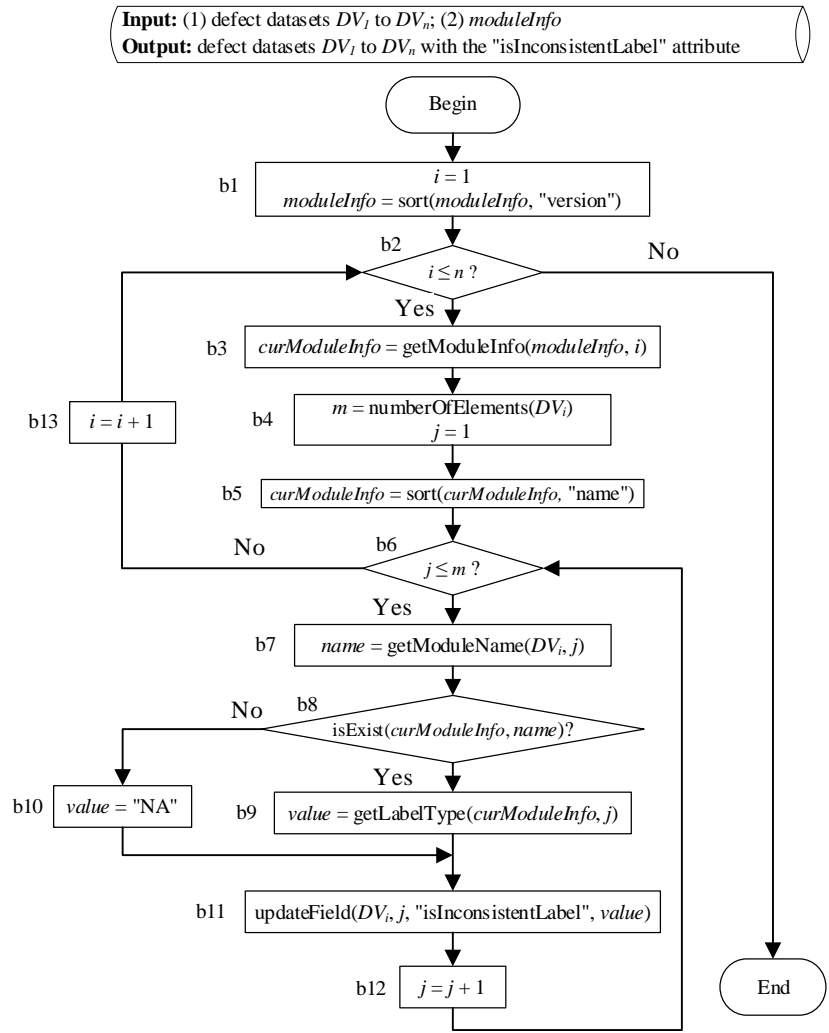

Fig. 18. The flowchart of augmenting the multi-version-project defect data set with inconsistent labels

## 5.4 Time complexity of TSILI

As mentioned above, for TSILI, the inputs are a multi-version-project defect data set (i.e. $DV_1, DV_2, …,$ and $DV_n$) and the corresponding source code databases (i.e. $SV_1, SV_2, …,$ and $SV_n$). In order to facilitate the complexity analysis of TSILI, we assume that: (1) $SV_i$ consists of $s_i$ modules and $DV_i$ consists of $d_i$ instances, $1 \leq i \leq n$; (2) $e_i$ is the number of instances in the intersection of $SV_i$ and $DV_i$, $1 \leq i \leq n$; (3) $s = \max(\{s_i | 1 \leq i \leq n\})$ and $d = \max(\{d_i | 1 \leq i \leq n\})$; and (4) $S = \sum_{i=1}^{n} s_i$ and $D = \sum_{i=1}^{n} d_i$. Let $E$ be the number of elements in the information table *moduleInfo*. Therefore, we have $E = \sum_{i=1}^{n} e_i$, $E \leq S$, and $E \leq D$. Furthermore, assume that $l$ is the number of characters of the module that has the most characters in the source code databases. Note that, for a given software version $V_i$, if the defect data collection process is accurate, we should have $e_i = s_i = d_i$, $1 \leq i \leq n$. In practice, however, due to a variety of unknown reasons, $s_i$ may be different from $d_i$, $1 \leq i \leq n$. In the following, for the simplicity of presentation, the former is called the "IDEAL" condition, while the latter is called the "REAL" condition.

The total time complexity of TSILI is equal to the sum of the time complexities of its three stages.

- Stage 1 has a time complexity of $O(S \times \log(S) + D \times \log(s))$. The first stage consists of three parts: initialize *moduleInfo* (b1), generate *moduleInfo* (b2~b10), and sort *moduleInfo* (b11). The first part has a time complexity of $O(1)$, while the third part has a time complexity of $O(E \times \log(E))$ (e.g. using quick sort). At the second part, each instance is examined one time, in which all statements but b4 and b7 have a complexity of $O(1)$. As for b4, each execution requires $s_i \times \log(s_i)$ comparisons. Since b4 is executed $n$ times, the total number of comparisons is: $s_1 \times \log(s_1) + … + s_n \times \log(s_n) < S \times \log(S)$. As for b7, since $SV_i$ has been sorted, each search is a binary search and requires at most $\log(s)$ comparisons. Since b7 is executed $D$ times, the total number of comparisons is $D \times \log(s)$. Therefore, the time complexity of the second part is at most $O(S \times \log(S) + D \times \log(s))$. As a result, stage 1 at most has a time complexity: $O(1) + O(E \times \log(E)) + O(S \times \log(S) + D \times \log(s)) = O(S \times \log(S) + D \times \log(s))$.

- Stage 2 has a time complexity of $O(E \times \log(n) \times l)$. The second stage consists of two parts: group instances (b1) and identify inconsistent labels (b2~b19). Since *moduleInfo* has already been sorted, the first part has a time complexity of $O(E)$. The second part iterates over each group: if the instances in the current group do not have the same label (b3~b5), then get the code (a filtered string) for each instance (b6~b10), partition instances into equivalence classes by the same code (b11), iterate over each equivalence class to examine the label (b12~b18). For the second part, "formatCode" in b9 and "getIndexPartition" in b11 dominate the time complexity. In "formatCode", the code is parsed to filter out comments, newlines, and extra spaces. For each execution, its time complexity is proportional to the code length (at most $l$). Since "formatCode" is executed at most $E$ times, the total time complexity of "formatCode" in b9 is at most $O(E \times l)$. In "getIndexPartition", the instances in the same group are compared by their codes to obtain equivalence classes. For a group with $x$ instances, the equivalence class partition can be performed by $x \times \log(x) \times l$ comparisons, i.e. each element needs $\log(x) \times l$ comparisons. Since $x$ is at most $n$ and the total number of instances in *moduleInfo* is $E$, the total time complexity of "getIndexPartition" in b11 is $O(E \times \log(n) \times l)$. As a result, stage 2 at most has a time complexity: $O(E) + O(E \times l) + O(E \times \log(n) \times l) = O(E \times \log(n) \times l)$.

- Stage 3 has a time complexity of $O(E \times \log(E) + D \times \log(d))$. The third stage consists of two parts: sort *moduleInfo* (b1) and augment the multi-version-project data set (b2~b13). The first part has a time complexity of $O(E \times \log(E))$. At the second part, each instance in the multi-version-project data set is examined one time, in which "sort" in b5 and "isExist" in b8 dominates the time complexity. For "sort" in b5, each execution requires $e_i \times \log(e_i)$ comparisons. Since b5 is executed $n$ times, the total number of comparisons is: $e_1 \times \log(e_1) + \ldots + e_n \times \log(e_n) < E \times \log(E)$. For "isExist" in b8, since *curModuleInfo* has already been sorted, each search is a binary search and requires at most $\log(d)$ comparisons. Since b8 is executed at most $D$ times, the total number of comparisons is at most $D \times \log(d)$. Therefore, the time complexity of the second part is at most $O(E \times \log(E) + D \times \log(d))$. As a result, stage 3 at most has a time complexity: $O(E \times \log(E)) + O(E \times \log(E) + D \times \log(d)) = O(E \times \log(E) + D \times \log(d))$.

Therefore, under the "REAL" condition, at the worst case, the total time complexity of TSILI is: $O(S \times \log(S) + D \times \log(s)) + O(E \times \log(n) \times l) + O(E \times \log(E) + D \times \log(d)) = O(S \times \log(S) + D \times \log(s) + E \times \log(n) \times l + D \times \log(d))$. Under the "IDEAL" condition, at the worst case, the total time complexity of TSILI will become $O(D \times \log(D) + D \times \log(n) \times l)$ (since $d < D$, $s = d$, and $E = S = D$).

# 6 Experimental design

In this section, we describe our experimental design. First, we list the research questions under investigation. Then, we introduce the used data sets. Finally, we present the data analysis method used to investigate the research questions.

**6.1 Research questions**

We attempt to investigate the following research questions to understand the degree of inconsistent labels in multi-version-project defect data sets and their impact on defect prediction.

***RQ1 (Degree of existence):** What is the degree of existence of inconsistent labels in multi-version-project defect data sets collected by typical defect data collection approaches?*

The purpose of RQ1 is to investigate whether inconsistent labels are common in multi-version-project defect data sets collected by typical defect collection approaches. On the one hand, the current three typical defect data collection approaches (i.e. time-window, affected-version, and SZZ-based) can theoretically lead to the existence of inconsistent labels (as shown in Section 4.2). On the other hand, we do find inconsistent label examples in multi-version-project data sets generated by these approaches (as shown in Section 3). However, it is unclear whether inconsistent labels are widespread. The answer to RQ1 not only enables us to understand the degree of existence of inconsistent labels in defect data sets collected by different approaches, but also provides a perspective for understanding the quality of existing open source defect data sets.

***RQ2 (Influence on prediction performance):** To what extent does the existence of inconsistent labels influence the prediction performance of a defect prediction model?*

The purpose of RQ2 is to investigate the influence of inconsistent labels in the defect data set on the prediction performance of a defect prediction model. As we analyzed in Section 4.3, for a pair of inconsistent labels, they are noise for a defect prediction model, regardless of whether they are mislabels or correct labels. The main reason is that when building a traditional defect prediction model, a pair of instances with inconsistent labels will lead to contradictory relationships between the features of an instance and its class label. Therefore, we want to investigate whether and to what extent inconsistent labels influence the prediction performance of a defect prediction model. The answer to RQ2 enables us to understand, from the viewpoint of prediction performance, whether it is necessary to deal with inconsistent labels before using a defect data set.

*RQ3 (Influence on detected defects): To what extent does the existence of inconsistent labels influence the actual defects detected by a defect prediction model?*

The purpose of RQ3 is to investigate the influence of inconsistent labels on the actual defects detected by a defect prediction model. The difference in prediction performance between two prediction models (built with clean data set and with noise data set) is not enough to fully depict the real influence of inconsistent labels on defect prediction. The reason is that even if they have the same prediction performance scores, the identified actual defective instances (i.e. true positive) may be quite different. Therefore, the influence on detected defective instances can be seen as a deep impact that is not easy to be captured by prediction performance scores. The answer to RQ3 enables us to understand whether dealing with or not dealing with inconsistent labels can influence the ability of a defect prediction model to actually detect defective instances.

*RQ4 (Influence on model interpretation): To what extent does the existence of inconsistent labels influence the interpretation of a defect prediction model?*

The purpose of RQ4 is to investigate the influence of inconsistent labels on the interpretation of a defect prediction model. The feature importance rank obtained from a defect prediction model is a basis for interpreting what features lead to defects, which is essential to derive knowledge for developing quality improvement plans. If the top features in the top importance rank of a defect prediction model are sensitive to inconsistent labels, this means that inconsistent labels will lead to a biased model interpretation. The answer to RQ4 enables us to know whether inconsistent labels should be excluded in future defect prediction studies if the objective is to derive knowledge from a defect prediction model for quality improvement.

**6.2 Multi-version-project defect datasets**

In our study, we used the five multi-version-project defect datasets to conduct the experiment: ECLIPSE-2007 [29], Metrics-Repo-2010 [27, 28], JIRA-HA-2019 [23], JIRA-RA-2019 [23], and MA-SZZ-2020. Of these datasets, the former two datasets are widely used benchmark datasets in the literature, the next two are two recently published data sets in ICSE 2019, while the last one is a dataset collected by ourselves. We took the following steps to obtain the inconsistent label information for each project in each multi-version-project defect dataset. First, for all involved versions of the target project, we downloaded their source codes from the official website. Then, we used "Understand"[16] to parse the codes to generate the source code databases. Finally, with the input of the source code databases and the corresponding defect data sets, the TSILI algorithm was run to identify the inconsistent label information for the target project.

Table 1 summarizes the five multi-version-project datasets after applying the TSILI algorithm. The first column reports the name of the dataset (the used defect label collection approach shown in parentheses, where AV denotes an affected version approach and TW denotes a time window approach). The second to sixth columns, respectively, report the project name, the versions, the number of instances (the number of instances with inconsistent labels identified by TSILI shown in parentheses), the percentage of defective instances, and the number of metrics. The metric information is composed of different software metrics [27, 28, 48-50]. Due to the limitation of space, we present the involved metrics in detail online[17]. Note that, these five multi-version-project datasets are appropriate for our study due to their diverse in the defect label data collection approaches: ranging from time-window, affected version, to (the state-of-the-art) SZZ-based.

---

[16] http://scitools.com
[17] http://github.com/sticeran/InconsistentLabels

- ECLIPSE-2007. This dataset corresponds to one project with three versions. According to [29], an affected version approach was employed to collect the defect label data. After identifying BFCs by matching regular expressions with comments of commits, the first release listed in the "*version*" field of the corresponding issue reports in BUGZILLA was used to link defective modules to versions. During this process, the analyzed issue reports were limited to those reported in the first six months after release.

- Metrics-Repo-2010. This dataset corresponds to 12 projects with 43 versions[18]. According to [27, 28], a time window approach was used to link defective modules to versions. For each target version of interest, a tool called BugInfo was employed to identify its BFCs by regular expression matching. During this process, the time window was set to the period between the release time of the target version and the release time of the next version.

- JIRA-HA-2019. This dataset corresponds to 9 projects with 32 versions. According to [23], a time window approach (Yatish et al. regarded it as a heuristic approach, abbreviated as HA) was used to link defective modules to versions. For each target version of interest, a collection of regular expressions was applied on commit logs to identify BFCs. During this process, the time window was set as a 6-month period after the version of interest was released.

- JIRA-RA-2019. This dataset was collected from the same projects as used in JIRA-HA-2019. However, an affected version approach (rather than a time window approach) was used to link defective modules to versions. In [23], Yatish et al. regarded it as a realistic approach (abbreviated as RA). For each target version of interest, they first retrieved these issue reports in JIRA whose "*affected version*" fields listed the target version as the earliest affected version. Then, they leveraged the traceable links (provided by JIRA) between issue reports and code commits to identify BFCs. During this process, the complete history of the target version after release was analyzed in order to reduce false negative modules.

- MA-SZZ-2020. This dataset corresponds to 5 projects with 50 versions. We first used MA-SZZ [13], the state-of-the-art SZZ variant, to collect BICs. When identifying BICs, the following changes were excluded: (1) non-semantic code changes (e.g. changes of annotations, spaces, and blank lines); (2) format changes (e.g. moving the bracket), and (3) meta-changes, including branch change (copy the project state from one branch to another), merge-change (apply change activity from one branch to another), property change (only impact file properties stored in the VCS). After that, we leveraged BFCs and their corresponding BICs to link defective modules to versions. Similar to [23, 90], we select Apache Java projects with Git VCS and JIRA ITS as the subject projects, because: (1) "Apache projects must have reached a certain level of maturity in order to be considered as a top-level project" [90]; and (2) they have a high (traceable) link rate between issue reports to commits[19]. In particular, they should have the property that a version with a smaller release number has an earlier release time. This property is important for the projects under consideration, as it helps accurately link defective modules to versions. To further ensure the maturity and popularity of projects, a project under consideration was required to have: (1) at least 10 versions; (2) at least 1000 stars on GitHub; and (3) at least 100 issue reports[20], each having a "Bug" *Type*, a "resolved" or "closed" *status*, and a "fixed" *Resolution*. Consequently, we obtained the following five projects: Zeppelin, Shiro, Maven, Flume, and Mahout. For each considered version of the five projects, we used a mature commercial tool called "Understand" to collect 44 code metrics.

---

[18] Note that the original Metrics-Repo-2010 data sets also include Ivy project with 1.1, 1.4, and 2.0 versions. However, we were unable to find their corresponding code from the official website and hence excluded them from our experiment.
[19] According to [102], "Apache developers are meticulous in their efforts to insert bug references in the change logs of the commits".
[20] All the issue reports had a "Created Date" less than "2019-12-03 12:00", as this was the time point we collected the data.

From Table 1, we have the following observations. First, the involved projects cover wide-ranging functions, including integration development environments (e.g. Eclipse), applications (e.g. JEdit), frameworks (e.g. Camel), libraries (e.g. Lucene), and database (e.g. HBase). Second, the number of instances varies greatly across the involved projects. The maximum number of instances is 10590, while the minimum number of instances is only 6. Third, the defect data show a highly imbalanced distribution in the involved projects. For most projects, the percentage of defective instances is well below 20%. Fourth, inconsistent labels in general exist, although the extent varies across the versions in the involved projects.

Table 1. The five multi-version-project defect data sets used in our study

| Dataset | Project | Versions | #Instances (#IL-instances) | %Defective | #Metrics |
|---|---|---|---|---|---|
| ECLIPSE-2007 (AV) | Eclipse | 2.0, 2.1, 3.0 | 6,727~10,590 (99~182) | 11%~15% | 198 |
| Metrics-Repo-2010 (TW) | Ant | 1.3, 1.4, 1.5, 1.6, 1.7 | 124~740 (2~14) | 11%~26% | 20 |
| | Camel | 1.0, 1.2, 1.4, 1.6 | 339~927 (28~123) | 4%~37% | |
| | Forrest | 0.6, 0.7, 0.8 | 6~30 (0~1) | 7%~17% | |
| | Jedit | 3.2, 4.0, 4.1, 4.2, 4.3 | 259~487 (1~18) | 2%~34% | |
| | Log4j | 1.0, 1.1, 1.2 | 103~193 (39~48) | 29%~95% | |
| | Lucene | 2.0, 2.2, 2.4 | 186~330 (55~89) | 49%~61% | |
| | Pbeans | 1.0, 2.0 | 26~51 (1~1) | 20%~77% | |
| | Poi | 1.5, 2.0, 2.5, 3.0 | 235~438 (49~240) | 12%~64% | |
| | Synapse | 1.0, 1.1, 1.2 | 157~256 (10~20) | 10%~34% | |
| | Velocity | 1.4, 1.5, 1.6 | 195~229 (41~81) | 34%~75% | |
| | Xalan | 2.4, 2.5, 2.6, 2.7 | 676~899 (314~481) | 16%~99% | |
| | Xerces | Init, 1.2, 1.3, 1.4 | 162~451 (25~212) | 15%~64% | |
| JIRA-HA-2019 (TW) /JIRA-RA-2019 (AV) | ActiveMQ | 5.0.0, 5.1.0, 5.2.0, 5.3.0, 5.8.0 | 1,884~3,420 (1~136/13~282) | 4%~8% / 6%~16% | 65 |
| | Camel | 1.4.0, 2.9.0, 2.10.0, 2.11.0 | 1,503~8,809 (3~140/4~131) | 2%~24% / 2%~19% | |
| | Derby | 10.2.1.6, 10.3.1.4, 10.5.1.1 | 1,963~2,704 (45~69/174~294) | 7%~9% / 14%~34% | |
| | Groovy | 1.5.7, 1.6.0.Beta 1, 1.6.0.Beta 2 | 756~883 (19~28/20~21) | 11%~14% / 3%~9% | |
| | HBase | 0.94.0, 0.95.0, 0.95.2 | 1,047~1,801 (0~19/0~104) | 6%~7% / 21%~27% | |
| | Hive | 0.9.0, 0.10.0, 0.12.0 | 1,319~2,512 (14~26/49~75) | 2%~4% / 8%~19% | |
| | JRuby | 1.1, 1.4, 1.5, 1.7.0.preview1 | 723~1,551 (10~79/4~17) | 10%~24% / 5%~19% | |
| | Lucene | 2.3.0, 2.9.0, 3.0.0, 3.1.0 | 803~1,802 (16~29/10~32) | 14%~20% / 6%~24% | |
| | Wicket | 1.3.0.beta1, 1.3.0.beta2, 1.5.3 | 1,669~2,570 (0~0/0~51) | 4%~17% / 4%~7% | |
| MA-SZZ-2020 (SZZ-based) | Zeppelin | 0.5.0, 0.5.5, 0.5.6, 0.6.0, 0.6.1, 0.6.2, 0.7.0, 0.7.1, 0.7.2, 0.7.3 | 129~413 (7~25) | 1~35% | 44 |
| | Shiro | 1.1.0, 1.2.0, 1.2.1, 1.2.2, 1.2.3, 1.2.4, 1.2.5, 1.2.6, 1.3.0, 1.3.1 | 381~493 (20~31) | 4~7% | |
| | Maven | 2.2.0, 2.2.1, 3.0.0, 3.0.1, 3.0.2, 3.0.3, 3.0.4, 3.0.5, 3.1.0, 3.1.1 | 318~703 (0~51) | 5~12% | |
| | Flume | 1.2.0, 1.3.0, 1.3.1, 1.4.0, 1.5.0, 1.5.1, 1.5.2, 1.6.0, 1.7.0, 1.8.0 | 272~574 (27~90) | 4~30% | |
| | Mahout | 0.4, 0.5, 0.6, 0.7, 0.8, 0.9, 0.10.0, 0.11.0, 0.12.0, 0.13.0 | 1027~1267 (32~88) | 13~29% | |

### 6.3 Data analysis method

In this section, we describe in detail the method to analyze the four research questions.

#### 6.3.1 Existence analysis of inconsistent labels (RQ1)

For each component defect data set in a multi-version-project dataset, we use the following three indicators to characterize the degree of existence of inconsistent labels:

- *ILinAll*: the ratio of the number of instances having inconsistent labels to the number of all instances in a data set;
- *ILinBuggy*: the ratio of the number of buggy instances having inconsistent labels to the number of all buggy instances in a data set;
- *ILinClean*: the ratio of the number of clean instances having inconsistent labels to the number of all clean instances in a data set.

For inconsistent labels, *ILinAll* characterizes their overall proportion, while *ILinBuggy* and *ILinClean*, respectively,

characterize their proportion on buggy and clean instances. In general, in a defect data set, the number of buggy instances is far less than the number of clean instances. If *ILinBuggy* has a large value, this means that the number of quality buggy instances may be not enough to build a good defect prediction model. In this sense, of the three indicators, *ILinBuggy* is the most important indicator to depict the degree of existence of inconsistent labels.

**6.3.2 Influence analysis on prediction performance (RQ2)**

**Defect prediction model construction**. Let $S$ be a data set containing the instances with inconsistent labels. How to evaluate the influence of the label noise introduced by inconsistent labels on the performance of a defect prediction model built on $S$? In our study, we first generate a cleaned data set $S'$ from $S$ by removing all the instances with inconsistent labels (denoted as $S' = $ clean($S$)). Then, we train two models: one on $S$ and another one on $S'$. For the simplicity of presentation, the former model will be called NC (built with **N**oise training data and applied to **C**lean test data), while the latter model will be called CC (built with **C**lean training data and applied to **C**lean test data). After that, we apply them to predict defective modules in the same inconsistent-label-free test data set $T$. The reason to keep the test data $T$ clean is to avoid the interference of label noise to performance evaluation. By comparing the performance scores of NC and CC, we can get an understanding to what extent inconsistent labels influence on the performance of defect prediction. The method we take to generate the clean model is the same as that used in literature [51, 52]. As stated in literatures [51-54], comparing model CC with model NC is a commonly used method to evaluate how label noises influence the performance of a defect prediction model.

In order to enable the building of NC and CC models, we need to generate such training and test data set pairs: ($S$, $T$) and ($S'$, $T$). In our study, to comprehensively analyze the influence of inconsistent labels, we build NC and CC models in the following contexts:

- Cross-version defect prediction (CVDP). In the CVDP context, prediction models are built on historical versions within a project and then applied to predict defective modules in the current version of this project. Assume that a project has $n$ versions. If the $k$th version is used as the test data set, then the recent $i$ historical version(s) will be used as the training data set, $1 \leq i \leq k-1$. Consequently, for this project, we will have $n \times (n-1)/2$ candidate pairs of training set and testing sets in total.

- Cross-project defect prediction (CPDP). In the CPDP context, defect prediction models are built on source projects and then applied to a target project, i.e. the training set and the test set are from different projects. In our study, we use a strict CPDP setting [63] to generate the candidate pairs of training set and test set: the current version of a target project is used as the test set, and all versions of other projects are used as the training set.

In both CVDP and CPDP, for each candidate pair (*CS*, *CT*), if *CS* contains inconsistent labels and both *CS* and *CT* have more than 10 defective instances, then we generate ($S$, $T$) and ($S'$, $T$) as follows: $S = CS$, $S' = $ clean(*CS*), and $T = $ clean(*CT*).

In our study, we use Random Forest (RF) with default parameters [68] to build defect prediction models. The reasons for this choice are two-fold. First, RF has been shown as a top-performing modeling technique in many studies [69-72]. Furthermore, Tantithamthavorn et al. reported that, of 26 classifiers with parameter optimization, the RF model with default parameters ranked in the second of a total of 14 ranking groups [73]. Second, RF can report the importance of each feature in a model, thus enabling us to analyze the influence of inconsistent labels on the interpretation of a defect prediction model (RQ4). In particular, we follow the established practices [31, 32, 54, 65] to build a RF model: (1) use log-transformation ($\log(x+1)$) to mitigate the skewed metric distributions [64]; (2)

employ CFS (correlation-based feature selection) [66] to remove redundant metrics, and (3) take SMOTE [67] to rebalance the imbalanced training data.

**Prediction performance evaluation.** We evaluate the prediction performance of a defect prediction model under two scenarios: classification and ranking. For a given test data set, assume that: (1) $n_0$ is the number of clean instances; (2) $n_1$ is the number of defective instances; (3) $N$ is the total number of instances; and (4) $q$ is the actual defect percentage. Then, we have $N = n_0 + n_1$ and $q = n_1/N$. For each instance in this test set, a RF defect prediction model can output a probability of being defective. In the classification scenario, a threshold (0.5 in default) is chosen to classify an instance as defective if the predicted probability is larger than the threshold and otherwise not defective. Consequently, there are four outcomes: *TP* (the set of modules correctly classified as defective), *TN* (the set of modules correctly classified as not defective), *FP* (the set of modules incorrectly classified as defective), and *FN* (the set of modules incorrectly classified as not defective). Clearly, $N = |TP| + |FP| + |TN| + |FN|$, and $q = (|TP| + |FP|) / N$. In our study, we use the following indicators to evaluate the classification performance of a RF model.

- $F_1$ [55]: the harmonic mean of precision (i.e. $p = |TP| / (|TP|+|FP|)$) and recall (i.e. $r = |TP| / (|TP|+|FN|)$), i.e. $2 \times p \times r / (p + r)$.
- *AUC* [56]: the area under ROC (Receiver Operating Characteristic) curve, which is defined on a 2-dimensional plot in which the *x*-axis is TPR (i.e. $|TP|/(|TP|+|FN|)$) and the *y*-axis is the FPR (i.e. $|FP|/(|TN|+|FP|)$).
- *ER* (Effort Reduction) [60]: the proportion of the reduced number of modules to be inspected (i.e., effort) compared with a random model that achieves the same recall. According to the prediction model, $|TP|+|FP|$ modules will be inspected. According to a random model that achieves the same recall, $|TP|/(|TP|+|FP|) \times N$ modules will be inspected. Therefore, the reduced effort is:

$$ER = \frac{|TP|/(|TP|+|FP|) \times N - (|TP|+|FP|)}{|TP|/(|TP|+|FP|) \times N} = 1 - \frac{q}{r}$$

- *RI* (Recall Increase) [60]: the proportion of the increased recall compared with a random model when inspecting the same number of instances (i.e., the same effort). According to a random model, inspecting $|TP|+|FP|$ modules will lead to a recall of $(|TP|+|FP|)/N$. Therefore, the recall increase is:

$$RI = \frac{r - (|TP|+|FP|)/N}{(|TP|+|FP|)/N} = \frac{p}{q} - 1$$

Of the above four indicators, $F_1$ and AUC are two popular classification performance indicators in the literature. However, they are non-effort-aware. In contrast, ER and RI are two effort-aware indicators. For each indicator, a larger value means a better performance.

In the ranking scenario, the instances in the test data set are ranked in descending order according to their predicted probability of being defective. In our study, we use the following indicators to evaluate the ranking performance of a RF model.

- *AP* (Average Precision): the average precision of defective instances. Let $p(k)$ be the precision calculated by considering only the ranked instances from rank 1 through *k*. Assume that *rel(k)* indicates if the *k*th instance is defective (*rel(k)* = 1) or not (*rel(k)* = 0). Then, $AP = \frac{\sum_{k=1}^{N} p(k) \times rel(k)}{n_1}$.
- *RR* (Reciprocal Rank): the reciprocal of the rank of the first defective instance.
- $P_{opt}$ [31, 61, 62]: the area under the cost-effectiveness curve normalized to the optimal and worst models in an Alberg diagraph. In such a diagraph, the cumulative percentage of SLOC of the top modules selected from the instance ranking (the *x*-axis) is plotted against the cumulative percentage of defects found (the *y*-axis). Formally, for a model *m*, $P_{opt}$ can be defined as:

$$P_{opt} = 1 - \frac{Area(optimal) - Area(m)}{Area(optimal) - Area(worst)}$$

Here, *Area*(*m*), *Area*(*optimal*) and *Area*(*worst*) represent the area under the curve corresponding to the prediction model, the optimal model, and the worst model, respectively. In the optimal model and the worst model, instances are, respectively, ranked in decreasing and ascending order according to their actual defect density.

- *ACC* [31]: the recall of defective instances when using 20% of the entire effort required to inspect all instances (i.e. 20% of the total SLOC) to inspect the top ranked instances.

Of the above four indicators, *AP* and *RR* are two non-effort-aware indicators, in which the order of defective instances in a ranking is considered. They are originally from the field information retrieval [57] but appropriate for evaluating the ranking performance of a defect prediction model. In particular, *RR* = 1 / (*IFA* + 1), where *IFA* is a recently proposed indicator [58] to characterize the number of Initial False Alarms encountered before the first defective instance is found in a rank. In contrast, $P_{opt}$ and *ACC* are two popular effort-aware indicators, in which the module size corresponding to an instance is used as the proxy of the effort to inspect the instance. For each indicator, a larger value means a better performance.

**Statistical performance comparison**. Let *perf*(NC) be the performance of NC and *perf*(CC) be the performance of CC. In our context, the performance measure *perf* can be $F_1$, *ACC*, *ER*, *RI*, *AP*, *RR*, $P_{opt}$, or *ACC*. For a pair of training and test set (*S*, *T*), a nature idea is to use the absolute value of *diff* to evaluate the influence of inconsistent labels in *S* on prediction performance [59]:

$$diff = \frac{perf(NC) - perf(CC)}{perf(CC)} \times 100\%$$

As can be seen, a positive *diff* means that inconsistent labels lead to an optimistic performance, while a negative *diff* means that inconsistent labels lead to a conservative performance. No matter whether the *diff* is positive or negative, it is clear that the smaller the absolute value of *diff* is, the less the influence is. However, the above analysis only considers the absolute performance *perf*. In practice, for a given test set *T*, it is easy to use a random model *random* to predict defect-proneness. For practitioners, the premise of using a model is that it should have a higher performance than *random*. In this sense, it is more important for practitioners to evaluate *perfGain*, the relative performance of a model with respect to *random* [74]. In our context, *perfGain*(NC) = *perf*(NC) − *perf*(*random*) and *perfGain*(CC) = *perf*(CC) − *perf*(*random*). Therefore, in this study, we also employ the absolute value of *pgr* (performance gain ratio) to evaluate the influence of inconsistent labels in *S* on prediction performance:

$$pgr = \frac{perfGain(NC) - perfGain(CC)}{perfGain(CC)} \times 100\%$$

Assume that a test set *T* consists of *N* instances, in which $n_1$ are defective. For a random model *random*, an instance will have an equal probability (i.e. 0.5) to be predicted as clean or defective. Consequently, we can conclude that[21]:

- $F_1(random) = \frac{2 \times n_1/N \times 0.5}{n_1/N + 0.5}$
- *AUC(random)* = 0.5
- *ER(random)* = 0
- *RI(random)* = 0
- $AP(random) = \frac{1}{N}\sum_{i=1}^{N}\sum_{k=1}^{i}\left(\frac{C_{i-1}^{k-1} \times C_{N-i}^{n_1-k}}{C_N^{n_1}} \times \frac{k}{i}\right)$

---

[21] Their proofs are available in an online appendix via http://github.com/sticeran/InconsistentLabels.

- $RR(random) = \sum_{i=1}^{N-n_1+1} \left( \frac{1}{i} \times \frac{C_{N-i}^{n_1-1}}{C_N^{n_1}} \right)$
- $P_{opt}(random) = 0.5$
- $ACC(random) = 0.2$

After obtaining |diff| and |pgr| for each pair of training and test set (S, T), we can analyze their distributions to understand the influence of inconsistent labels on prediction performance. In particular, we use bootstrapping [77] (the number of bootstrap replicates is set to 1000), a nonparametric technique, to obtain the 95% confidence interval for the mean |diff| and |pgr|. This will help determine whether the influence of inconsistent labels is statistically significant or not: if 0 is not in the interval, it is significant and otherwise not significant. In the context of effort estimate, Briand et al. used the bootstrapping technique to obtain the 95% confidence interval for MRE (Magnitude of Relative Error, similar to |diff|) [78].

Note that, it seems that we may follow previous studies [23, 32] to employ the following two-step approach to analyze the influence of label noise (i.e. inconsistent labels in our context). At the first step, use the Wilcoxon signed-rank test with a Benjamini-Hochberg corrected p-value [75] to examine whether the performance difference between NC and CC is statistically significant at the 95% confidence level (i.e., p-value ≤ 0.05). At the second step, use the Cliff's $\delta$ to compute the effect size [76] to examine whether the performance difference is practically important ($|\delta| < 0.147$: negligible, $0.147 \leq |\delta| < 0.33$: small, $0.33 \leq |\delta| < 0.474$: moderate, $|\delta| \geq 0.474$: large). However, as pointed out by Twomey and Viljoen [79], such an analysis approach may be unable to uncover the real difference that we purport to investigate. The reason is that, in nature, the Wilcoxon signed-rank test and the Cliff's $\delta$ examine whether and to what extent the median of the differences, rather than individual difference, is different from zero. For example, assume that, on 20 pairs of training and test sets, NC has the following 20 $F_1$ values:

(0.9, 0.9, 0.9, 0.9, 0.9, 0.9, 0.9, 0.9, 0.9, 0.9, 0.1, 0.1, 0.1, 0.1, 0.1, 0.1, 0.1, 0.1, 0.1, 0.1),

while CC has the following 20 $F_1$ values:

(0.1, 0.1, 0.1, 0.1, 0.1, 0.1, 0.1, 0.1, 0.1, 0.1, 0.9, 0.9, 0.9, 0.9, 0.9, 0.9, 0.9, 0.9, 0.9, 0.9).

Clearly, NC and CC have a large performance difference on each pair of training and test sets, indicating that that inconsistent labels have a large influence. However, the Wilcoxon signed-rank test will report a *p-value* of 1, while the Cliff's $\delta$ will report an effect size of 0. In other words, the real performance difference is unable to be discovered. Indeed, Smucker et al. further pointed out that the Wilcoxon signed-rank test not only had a poor ability to detect significance but also had the potential to lead to false detections of significance [80]. As a result, they recommend that its use should be discontinued for measuring the significance of a difference. In summary, it is inappropriate to use the above two-step approach for our purpose.

### 6.3.3 Influence analysis on detected defects (RQ3)

After applying a prediction model to a target test set *T*, a set of instances, denoted as PAD (Predicted As Defective), will be recommended to developers for quality assurance. Under the classification scenario, PAD consists of the instances whose predicted probabilities of being defective are larger than a given classification threshold. Under the ranking scenario, PAD consists of the top ranked instances whose ranks are above a given cut-off threshold. In either scenario, the instances in PAD consist of true positive (TP) and false positive (FP), i.e. PAD = TP ∪ FP.

In order to analyze the influence of inconsistent labels on detected defective instances, we compare the identified defective instances by NC and CC. Specifically, after applying NC and CC to a target test set *T*, we have two corresponding true positive sets TP(NC) and TP(CC). Based on TP(NC) and TP(CC), we use, DTP (Difference in

True Positives), to quantify the extent of the difference in identified defective instances:

$$DTP = 1 - \frac{|TP(CC) \cap TP(NC)|}{|TP(CC) \cup TP(NC)|}$$

As can be seen, the range of DTP is between 0 and 1. On the one hand, if DTP has a value of 0, this will mean that TP(NC) is identical to TP(NC), i.e. inconsistent labels have no influence on the identified defective instances. On the other hand, if DTP has a value of 1, this will mean that TP(CC) and TP(NC) have an empty intersection, i.e. inconsistent labels lead to the identification of totally different defective instances. Furthermore, for a given DTP, we employ bootstrapping [77] (the number of bootstrap replicates is set to 1000) to obtain the corresponding 95 confidence interval. This will help determine whether the difference in true positives between NC and CC is statistically significant or not. If it is significant, this will mean that inconsistent labels have an important influence on detected defects and otherwise not.

In our study, under the classification scenario, PAD is obtained by a default classification threshold 0.5. Under the ranking scenario, we consider PADs obtained by the following representative cut-off thresholds: (1) the top ranked instances account for 20% of the total SLOC; (2) the top 10% ranked instances; (3) the top 20% ranked instances; and (4) the top 30% ranked instances. In total, we will report DTPs and their significance under five cases: one for classification and four for ranking. For the simplicity of presentation, the five cases in RQ3 are, respectively, abbreviated as: binary, 20% size, top 10%, top 20%, and top 30%. This will enable to get a comprehensive understanding on the influence of inconsistent labels on the defective instances detected.

**6.3.4 Influence analysis on model interpretation (RQ4)**

In order to evaluate the influence of inconsistent labels on model interpretation, we analyze how the ranks of the top 3 important features in a CC model change compared with those in the corresponding NC model. Since CC and NC are built with random forest, each involved feature will have a rank according to its Breiman's variable importance score. In CC, the top 3 features are the most important features for understanding how the inputs affect the output. In this context, the rank shift of the top 3 features in CC with respect to NC can depict to what extent inconsistent labels affect model interpretation. Similar to [54], for a feature $f$ at rank $k$ in CC, we take the following way to compute the corresponding rank shift:

$$\text{shift}(k) = k - \text{rank}(f, NC)$$

where rank($f$, NC) denotes the rank of $f$ in NC. For example, if a feature $f$ has a rank of 1 in CC and has a rank of 3 in NC, we will have shift(1) = 1 − 3 = −2. Note that, due to the use of feature selection in model building, it is possible that a feature $f$ appears at rank $k$ in CC but does not appear in NC. In this case, we assign a value of "OOM" (out of model) to shift($k$). Given multiple pairs of NC and CC, we can analyze the distribution of shift($k$) ($k$ = 1, 2, and 3) to understand to what extent inconsistent labels influence the interpretation of defect prediction models.

Recently, Tantithamthavorn et al. suggested that feature selection and class rebalancing techniques should be avoided when interpreting defect prediction models [81]. However, it is the current practice that interpreting the contributions of features to defect prediction based on a model trained with feature selection and/or class rebalancing techniques [32, 82-85]. From the viewpoint of practical application, feature selection and/or class rebalancing techniques can help build a highly accurate model. After obtaining a highly accurate model, it is natural for practitioners to ask what features in this model make important contributions. For example, which features drive predictions? What features are not worth the money and time to collect? In this sense, there is a need to interpret this model for understanding the mechanism behind the data [86]. With such knowledge in hand, practitioners may derive a quality improvement plan. If we follow Tantithamthavorn et al.'s suggestion, another model built without feature selection and class rebalancing techniques will be used to interpret the relationships between features and defect prediction. Consequently, two separate models will be used for prediction and model interpretation, which is inconsistent with the current practice in defect prediction community. In this case, the learned relationships may not interpret why the prediction model has a high accuracy. In the light of the above facts, our study conducts a rank shift analysis with the prediction models CC and NC.

## 7. Experimental results

In this section, we report in detail the experimental results with respect to the four research questions.

**7.1 RQ1: Degree of existence of inconsistent defect labels**

Fig. 19 reports the distribution of three types of inconsistent label ratios (i.e. *ILinAll*, *ILinBuggy*, and *ILinClean*) in five multi-version-project defect data sets, in which each point corresponds to a version. Table 2 summarizes the detailed statistical information. In Table 2, the 2nd, 4th, and 6th columns, respectively, report the ratio of the number of versions with *ILinAll* ≠ 0, *ILinBuggy* ≠ 0, and *ILinClean* ≠ 0 to the total number of versions. The 3rd, 5th, and 7th columns, respectively, report the average *ILinAll*, *ILinBuggy*, *ILinClean* as well as their range (shown in parentheses) over all the versions in a multi-version-project defect data set. In particular, the row "Average ratio" reports the average inconsistent label ratio and the row "All versions" reports the ratio of the number of versions with inconsistent labels over all the versions in the five multi-version-project defect data sets.

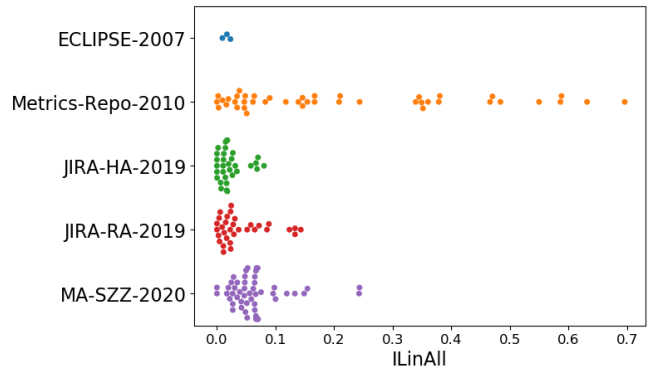

(a) Inconsistent labels in all instances

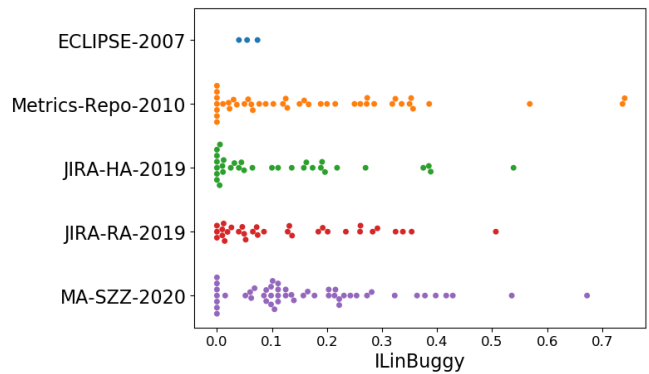

(b) Inconsistent labels in buggy instances

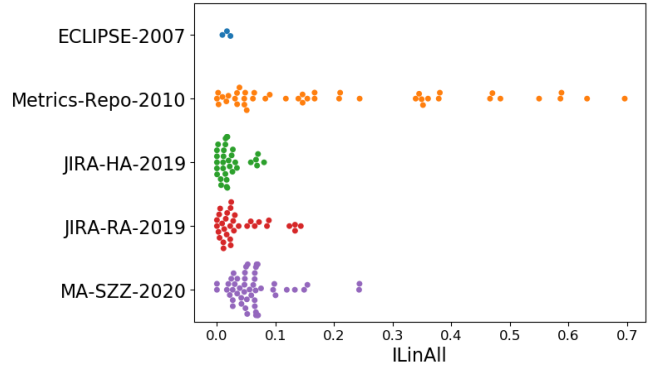

(c) Inconsistent labels in clean instances

Fig. 19. The distribution of inconsistent labels in five multi-version-project defect data sets

Table 2. Statistical information of inconsistent labels for five multi-version-project defect data sets

| Dataset | $\frac{\text{\#versions(ILinALL} \neq 0)}{\text{\#versions}}$ | Average ILinAll (range) | $\frac{\text{\#versions(ILinBuggy} \neq 0)}{\text{\#versions}}$ | Average ILinBuggy (range) | $\frac{\text{\#versions(ILinClean} \neq 0)}{\text{\#versions}}$ | Average ILinClean (range) |
|---|---|---|---|---|---|---|
| ECLIPSE-2007 | $\frac{3}{3}$ | 2% (1%~2%) | $\frac{3}{3}$ | 6% (4%~7%) | $\frac{3}{3}$ | 1% (0.1%~2%) |
| Metrics-Repo-2010 | $\frac{42}{43}$ | 22% (0~70%) | $\frac{36}{43}$ | 18% (0~74%) | $\frac{38}{43}$ | 22% (0~75%) |
| JIRA-HA-2019 | $\frac{27}{32}$ | 2% (0~8%) | $\frac{26}{32}$ | 12% (0~54%) | $\frac{27}{32}$ | 1% (0~8%) |
| JIRA-RA-2019 | $\frac{30}{32}$ | 4% (0~14%) | $\frac{29}{32}$ | 14% (0~51%) | $\frac{29}{32}$ | 3% (0~13%) |
| MA-SZZ-2020 | $\frac{48}{50}$ | 6% (0~24%) | $\frac{43}{50}$ | 17% (0~67%) | $\frac{48}{50}$ | 4% (0~22%) |
| Average ratio | | 9% | | 16% | | 8% |
| All versions | $\frac{150}{160}$ | | $\frac{137}{160}$ | | $\frac{145}{160}$ | |

From Fig. 19 and Table 2, we have the following observations. *First, inconsistent labels exist on almost all the versions*. Overall, 94% (150/160) versions contain inconsistent labels (i.e. *ILinAll* ≠ 0). On these versions, the average ratio of inconsistent labels is 9%, while the largest ratio of inconsistent labels is 70%. If only buggy instances are considered, we can find that 86% (137/160) versions contain inconsistent labels (i.e. *ILinBuggy* ≠ 0). On average, each version has 16% instances with inconsistent labels. In particular, the largest inconsistent label ratio reaches 74%. If only clean instances are considered, a similar phenomenon can be observed. The above results indicate that inconsistent labels are ubiquitous in multi-version-project defect data sets, regardless of whether buggy or clean instances are considered. *Second, all the investigated defect data collection approaches lead to inconsistent labels*. From ECLIPSE-2007 and JIRA-RA-2019, we can see that the affected version approach introduces inconsistent labels to almost all the versions. Under the worst case, the ratio of inconsistent labels is up to 51% on buggy instances and is up to 13% on clean instances. Similar phenomena can be observed for the time window approach (as shown in Metrics-Repo-2010 and JIRA-HA-2019) and for the SZZ-based approach (as shown in MA-SZZ-2020). This shows that all the existing common defect collection approaches cannot avoid the generation of inconsistent labels. *Third, buggy instances have a higher proportion of inconsistent labels than clean instances*. On average, for each version, 16% of buggy instances contain inconsistent labels, while 8% of clean instances contain inconsistent labels. In other words, the degree of inconsistent labels on buggy instances is twice as high as that on clean instances. If we look further into the matter, we can find that, for all the multi-version-project defect datasets except Metrics-Repo-2010, the largest inconsistent label ratio on buggy instances is 3~6 times as high as that on clean instances. This shows that inconsistent labels have a larger influence on buggy instances compared with clean instances. This is understandable, if we take into account the fact that the number of buggy instances in a version is often much less than the number of clean instances.

In addition to the above observations, we have another interesting finding: JIRA-RA-2019 exhibits a higher inconsistent label ratio than JIRA-HA-2019, regardless of whether *ILinAll*, *ILinBuggy*, or *ILinClean* is considered. This reveals that, an affected version approach does not necessarily lead to a higher quality defect data set than a time window approach. In [23], Yatish et al. believed that an affected version approach was a realistic approach and hence was more accurate than a time-window approach. In particular, they used JIRA-RA-2019 (generated by an affected version approach) as the ground truth to investigate the label noises in JIRA-HA-2019 (generated by a time window approach). Consequently, they concluded that JIRA-HA-2019 contained considerable label noises. However, according to Table 2, there are a large number of label noises (i.e. inconsistent labels) in JIRA-RA-2019. Therefore, it is questionable to use JIRA-RA-2019 as the ground truth to investigate the label noises in JIRA-HA-2019. In this sense, the reliability of an affected version approach may be overestimated with respect to a time window approach, which needs to be further investigated in the future.

***Conclusion.*** *Almost all the versions in multi-version-project defect data sets contain inconsistent labels, regardless of which defect data collection approach is considered. For inconsistent labels in buggy instances, the average ratio is 16%, while the largest ratio is around 70%. This unexpected high ratio raises threats to the validity and the reliability of any prediction models built with these defect data sets.*

## 7.2 RQ2: Influence of inconsistent defect labels on prediction performance

Table 3 and Table 4, respectively, report the distributions of |*diff*| and |*pgr*| between model NC and model CC with respect to eight performance evaluation indicators (i.e., $F_1$, $AUC$, $ER$, $RI$, $AP$, $RR$, $P_{opt}$, and $ACC$) in the CVDP context. Table 5 and Table 6, respectively, report the same in the CPDP context. In each table, the first column lists the multi-version-project defect data set under investigation. The 2nd to 9th columns, respectively, list the evaluation indicator used in calculating *diff* or *pgr*. For each data set, there are four rows of results: the 1st (Mean) and 2nd (StdDev) rows, respectively, report the mean value and standard deviation of *diff* (or *pgr*) under 1000 bootstrap sampling; the 3rd and 4th rows (95% CI [LL, UL]), respectively, report the 95% confidence interval of *diff* (or *pgr*), where LL is the lower limit of the confidence interval and UL is the upper limit of the confidence interval. In particular, the values in shown in the yellow background in the table indicate the maximum mean value of *diff* or *pgr* in five multi-version-project defect data sets. The last row (Average mean) in the table is the average of mean value of *diff* or *pgr* in five multi-version-project defect data sets. Note that since Eclipse-2007 only consists of the data from three versions from one project, CPDP is not applicable.

Table 3. Distribution of the prediction performance |*diff*| in CVDP

| Dataset | | Classification scenario | | | | Ranking scenario | | | |
|---|---|---|---|---|---|---|---|---|---|
| | | $F_1$ | $AUC$ | $ER$ | $RI$ | $AP$ | $RR$ | $P_{opt}$ | $ACC$ |
| Eclipse-2007 | Mean | 1.94 | 0.31 | 0.61 | 1.72 | 2.48 | **65.92** | 1.1 | 5.18 |
| | StdDev | 1.71 | 0.29 | 0.23 | 0.62 | 1.4 | 15.55 | 0.58 | 4 |
| | 95% CI | 0.82 | 0.04 | 0.48 | 1.34 | 1.63 | 50 | 0.43 | 1.66 |
| | [LL, UL] | 2.97 | 0.5 | 0.75 | 2.09 | 3.3 | 76.28 | 1.45 | 7.8 |
| Metrics-Repo-2010 | Mean | **8.51** | **5.8** | **17.48** | **19.95** | 7.34 | 34.34 | **7.78** | **28.73** |
| | StdDev | 12.36 | 5.88 | 41.76 | 24.84 | 8.06 | 84.85 | 7.52 | 34.52 |
| | 95% CI | 6.23 | 4.51 | 10.9 | 14.68 | 5.54 | 18.36 | 6.11 | 21.76 |
| | [LL, UL] | 13.87 | 7.51 | 40.02 | 28.02 | 9.63 | 67.29 | 10.03 | 40.97 |
| JIRA-HA-2019 | Mean | 9.35 | 1.57 | 3.53 | 15.43 | **12.77** | 43.14 | 3.93 | 17.76 |
| | StdDev | 7.73 | 1.42 | 3.23 | 13.1 | 17.48 | 93.47 | 3.19 | 19.23 |
| | 95% CI | 7.5 | 1.2 | 2.73 | 12.15 | 9.43 | 21.94 | 3 | 12.77 |
| | [LL, UL] | 12.71 | 2.13 | 4.75 | 20.6 | 22.74 | 84.65 | 4.96 | 25 |
| JIRA-RA-2019 | Mean | 5.95 | 1.43 | 3.07 | 11.64 | 7.89 | 59.54 | 3.17 | 9.15 |

| Dataset | | | | | | | | | |
|---|---|---|---|---|---|---|---|---|---|
| | | StdDev | 5.3 | 1.23 | 3.31 | 11.19 | 6.15 | 124.37 | 3.72 | 8.09 |
| | | 95% CI | 4.46 | 1.1 | 2.32 | 8.86 | 6.15 | 34.37 | 2.29 | 7.26 |
| | | [LL, UL] | 7.74 | 1.86 | 4.4 | 15.71 | 9.85 | 121.86 | 4.65 | 12.69 |
| MA-SZZ-2020 | | Mean | 5.48 | 0.91 | 1.36 | 9.12 | 5.24 | 9.53 | 2.39 | 9.8 |
| | | StdDev | 6.98 | 2 | 1.48 | 11.13 | 9.74 | 27.84 | 2.94 | 12.03 |
| | | 95% CI | 4.66 | 0.7 | 1.17 | 7.68 | 4.1 | 6.16 | 2.05 | 8.17 |
| | | [LL, UL] | 6.49 | 1.3 | 1.57 | 10.69 | 6.82 | 13.67 | 2.82 | 11.57 |
| Average mean | | | 6.25 | 2 | 5.21 | 11.57 | 7.14 | 42.49 | 3.67 | 14.12 |

Table 4. Distribution of the absolute performance gain ratio |*pgr*| in CVDP

| Dataset | | Classification scenario | | | | Ranking scenario | | | |
|---|---|---|---|---|---|---|---|---|---|
| | | $F_1$ | AUC | ER | RI | AP | RR | $P_{opt}$ | ACC |
| Eclipse-2007 | Mean | 5.29 | 0.89 | 0.61 | 1.72 | 2.62 | 113.24 | 6.57 | 9.5 |
| | StdDev | 4.5 | 0.83 | 0.23 | 0.62 | 1.45 | 31.11 | 3.27 | 8.36 |
| | 95% CI | 1.71 | 0.12 | 0.48 | 1.34 | 1.75 | 91.1 | 2.87 | 2.6 |
| | [LL, UL] | 8.17 | 1.44 | 0.75 | 2.09 | 3.48 | 132.47 | 8.63 | 14.9 |
| Metrics-Repo-2010 | Mean | **49.47** | **17.75** | **17.48** | **19.95** | 10.37 | 100.64 | **139.03** | 112.37 |
| | StdDev | 90.72 | 18.42 | 41.76 | 24.84 | 10.49 | 233.46 | 556.91 | 165.87 |
| | 95% CI | 31.47 | 13.62 | 10.9 | 14.68 | 8.04 | 55.56 | 53.93 | 81.58 |
| | [LL, UL] | 82.35 | 23.26 | 40.02 | 28.02 | 13.63 | 177.61 | 516.17 | 179.31 |
| JIRA-HA-2019 | Mean | 14.49 | 3.85 | 3.53 | 15.43 | **12.97** | 81.83 | 30.43 | **140.81** |
| | StdDev | 12.48 | 3.55 | 3.23 | 13.1 | 17.58 | 220.75 | 35.87 | 381.65 |
| | 95% CI | 11.8 | 2.9 | 2.73 | 12.15 | 9.59 | 35.41 | 21.41 | 60.25 |
| | [LL, UL] | 20.47 | 5.25 | 4.75 | 20.6 | 22.94 | 201.88 | 43.71 | 358.93 |
| JIRA-RA-2019 | Mean | 9.71 | 3.57 | 3.07 | 11.64 | 8.04 | **181.09** | 23.62 | 42.15 |
| | StdDev | 8.51 | 3.19 | 3.31 | 11.19 | 6.21 | 629.16 | 40.5 | 61.37 |
| | 95% CI | 7.45 | 2.73 | 2.32 | 8.86 | 6.26 | 70.3 | 15 | 29.3 |
| | [LL, UL] | 12.8 | 4.68 | 4.4 | 15.71 | 10.01 | 591.99 | 41.55 | 73.49 |
| MA-SZZ-2020 | Mean | 7.46 | 1.94 | 1.36 | 9.12 | 5.37 | 58.66 | 6.38 | 20.85 |
| | StdDev | 9.16 | 4.19 | 1.48 | 11.13 | 9.88 | 368.66 | 8.85 | 31.7 |
| | 95% CI | 6.33 | 1.49 | 1.17 | 7.68 | 4.2 | 22.9 | 5.33 | 16.95 |
| | [LL, UL] | 8.8 | 2.75 | 1.57 | 10.69 | 6.97 | 150.44 | 7.85 | 26.13 |
| Average mean | | 17.28 | 5.6 | 5.21 | 11.57 | 7.87 | 107.09 | 41.21 | 65.14 |

Table 5. Distribution of the prediction performance |*diff*| in CPDP

| Dataset | | Classification scenario | | | | Ranking scenario | | | |
|---|---|---|---|---|---|---|---|---|---|
| | | $F_1$ | AUC | ER | RI | AP | RR | $P_{opt}$ | ACC |
| Metrics-Repo-2010 | Mean | 7.44 | **3.62** | 18.36 | 25.43 | **9.3** | 19.17 | **10.49** | 16.23 |
| | StdDev | 7.63 | 4.9 | 26.37 | 30.44 | 25.83 | 58.78 | 14.88 | 16.65 |
| | 95% CI | 5.52 | 2.53 | 12.31 | 18.03 | 4.47 | 6.25 | 7.01 | 11.62 |
| | [LL, UL] | 10.4 | 6 | 30.25 | 37.85 | 26.88 | 50.42 | 17.16 | 22.02 |
| JIRA-HA-2019 | Mean | 4.57 | 1.12 | 2.04 | 7.6 | 4.99 | 19.27 | 2.59 | 9.2 |
| | StdDev | 3.09 | 0.83 | 1.58 | 4.99 | 4.65 | 44.02 | 2.16 | 11.06 |
| | 95% CI | 3.68 | 0.86 | 1.6 | 6.01 | 3.57 | 8.33 | 1.91 | 6.68 |
| | [LL, UL] | 5.75 | 1.43 | 2.65 | 9.33 | 6.97 | 43.92 | 3.42 | 15.81 |
| JIRA-RA-2019 | Mean | 2.78 | 0.87 | 1.39 | 4.99 | 3.95 | 15.89 | 4.57 | 12.42 |
| | StdDev | 2.44 | 1.14 | 1.3 | 4.15 | 3.44 | 39.73 | 8.3 | 18.18 |
| | 95% CI | 2.11 | 0.58 | 1 | 3.71 | 2.97 | 6.77 | 2.44 | 8.04 |
| | [LL, UL] | 3.83 | 1.52 | 1.91 | 6.52 | 5.29 | 38.11 | 8.05 | 20.52 |
| MA-SZZ-2020 | Mean | **18.16** | 2.83 | **31.1** | **35.22** | 8.69 | **29.77** | 3.92 | **29.05** |
| | StdDev | 20.51 | 1.74 | 55.55 | 48.86 | 7.15 | 33.04 | 2.68 | 33.88 |
| | 95% CI | 13.34 | 2.34 | 19.49 | 24.8 | 6.69 | 21.48 | 3.24 | 21.56 |
| | [LL, UL] | 25.78 | 3.32 | 53.47 | 55.26 | 10.88 | 40.15 | 4.82 | 43.23 |
| Average mean | | 8.24 | 2.11 | 13.22 | 18.31 | 6.73 | 21.03 | 5.39 | 16.73 |

Table 6. Distribution of the absolute performance gain ratio |*pgr*| in CPDP

| Dataset | | Classification scenario | | | | Ranking scenario | | | |
|---|---|---|---|---|---|---|---|---|---|
| | | $F_1$ | AUC | ER | RI | AP | RR | $P_{opt}$ | ACC |
| Metrics-Repo-2010 | Mean | **110.53** | **30.61** | 18.36 | 25.43 | **13** | 38.92 | **99.63** | 55.51 |
| | StdDev | 409.22 | 88.43 | 26.37 | 30.44 | 29.09 | 130.6 | 249.9 | 95.6 |
| | 95% CI | 40.6 | 12.98 | 12.31 | 18.03 | 6.55 | 10.08 | 52.96 | 34.61 |
| | [LL, UL] | 377.11 | 80.6 | 30.25 | 37.85 | 30.96 | 110.67 | 268.05 | 102.4 |
| JIRA-HA-2019 | Mean | 8.51 | 2.97 | 2.04 | 7.6 | 5.11 | 229.91 | 82.12 | 54.49 |
| | StdDev | 5.27 | 2.26 | 1.58 | 4.99 | 4.69 | 1186.17 | 231.46 | 191.96 |
| | 95% CI | 6.94 | 2.27 | 1.6 | 6.01 | 3.7 | 15.93 | 31.26 | 19.27 |
| | [LL, UL] | 10.39 | 3.82 | 2.65 | 9.33 | 7.12 | 1072.55 | 227.54 | 256.53 |
| JIRA-RA-2019 | Mean | 4.79 | 2.22 | 1.39 | 4.99 | 4.09 | 22.56 | 53.44 | 125.22 |
| | StdDev | 3.73 | 3 | 1.3 | 4.15 | 3.44 | 54.62 | 104.22 | 274.82 |
| | 95% CI | 3.71 | 1.44 | 1 | 3.71 | 3.09 | 9.69 | 29 | 56.08 |
| | [LL, UL] | 6.35 | 3.98 | 1.91 | 6.52 | 5.42 | 51.71 | 109.24 | 285.37 |
| MA-SZZ-2020 | Mean | 54.9 | 13.09 | **31.1** | **35.22** | 9.33 | **233.1** | 43.52 | **140.79** |
| | StdDev | 88.36 | 18.58 | 55.55 | 48.86 | 7.47 | 1111.39 | 65.88 | 221.1 |

| | 95% CI | 37.05 | 9.12 | 19.49 | 24.8 | 7.25 | 53.01 | 29.78 | 90.95 |
| | [LL, UL] | 102 | 20.93 | 53.47 | 55.26 | 11.66 | 1037.41 | 72.45 | 221.29 |
| Average mean | | 44.68 | 12.22 | 13.22 | 18.31 | 7.88 | 131.12 | 69.68 | 94 |

From Table 3~6, we have the following observations. *First, both |diff| and |pgr| are significantly different from zero.* In both CVDP and CPDP context, |diff| and |pgr| have a 95% confidence interval that does not contain 0, regardless of which evaluation indicator or data set is considered. This indicates that the influence of inconsistent labels is statistically significant. *Second, there is a trend that, the higher the inconsistent label ratio (i.e., ILinAll) is, the greater both |diff| and |pgr| are.* Of the five multi-version-project data sets, Metrics-Repo-2010 and MA-SZZ-2020, respectively, have the first and second highest inconsistent label ratios. Consistently, we observe that Metrics-Repo-2010 and MA-SZZ-2020, respectively, are the first and second most to have the largest mean |diff| and |pgr| (count the cells shown in yellow background). Specifically, the former has 18 times to be the largest mean values, while the latter has 9 times. *Third, in terms of the average mean, |pgr| is 1~11 times in the CVDP context and 1~13 times in the CVDP context as high as than |diff|.* This indicates that after eliminating the influence of the actual prediction difficulty (i.e., minus the performance value of a random model), inconsistent labels actually exhibit a more considerable influence on the prediction performance of a defect prediction model. *Fourth, the values of |diff| and |pgr| vary with different performance evaluation indicators.* Under the classification scenario, it appears that |diff| and |pgr| exhibit larger means on $F_1$ and $RI$ than $AUC$ and $ER$. Under the ranking scenario, it appears that |diff| and |pgr| exhibit larger means on $RR$ and $ACC$ than $AP$ and $P_{opt}$. This indicates that the influence of inconsistent labels varies with the performance evaluation indicators.

**Conclusion.** *The performance of a defect prediction model is significantly influenced by inconsistent labels, with the trend that a higher inconsistent label ratio has a larger influence. Furthermore, all the commonly used performances indicators are sensitive, with varying degrees, to inconsistent labels, regardless of whether the classification or ranking scenario is considered.*

### 7.3 RQ3: Influence of inconsistent defect labels on detected actual defects

Table 7 summarizes the percentage of significant DTPs under the classification scenario and the ranking scenario (i.e., 20% size, top 10%, top 20%, and top 30%) for each multi-version-project defect data set. Specifically, for each DTP, we first use bootstrap to calculate its 95% confidence interval. Then, we consider it as significant if 0 is not included in the 95% confidence interval. A significant DTP indicates that there is a significant difference in the true positives detected by model NC and model CC, i.e. the actual defects detected by NC and CC are significantly different. Finally, the percentage of significant DTPs is obtained by dividing the number of significant DTPs to the total number of DTPs. Fig. 20 shows the distribution of the magnitude of the difference in true positives between model NC and model CC in each multi-version-project defect data set.

Table 7. The percentage of significant DTPs in five cases

| Dataset | CVDP | | | | | CPDP | | | | |
| --- | --- | --- | --- | --- | --- | --- | --- | --- | --- | --- |
| | Classification | Ranking | | | | Classification | Ranking | | | |
| | | 20% size | top 10% | top 20% | top 30% | | 20% size | top 10% | top 20% | top 30% |
| ECLIPSE-2007 | 100% | 100% | 100% | 100% | 100% | - | - | - | - | - |
| Metrics-Repo-2010 | 75% | 84% | 84% | 88% | 84% | 73% | 43% | 73% | 73% | 75% |
| JIRA-HA-2019 | 97% | 97% | 95% | 95% | 85% | 84% | 81% | 91% | 84% | 84% |
| JIRA-RA-2019 | 90% | 93% | 100% | 95% | 90% | 81% | 88% | 88% | 91% | 78% |
| MA-SZZ-2020 | 27% | 75% | 63% | 38% | 19% | 69% | 52% | 65% | 73% | 77% |

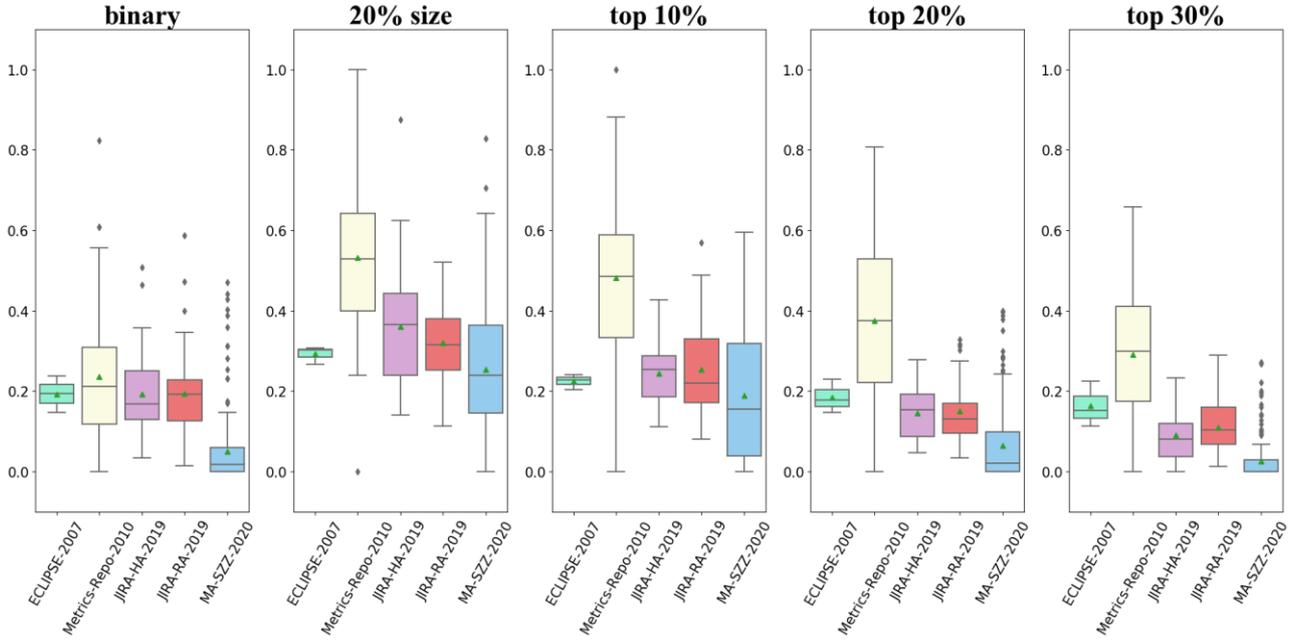

(a) Distribution in the CVDP context

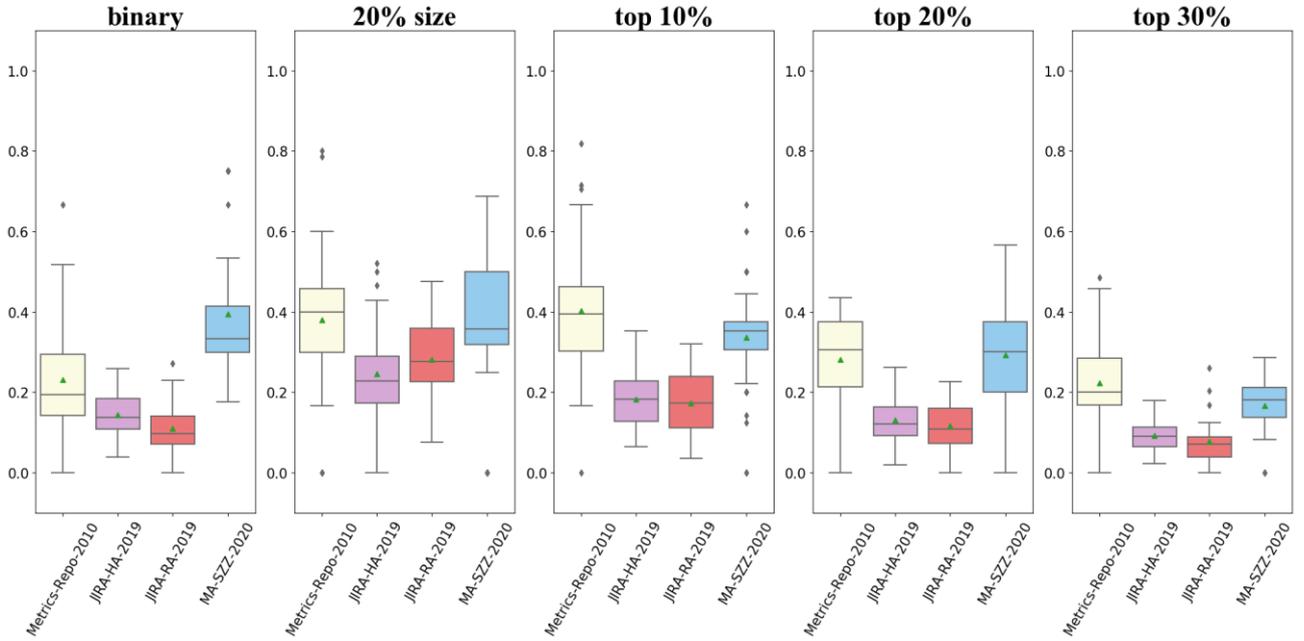

(b) Distribution in the CPDP context

Fig. 20. The distribution of the magnitude of DTPs in five multi-version-project defect data sets

From Table 7 and Fig. 20, we have the following observations. *First, in most cases, there is a statistically significant difference between the actual defects detected by NC and CC.* As show in Table 7, for all data sets expect MA-SZZ-2020, more than 70% DTPs are statistically significant, regardless of which prediction context (CVDP or CPDP) or whether classification or ranking scenario is considered. The only exception is the 20% size cut-off under the ranking scenario in the CPDP context on Metrics-Repo-2010, where 43% DTPs are significant. For MA-SZZ-2020, 19%~75% DTPs are significant in CVDP and 52%~77% DTPs are significant in CPDP. This means that, in most cases, inconsistent labels lead to a significant difference in true positives between model NC and model CC. *Second, the actual defects detected by NC can be substantially different from those detected by CC.* As shown in Fig. 20, under the classification scenario, DTP in general has a median ranging from 0.2 to 0.3. Under the ranking scenario, the median DTP is up to around 0.5 at the 20% size cut-off and the top 10% cut-off. When looking at individual DTPs, we can see that a DTP can be larger than 0.6 or even achieve 1 (see Metrics-Repo-2010 in the "top 10%" subgraph in CVDP). This means that inconsistent labels may lead to a substantially different set of actual defects detected. *Third, DTP in general exhibits a higher value under the ranking scenario.* From Fig. 20, we can see that, DTP tends to have a larger median under the ranking scenario (see "20% size" cut-off and "top 10%" cut-off) compared with the classification scenario. For example, for Metrics-Repo-2010 in CVDP (CPDP), DTP has a median around 0.2 in the "binary" subgraph but has a median around 0.5 (0.4) in the "20% size" cut-off and "top 10%" subgraphs. Similar phenomena can be observed for the other data sets. This indicates that inconsistent labels have a larger influence when the rank of actual defects is considered.

***Conclusion.*** *Inconsistent labels in general lead to a significantly different set of actually defects detected. In many cases, this may result in a substantially different set of actually defects detected. Compared with the classification scenario, inconsistent labels exhibit a larger influence under the ranking scenario.*

**7.4 RQ4: Influence of inconsistent defect labels on model interpretability**

Fig. 21 reports the distribution of shift($k$) for the top 3 important features in CC with respect to NC in each multi-version-project defect data set. For the feature in the first rank in CC, its rank in NC has one of the following values: 1, 2, 3, ≥ 4, or OOM (out of model). Consequently, shift(1) = 0 means an unchanged rank, shift(1) = −1 or −2 means that it is still in top 3, and shift(1) ≤ −3 means that it is out of the top 3 features in NC. Furthermore, shift(1) = "OOM" means that it is not selected in NC. The rank shift has a similar value pattern when considering the feature in the second rank and the feature in the third rank in CC. Fig. 21 visualizes the proportions of shift($k$) over different values for the top 3 ranked features. In particular, the percentage above the $x$-axis represents how often the features in the $k$th rank in CC is ranked top 3 in NC. In contrast, the percentage below the $x$-axis represents how often the features in $k$th rank in CC is out of top 3 in NC. By observing the proportion of the top three features in different shift($k$) values, we can analyze the influence of inconsistent labels on the ranks of the most important features of a defect prediction model and hence the influence on model interpretability.

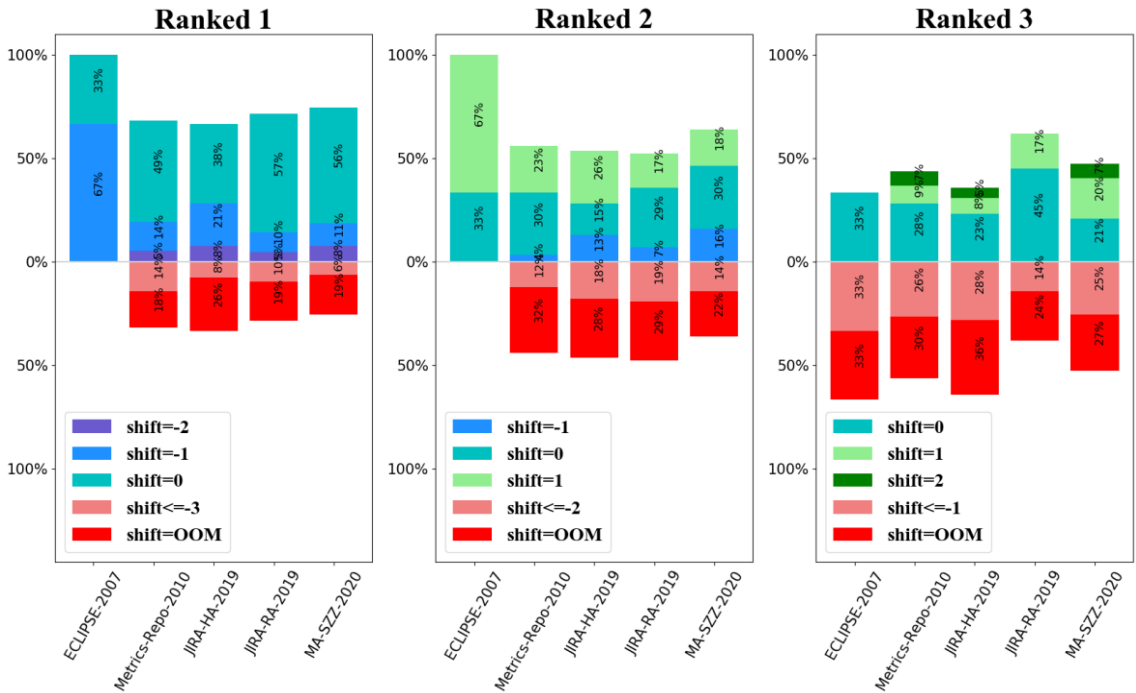

(a) Distribution in the CVDP context

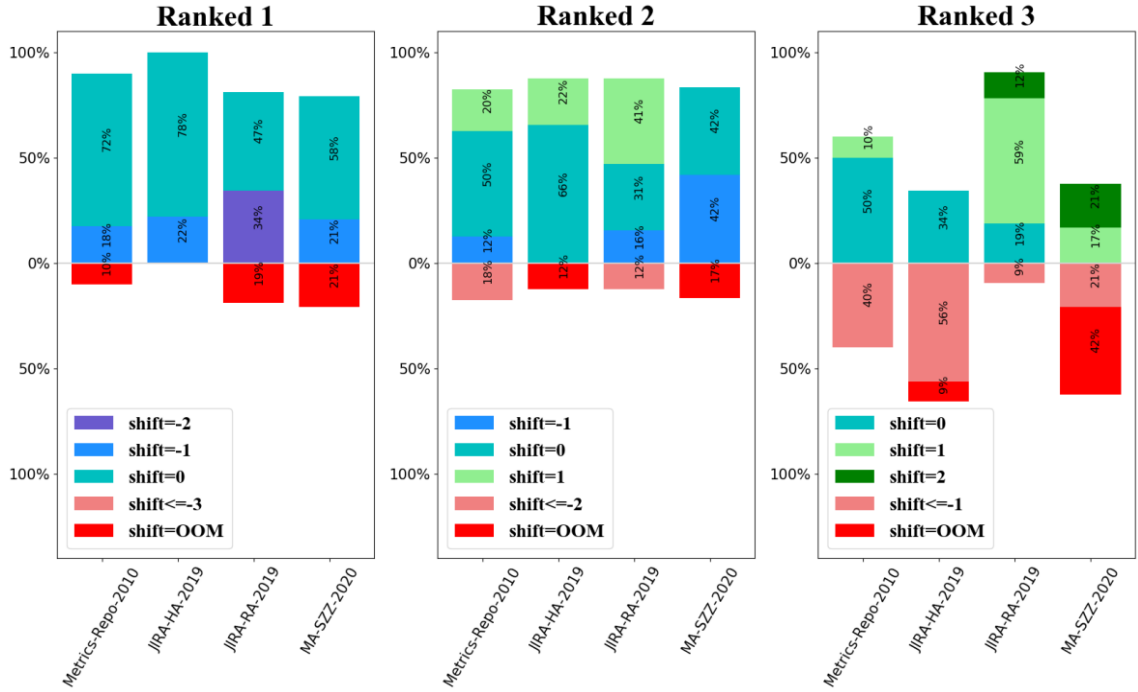

(b) Distribution in the CPDP context

Fig. 21. The distribution of shift($k$) over different values for the top 3 important features in CC

From Fig. 21, we have the following observations. *First, for the top three importance features in CC, their ranks shift substantially in NC due to the existence of inconsistent labels.* In CVDP (CPDP), 43%~67% (22%~53%) of the features in the first rank in CC drop one or more ranks in NC; 67%~85% (34%~69%) of the features in the second rank in CC do not appear in the second rank in NC; 55%~79% (50%~100%) of the features in the third rank in CC do not appear in the third rank in NC. Overall, for the top 3 important features, the proportion of shifted rank is 22%~67%, 34%~85%, and 50%~100%, respectively. *Second, the features in the second and third ranks in CC are less stable to inconsistent labels compared with the features in the first rank.* On each multi-version-project data set, the proportion of shift(2) = 0 (or shift(3) = 0) is smaller than that of shift(1) = 0. This suggests that the features in the first rank are more resistant to inconsistent labels. However, the first observation reveals that the features in the first rank are indeed not robust to inconsistent labels. Therefore, this suggests that inconsistent labels will mislead the interpretation of a defect prediction model. *Third, the ranks of the top three features in CC shift more dynamically in the CVDP context compared with in the CPDP context.* On the one hand, a defect prediction model in CVDP has a smaller shift($k$) =0 ($k$ = 1, 2, 3) than in CPDP. On the other hand, when looking at the proportion below the *x*-axis, we can find that a defect prediction model in CVDP in general has a larger proportion.

**Conclusion.** *For the top three important features, the proportions of shifted features in the first, second, and third ranks are 22%~67%, 34%~85%, and 50%~100%, respectively. Compared with in the CPDP context, the top three features are less stable in the CVDP context. It suggests that inconsistent labels can substantially change feature importance rank in a model and hence may lead to a biased interpretation.*

## 8. Discussion

In previous sections, we show that inconsistent labels are ubiquitous in multi-version-project defect data sets collected by *common automatic defect data collection approaches*. In this section, we first analyze what factors would influence the number of detected inconsistent labels (Section 8.1). Then, we discuss whether it is feasible to use software metrics as the proxy of source code to identify inconsistent labels (Section 8.2). Next, we investigate whether inconsistent labels still exist in a recently published multi-version-project defect data set collected by a semi-automatic approach (i.e., *an improved SZZ-based approach combined with manual validation*, section 8.3). Finally, we examine how many previous studies might be potentially influenced by the multi-version-project defect data sets that are found to have inconsistent labels in our study (Section 8.4).

**8.1 What factors would influence the number of inconsistent labels identified by the TSILI algorithm?**

Sections 7.1 shows that our TSILI algorithm is effective in identifying inconsistent labels. The rationale of TSILI is simple: it only needs to compare the source code and defect labels of cross-version instances, without any other complex procedures and additional information. Therefore, the following question naturally arises: Can TSILI identify all identify inconsistent labels in a defect data set? The answer is No. Because, in practice, there are three factors that can influence the number of inconsistent labels that the TSILI algorithm can identify. First, TSILI cannot be applied to cross-version instances that have no source code. For example, for the Camel project in the JIRA-HA-2019 and JIRA-RA-2019 data sets, the cross-version instance "package-info.java" contains only comment statements. In addition, due to unknown reasons, in ECLIPSE-2007, METRICS-REPO-2010, JIRA-HA-2019, and JIRA-RA-2019, about 1%~7% instances in 77% (85/110) versions cannot be found in the source codes downloaded from the official websites. In this case, for these instances, it is not possible to apply TSILI to identify inconsistent labels. Second, if the path name or file name of a cross-version instance has changed between versions, the TSILI algorithm is unable to detect inconsistent labels, even if its source code remain no change. In our implementation, TSILI identifies a cross-version instance (or module) among versions based on its full name (path name + file name). If the full name is changed, a cross-version instance will be treated as two different instances. In this case, TSILI may miss inconsistent labels. In our study, we did observe such a phenomenon in the investigated multi-version-project defect data sets, although it did not happen often. Third, the number of versions used will influence the number of inconsistent labels that TSILI can identify. As reported in [106, 107], for a project, it was common that a considerable proportion of bugs in a low version would not be discovered until in high versions. In our study, for ECLIPSE-2007, METRICS-REPO-2010, JIRA-HA-2019, and JIRA-RA-2019, each project contains only three to five versions. Nonetheless, TSILI still found a large number of inconsistent labels. Therefore, it is reasonable to believe that more inconsistent labels can be identified if more versions are analyzed, i.e., the multi-version-project defect data sets investigated in our study should contain more inconsistent labels than reported in Section 7.1. This means that the influence of inconsistent labels may be underestimated in our study.

**8.2 Can software metrics be used as a proxy of source code to identify inconsistent labels?**

In TSILI, inconsistent labels are regarded as found if a cross-version instance has the same source code but different labels in different versions. During this process, there is a need to compare source code to identify cross-version instances. However, in practice, it is common to see that a data set only provides for each instance a number of software metrics (i.e. features) and a label indicating whether it is defective. In other words, source code is external information for a defect data set, which needs to be acquired additionally. In this context, an interesting question naturally arises: Can we use software metrics as a proxy of source code to identify cross-version instances? Indeed, in previous studies [110, 111, 124], it is not uncommon to see that software metric information is used to identify "inconsistent instances" in a defect data set. In [110, 124], if two instances in a version had identical values for all features but different labels, they were called "inconsistent instances". In [111], inconsistent cross-version instances were also examined. In their view, inconsistent instances are problematic in the context of machine learning and hence should be excluded when building and evaluating a defect prediction model. As can be seen, the concept of their inconsistent cross-version instances is very similar to the concept cross-version instances with "inconsistent labels" in our study. The difference is that the former uses software metrics rather than source code to identify cross-version instances. At a glance, it seems that we could use software metrics to replace source code to identify cross-version instances in our study. In fact, such a replacement is problematic due to the following two-fold reasons. On the one hand, the fact that two instances have identical source code does not necessarily mean that they have identical values for all features (i.e. metrics). The reason is that many metrics depend on the use context of an instance rather than only on its source code. For example, two functions with identical code may have different values for the "called-by number" feature. On the other hand, the fact that two instances have identical values for all features does not necessarily mean that they have identical source code. For example, it is possible that two instances with different source code have identical values for all features. As a result, if we use software metrics as a proxy to identify cross-version instances, it is possible to miss real "inconsistent labels" or report false "inconsistent labels", thus leading to a low accuracy of inconsistent label identification. Given this situation, we should not use software metrics as a proxy to identify inconsistent labels.

**8.3 Do inconsistent labels still exist in a defect data set collected by a semi-automatic approach?**

Our analysis in Section 4.2 and the experimental results in Section 7.1 show that the introduction of inconsistent labels cannot be avoided by the current three types of automatic defect collection approaches (SZZ-based, time-window, and affected version). Therefore, the following question naturally arises: are inconsistent labels still prevalent in defect data sets collected by other (more accurate) approaches? In this section, we answer this question by examining inconsistent labels in a recently published multi-version-project defect data set collected by the state-of-the-art defect data collection approach IND-JLMIV+R.

The IND-JLMIV+R-2020 data set [90] consists of 395 versions of 38 projects[22]. Table 8 summarizes the specific information after applying the TSILI algorithm. The 4th column (Buggy versions: Clean versions) lists the number of buggy versions and the number of clean versions. According to [90], this data set is collected by a semi-automatic defect collection approach, which is called IND-JLMIV+R. In nature, IND-JLMIV+R is an improved SZZ-based approach combined with manual validation. It uses manual validation to identify BFCs which can be used as the ground truth and uses a variety of improved heuristic methods to improve the accuracy of SZZ approach in identifying BICs. Specifically, this approach reduces the introduction of mislabels in five ways. First, this approach uses manual classification to validate types of issue reports and correct type errors. Second, use "JIRA link" to establish the link between a BFC and an issue report. "JIRA link" refers to the fixed format (<project-ID>) used for the identifier of issue report in JIRA, and the corresponding BFC also uses the same format to indicate the fixed bug. Using "JIRA link" helps to avoid misidentification of BFCs. Third, the heuristic rules of "JIRA link" and original SZZ approach are used to search for BFCs, and then manual validation is carried out to select correct BFCs that can be used as the ground truth. Fourth, use the RA-SZZ approach [14], one of the latest SZZ variants, to identify BICs. The accuracy of RA-SZZ approach is further improved by using the RefactoringMiner [108] tool instead of RefDiff [109]. Fifth, according to the creation time of pairs of BICs and BFCs, link defective modules to versions (SZZ-based approach described in Section 2.1). After the above steps, the IND-JLMIV+R approach can be expected to have a high accuracy in defect data collection. In this sense, the IND-JLMIV+R-2020 data set collected by IND-JLMIV+R is expected to contain little or even no inconsistent labels.

---

[22] Note: the original dataset contains 398 versions. We did not use three (santuario-java-1.5.9, parquet-mr-1.8.0, and parquet-mr-1.9.0) of them because of the lack of corresponding versions on the official website or GitHub, or because the version code cannot be parsed with the Understand tool. Considering the large number of versions of this data set, we put the details of this data set on the online appendix: http://github.com/sticeran/InconsistentLabels.

Table 8. The IND-JLMIV+R-2020 defect data set used in our study

| Dataset | #Project | #Versions | #Buggy-versions: #Clean-versions | #Instances (#IL-instances) | %Defective |
|---|---|---|---|---|---|
| IND-JLMIV+R-2020 | 38 | 395 | 361: 34 | 3~1708 (0~58) | 0~36% |

We repeat the experiment in RQ1 using the IND-JLMIV+R-2020 data set. Fig. 22 reports the distributions of three types of inconsistent label ratios (i.e. *ILinAll*, *ILinBuggy*, and *ILinClean*) in this defect data set, in which each point corresponds to a version. In Table 9 (a), except for the 4th column, the meaning of each column is the same as that in Table 2. The 4th column report the ratio of the number of buggy versions with *ILinBuggy* ≠ 0 to the total number of buggy versions. In Table 9 (b), the 2nd and 3rd columns, respectively, report the ratio of the number of versions with *ILinAll* = 0 and *ILinBuggy* < 5% to the total number of versions. The 4th and 5th columns, respectively, report the ratio of the number of buggy versions with *ILinBuggy* = 100% and clean versions with *ILinClean* ≠ 0 to the total number of buggy versions and the total number of clean versions.

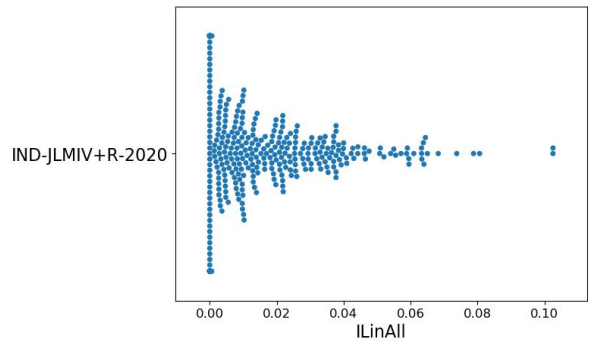

(a) Inconsistent labels in all instances

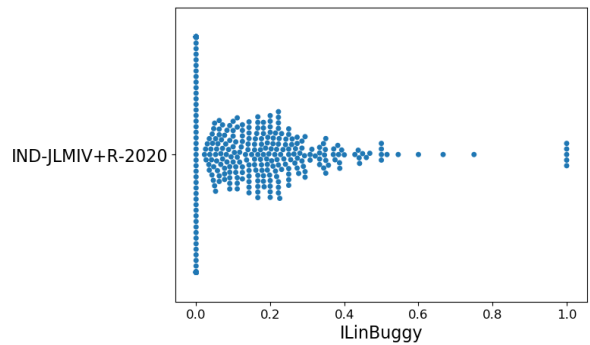

(b) Inconsistent labels in buggy instances

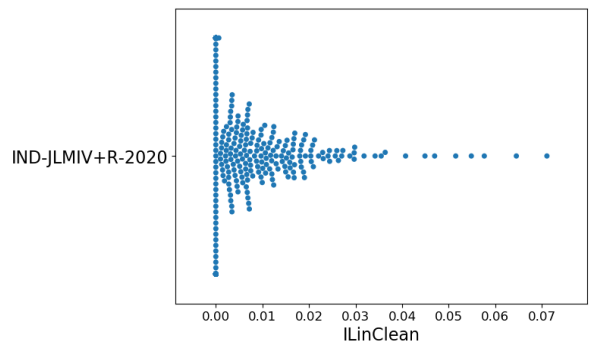

(c) Inconsistent labels in clean instances

Fig. 22. The ratios of inconsistent labels (*ILinAll*, *ILinBuggy*, and *ILinClean*) on the IND-JLMIV+R-2020 data set

Table 9. Statistical information of inconsistent labels of the IND-JLMIV+R-2020 data set

(a) The ratios of inconsistent labels ($ILinAll \neq 0$, $ILinBuggy \neq 0$, and $ILinClean \neq 0$)

| Dataset | $\frac{\#versions(ILinALL \neq 0)}{\#versions}$ | Average ILinAll (range) | $\frac{\#buggy\text{-}versions(ILinBuggy \neq 0)}{\#buggy\text{-}versions}$ | Average ILinBuggy (range) | $\frac{\#versions(ILinClean \neq 0)}{\#versions}$ | Average ILinClean (range) |
|---|---|---|---|---|---|---|
| IND-JLMIV+R-2020 | $\frac{274}{395}$ | 2% (0%~10%) | $\frac{242}{361}$ | 15% (0%~100%) | $\frac{139}{395}$ | 1% (0%~7%) |

(b) The ratios of inconsistent labels in other cases

| Dataset | $\frac{\#versions(ILinALL = 0)}{\#versions}$ | $\frac{\#versions(ILinALL < 5\%)}{\#versions}$ | $\frac{\#buggy\text{-}versions(ILinBuggy = 100\%)}{\#buggy\text{-}versions}$ | $\frac{\#clean\text{-}versions(ILinClean \neq 0)}{\#clean\text{-}versions}$ |
|---|---|---|---|---|
| IND-JLMIV+R-2020 | $\frac{121}{395}$ | $\frac{370}{395}$ | $\frac{5}{361}$ | $\frac{2}{34}$ |

From Fig. 22 and Table 9, we have the following observations. *First, overall, the inconsistent label ratio (ILinAll) of IND-JLMIV+R-2020 data set is relatively low*. Inconsistent labels are not detected in 31% (121/395) of all versions, and the inconsistent label ratio (*ILinAll*) is less than 5% in 94% (370/395) of all versions. *Second, buggy instances have a higher proportion of inconsistent labels than clean instances*. On average, for each version, 15% of buggy instances contain inconsistent labels, while 1% of clean instances contain inconsistent labels. *Third, there is a very high inconsistent label ratio (ILinBuggy) in many buggy versions*. For example, the inconsistent label ratio (*ILinBuggy*) is even equal to 100% in five buggy versions. After checking, we find that the number of buggy instances in these five buggy versions is equal to 1 or 2, and all of them are found to have inconsistent labels. Therefore, the inconsistent label ratio (*ILinBuggy*) on these versions reaches 100%. Similar phenomena appear in other buggy versions with extremely high inconsistent label ratio (*ILinBuggy*). *Fourth, inconsistent labels are found in two clean versions*. This observation informs us that a clean version generated by the IND-JLMIV+R approach indeed may contain buggy instances.

The above-mentioned results show that, when all instances are considered, there is a low average inconsistent label ratio (as shown by *ILinAll*). However, there is still a relatively high average inconsistent label ratio on buggy instances (as shown by *ILinBuggy*). We believe that the reasons are three-folds. First, in the identification of BICs, the IND-JLMIV+R approach used the RA-SZZ approach. The RA-SZZ approach itself cannot guarantee 100% accuracy and hence may produce false positives and false negatives (even if the proportion is small) [14]. Second, as we analyzed in Section 4.2, the SZZ-based approach itself does not consider rollback changes, so it is possible to cause inconsistent labels. Third, there may be a certain number of extrinsic bugs correctly/mistakenly identified in the data set, which leads to the existence of inconsistent labels. To sum up, our experimental results show that inconsistent labels are a common phenomenon in multi-version-project defect data sets, even if the data sets are collected by defect collection approaches that are believed to be accurate.

### 8.4 To what extent might previous studies be potentially influenced by inconsistent labels?

Our experimental results show that existing multi-version-project defect data sets (such as ECLIPSE-2007, METRICS-REPO-2010, JIRA-HA-2019, and JIRA-RA-2019) contain inconsistent labels. In particular, for a defect prediction model, the existence of such inconsistent labels may considerably change its prediction performance scores, the identified actual defects, and the model interpretation. This raises concerns on the reliability of the experimental results or conclusions reported in previous studies that used these data sets. Therefore, the following question naturally arises: how many previous studies might be potentially influenced by these multi-version-project defect data sets? In this section, we answer this question by investigating the number of previous studies that used them as the subject data sets to conduct their experiments.

Table 10 summarizes the number of previous studies potentially influenced by inconsistent labels. The first column lists five existing multi-version-project defect data sets investigated in our study. The second column lists the original literature publishing each multi-version-project defect data set. The third column reports how many studies cite the original literature (reported by Google scholar, September 28, 2020). The fourth column reports the total number of studies (written in English) that use the corresponding data sets to conduct their experiments (inspected by the first author and confirmed by the seventh author). Note that the Metrics-Repo-2010 data set was first published in [27]. However, most literature cite [28] and [125] as its source. The reason was that [28] was published in a well-known international conference on Predictive Models in Software Engineering (PROMISE), aiming to share publicly accessible data sets. In particular, Metrics-Repo-2010 was put on the corresponding promise repository website [125]. Given this situation, we use them (i.e. [27], [28], and [125]) as three sources to count the number of (different) citations. As can be seen, JIRA-RA-2019 (JIRA-HA-2019) and IND-JLMIV+R-2020 data sets were used by few studies, as they are two recently published multi-version-project defect data sets. However, ECLIPSE-2007 and Metrics-Repo-2010 data sets were widely used in previous studies. This indicates that inconsistent labels have a potentially wide influence on previous studies.

Table 10. Literatures potentially influenced by target multi-version-project defect data sets

| Defect data sets | Source | Total number of citations | Total number of citations using the corresponding data set for experiments |
| --- | --- | --- | --- |
| ECLIPSE-2007 | [29] | 744 | 139 |
| Metrics-Repo-2010 | [27, 28, 125] | 414 | 238 |
| JIRA-HA-2019 / JIRA-RA-2019 | [23] | 7 | 2 |
| IND-JLMIV+R-2020 | [90] | 5 | 5 |

## 9. Implications

Our study has important implications for both practitioners and researchers. Our work contributes an automatic approach TSILI to detect inconsistent labels in defect data sets. In practice, inconsistent labels can be used as a risk indicator to reflect the quality of defect data sets. The detailed implications are as listed follows:

(1) **Our work warns that existing publicly available multi-version-project defect data sets should be used with caution.** We uncover that there are non-ignorable inconsistent labels in existing multi-version-project defect data sets, including not only the widely used ECLIPSE-2007 and Metrics-Repo-2010 but also the recent published JIRA-HA-2019, JIRA-RA-2019, and IND-JLMIV+R-2020. In particular, we find that, on the one hand, of all data sets, the most widely used Metrics-Repo-2010 has the overall highest inconsistent ratio. On the other hand, JIRA-RA-2019 and IND-JLMIV+R-2020, believed as high quality by their authors, also contain many inconsistent labels. The fact cautions us that, in the future studies, these data sets should be used and explained with caution, especially for Metrics-Repo-2010.

(2) **Our work provides a simple but effective approach to examine the quality of multi-version-project defect data sets.** In practice, there are two approaches to examining the quality of a data set. The first approach is to apply a noise detection approach such as CLNI [53] to identify noisy instances, while the second approach uses manually-curated data in VCS/ITS to identify noisy instances [54]. However, for the former, it is inevitable to report false noisy instances, i.e. actually non-noisy instances identified as noisy instances. For the latter, it is time consuming to manually curate the data. Compared with the first approach, Our TSILI approach does not report false noisy instances (i.e. false inconsistent labels). Compared with the second approach, Our TSILI approach is a light-weight approach and can be automated, as no manual curation is needed. As a result, we suggest using TSILI to examine the quality of multi-version-project defect data sets in practice.

(3) **Our work highlights the need to re-examine the conclusions drawn from the investigated multi-version-project data sets in previous studies.** Our experimental results show that the influence of inconsistent labels on a prediction model is all-round: inconsistent labels can considerably influence the prediction performance, the identified actual defects, and the model interpretation. This is in line with the existing research findings: mislabels and extrinsic bugs have a significant negative impact on the performance of a prediction model [51, 54]. In particular, we uncover non-ignorable inconsistent labels on existing multi-version-project data sets, especially on the widely used Metrics-Repo-2010. This raises concerns on the reliability of the experimental conclusions reported in previous studies that used these data sets with inconsistent labels. As a result, there is a need to re-examine these conclusions using accurate data sets.

(4) **Our work makes prominent that existing defect label collection approaches are vulnerable to inconsistent labels and hence need to be improved.** We have detected inconsistent labels in defect data sets collected by three common types of defect collection approaches (SZZ-based, time-window, and affected version). The results are surprising: even if the affected version approach is considered to be accurate by previous studies [13, 23], it still leads to a considerable proportion of inconsistent labels; in particular, on average, the inconsistent label ratio generated by the affected version approach is higher than that of the time-window approach. Moreover, we find that the state-of-the-art defect data collection approach IND-JLMIV+R [90] also produces many inconsistent labels. This indicates that the accuracy of all existing defect label collection approaches is not satisfactory and hence needs to be further improved.

(5) **Our work discloses the possible reason why existing defect label collection approaches lead to inconsistent labels and hence provides an opportunity to improve their effectiveness.** From the perspective of causing inconsistent labels, we systematically analyze the shortcomings of the mechanism of the commonly used three types of defect label collection approaches: SZZ-based, time window, and affected version. For example, for the SZZ approach, we point out a shortcoming of the SZZ mechanism that has not been noticed yet by existing SZZ related studies: SZZ cannot recognize rollback changes, which may lead to the failure to recognize BICs, resulting in inconsistent labels. Based on the specific problems with each approach we pointed out, researchers can make targeted improvements to improve their effectiveness.

(6) **Our work provides a means of generating (partial) noise ground truth, which can assist researchers in evaluating the effectiveness of existing noise identification approaches.** In the literature, multiple approaches have been proposed to detect the noise in a defect data set [53, 110, 111, 124]. However, it is difficult to evaluate their effectiveness due to the following two reasons. On the one hand, it is time consuming to obtain the ground truth manually [105, 112]; on the other hand, it is difficult to obtain the ground truth for the data sets lacking background information (such as not knowing the tools, approaches, issue reports, and commits used). In this context, we recommend using TSILI to automatically identify inconsistent labels. As mentioned earlier, inconsistent labels mean noise data. Therefore, the identified inconsistent labels can be used as the (partial) noise ground truth to assist in the evaluation of noise identification approaches.

## 10. Related work and comparison

In this section, we first introduce the most related work in the literature from two aspects: root-cause analysis of mislabels and label noise detection in defect prediction. Then, we analyze the similarities and differences between our work and the related work.

**10.1 Root-cause analysis of mislabels**

The primary causes of mislabels in defect data sets mainly come from two sources: incorrect basic information (issue reports and commits) and inaccurate defect label collection mechanism.

*Incorrect basic information.* Previous studies found that there were many kinds of errors in or between BFCs (bug-fixing commits) and issue reports. Herzig et al. [105] found that more than one third of "bug" type issue reports did not contain bugs through manual analysis. Aranda et al. [94], Bird et la. [95], Bettenburg et al. [96], and Nguyen et al. [97] found that developers always forgot to write specific keywords or leave links for commits in VCS and issue reports in ITS. Furthermore, it was reported that, when establishing the links from BFCs to issue reports, it might lead to many false or missing BFCs due to insufficient information and inaccurate heuristic rules [98-102]. In particular, Nguyen et al. [103] found that 11%~39% of BFCs contained non-fixing changes, and more than 50% of committed files in BFCs were not used for bug fixing. Mills et al. [104] found that about 64% of file changes made in BFCs were not part of the bug fixes but other changes. All the above factors will lead to the identification of false BFCs, thus resulting in mislabels. In addition, Chen et al. [106] and Ahluwalia et al. [107] reported that there were dormant bugs in a software system. Dormant bugs refer to bugs introduced in a lower version of in a software system, which is not found until in a higher version. More specifically, Chen et al. [106] found that 33% of the bugs introduced in a version were not reported until much later (i.e., they were reported in higher versions) and 18.9% of the reported bugs in a version were not even introduced in that version (i.e., they were dormant bugs from prior versions). Ahluwalia et al. [107] investigated the version interval between the bug-introducing and bug-fixing versions based on SZZ. They found that most bugs in a software system slept for over 20% of the existing versions. This indicated that a number of instances that were marked as "clean" on lower versions would be found to be mislabeled (false negative) over time.

*Inaccurate defect collection mechanism.* Many studies [12-14, 39, 92] pointed out that the code backtracking mechanism of the original SZZ approach [11] had deficiencies. Kim et al. [12] observed that SZZ might consider non-semantic code changes (such as changes of annotations, spaces, and blank lines) and format changes (such as moving the bracket). In nature, these lines of code should not be considered to introduce bugs since they do not affect the behavior of the code. Da Costa et al. [13] further noticed that SZZ might marked meta-changes that were not related to code modification as potential BICs. They found three types of meta-changes: branch changes (i.e., changes that copy code from one branch to a new branch), merge changes (i.e., changes that apply code modifications from one branch to another) and property changes (i.e., changes that only impact file properties). Neto et al. [14] further found that SZZ might identify incorrect BICs due to the impact of refactoring changes (such as the modification of method name). Rodríguez-Pérez et al. [39] and Wen et al. [92] found that the hypothesis "a bug was introduced by the lines of code that were modified to fix it" was not always true. Specifically, Wen et al. [92] observed that the BFC and the BIC did not always modify the same code elements / files, while Rodríguez-Pérez et al. [39] found that there was no corresponding BICs for the BFC fixing extrinsic bugs. Furthermore, Yatish et al. [23] recently demonstrated that the time-window approach [24-28] was inaccurate when collecting the defect label data. For one version of a software, the time-window approach might miss BFCs outside the window period and might mistakenly identify BFCs that fix bugs on other versions as the BFCs of the current version. More recently,

Herbold et al. [90] pointed out that the reliability of the affected version approach was overestimated. The reason was that there were many missing and/or incorrect version records in the *affected version* field of issue reports. The above works show that mislabels can be introduced by inaccurate defect collection approaches.

**10.2 Label noise detection in defect prediction**

The mislabels in the defect data set may be acceptable if they do not affect the defect prediction model. However, previous studies [23, 46, 47, 53, 54, 105, 113-123] reported that label noise data in data sets can lead to biased analysis and mislead model evaluation. Therefore, researchers tend to use different approaches to detect label noise in defect data sets before constructing defect prediction models. For issue reports that are necessary to generate a defect data set, Herzig et al. [105] and Herbold et al. [90] detected and corrected the misclassified issue reports by manual analysis to ensure the correctness of the basic information. Although the manual analysis method can produce accurate results, it is costly and cannot analyze the mislabels in the defect data sets which lack background information. Kim et al. [53] proposed the Closest List Noise Identification (CLNI) algorithm to automatically detect label noise from defect data sets. The CLNI algorithm calculates the feature similarity of different instances in the same defect data set. For an instance, if its defect label is different from those of $n$ instances similar to it ($n$ is an empirical threshold), then its label will be considered as a label noise. Since CLNI can be automated, it can help us identify label noises from a data set quickly. In their original study [53], Kim et al. used randomly generated label noises to evaluate the effectiveness of CLNI. As a result, they reported that CLNI had a high recall and a higher precision. However, Tantithamthavorn et al. [54] pointed out that label noises in real defect data sets were not random. In this sense, there is a need to use real label noises to re-investigate the effectiveness of CLNI in the future. Different from the above work, Gray et al. [110] and Sun et al. [111] regarded instances with identical values for all features but different labels as problematic instances. In their viewpoint, such instances should be excluded in defect prediction. In particular, Sun et al. reported that they had a potentially important influence on the performance of a CPDP model [111]. Similarly, Shepperd et al. [124] stressed that we should invest effort in understanding the quality of data prior to applying machine learners.

**10.3 Comparison of our work with existing work**

The above two aspects of existing work are related to mislabels, which are collectively referred to as mislabel-related studies. The purpose of our work is to detect inconsistent labels in data sets, so our work can be referred to as an inconsistent label study. Our study is similar to mislabel-related studies, both of which are (directly or indirectly) concerned with label noises in data sets. However, these two types of studies are not equivalent, because there are at least two obvious differences. On the one hand, inconsistent labels are not equivalent to mislabels. Inconsistent labels are a phenomenon existing in cross-version instances, which is caused by mislabels and extrinsic bugs. On the other hand, their detection approaches are different. To the best of our knowledge, mislabels can be detected by the following approaches: manual analysis based on basic information (such as issue reports and commits) [39, 90, 92, 94-107], comparison between defect collection approaches using a more accurate approach as a benchmark [12-14, 23, 32, 90], and noise detection using feature information of data sets [53, 110, 111, 124]. Different from these approaches, our TSILI uses source code and instance labels to detect inconsistent labels and does not care about the background information (i.e., basic information, defect collection approach, and features) in a data set. In general, these two types of research are complementary in that label noises in defect data sets can be understood and detected from different perspectives. Table 11 further lists the similarities and differences between our inconsistent label study and existing mislabel-related studies.

Table 11. A comparison between our inconsistent label study and existing mislabel studies

| Study | The found causes of label noises (i.e. mislabels) | Does it directly detect label noises in data sets? | Can it be applied to an existing data set? | Advantages | Disadvantages |
|---|---|---|---|---|---|
| Errors in basic information [94-105] | Errors in issue reports / commits / and their links | No | Depends | Manual analysis may be the most accurate | Manual analysis is costly and time-consuming |
| Dormant bug [106, 107] | A bug introduced in one low version was found until in a high version | No | No | It indicates that a software version may contain many undiscovered bugs | Information of future versions are required |
| SZZ variants [12-16, 90, 92] | The mechanism of SZZ is not accurate | No | No | SZZ variants point out that there are shortcomings in the mechanism of SZZ | Extrinsic bugs have no BICs, so they cannot be identified by SZZ variants [39] |
| Affected version approach [23] | The mechanism of time-window approach is not accurate | No | No | The *affected version* field is recorded by developers, so this approach may be more accurate | There are many missing and/or incorrect version records in the *affected version* field [90] |
| The CLNI approach: detecting label noises based on features [53] | Outliers are considered as label noises | Yes | Yes | This approach is automatic and has some rationality | Outliers are not necessarily mislabeled, and the empirical threshold set manually may be difficult to generalize to other defect data sets |
| Detecting inconsistent/duplicate instances based on features [110, 111, 124] | Instances with identical values for all features are considered problematic | No | Yes | This approach is automatic and has some rationality | Literature [124] pointed out that whether this approach is reasonable depends on the objective to be investigated |
| Inconsistent labels (our study) | Inconsistent labels are generated by mislabels and the labels corresponding to extrinsic bugs | Yes | Yes | No false inconsistent labels will be generated. The approach is automatic and insensitive to background information | The source code of the module (external information) is required |

## 11. Threats to validity

We consider the most important threats to construct, internal, and external validity of our study. Construct validity is the degree to which the independent and dependent variables accurately measure the concept they purport to measure. Internal validity is the degree to which conclusions can be drawn about the causal effect of independent variables on the dependent variable. External validity is the degree to which the results of the research can be generalized to the population under study and other research settings.

**11.1 Construct validity**

For RQ1 (degree of existence), the dependent variable is a binary value that indicates whether the defect label of a cross-version instance on a version is an inconsistent label. According to our TSILI algorithm, all reported inconsistent labels are real inconsistent labels. However, one possible threat to the construct validity of the dependent variable is that there are bugs in our TSILI implementation. In our study, we double checked the code of TSILI to reduce this threat. The independent variables are defect labels and the source code of each cross-version instance. The defect labels come from multi-version-project defect data sets collected by different approaches. Therefore, the threats to the construct validity of the dependent variable are from two-fold. First, it is possible that the defect labels are not collected in accordance with the data label collected approaches as stated completely. Second, we found that few instances (less than 3%) in the defect data sets did not appear in the source codes downloaded from the corresponding official websites. Considering the relatively small occurrence possibility of such cases, we believe that the above threats should not have a large influence.

For RQ2~RQ4 (influence on prediction performance, detected defects, and model interpretation), the dependent variable is the defect label indicating whether an instance is defective, while the independent variables are the software metrics provided in the investigated data sets. On the one hand, the defect labels in the investigated multi-version-project data sets were mined from software repositories. As a result, the discovered defect labels may be incomplete, which is a potential threat to the construct validity of the dependent variable. Indeed, this is an inherent problem to most, if not all, studies that discover defect labels by mining software repositories, not unique to us. Nonetheless, this threat needs to be eliminated by using complete defect data in the future work. On the other hand, the software metrics available in the investigated multi-version-project data sets are mainly the most commonly used structural metrics such as size, complexity, coupling, and/or cohesion metrics. For their construct validity, previous research has investigated the degree to which they accurately measure the concepts they purport to measure [127, 128]. In particular, they have a clear definition and can be easily collected. In this sense, the construct validity of the independent variable should be acceptable in our study.

**11.2 Internal validity**

The first threat to the internal validity (for RQ2~RQ4) is that the unknown effect of uncovered inconsistent labels. In our study, in order to investigate RQ2~RQ4, we built two types of models: NC (built with Noise training data and applied to Clean test data) and CC (built with Clean training data and applied to Clean test data). Given a noisy data set, we remove all the found instances with inconsistent labels to obtain the corresponding clean data set. For each involved project, TSILI used all the available versions in the corresponding data set to identify inconsistent labels. Consequently, it is possible that there are uncovered instances with inconsistent labels. In other words, the generated "clean" data sets in our study may not be completely clean. In fact, if more versions are compared, more inconsistent labels would be expected to be found. In this case, the difference between the noisy and clean training sets should be larger. This will strengthen, rather than weaken, the conclusions for RQ2~RQ4. Therefore, this threat does not negatively influence the internal validity of RQ2~RQ4.

The second threat to the internal validity (for RQ2~RQ4) is that the differences between the results of model NC and model CC could be influenced by the difference in the sizes of the noisy and clean training sets. To mitigate this threat, we followed the method in the literature [52] to conduct the following experiment. First, we randomly removed the instances from the noisy training set. The size of the removed instances was equal to the size of the instances identified as inconsistent labels. As such, the resulting noisy training set used for NC had the same size as the clean training set used for CC. Then, we used them to re-analyze RQ2~RQ4. We repeated the above process 10 times, each with the removal of a different random set of samples (i.e., instances). Consequently, we found that the overall trend of the difference between NC and CC did not change. Therefore, it can be considered that the difference between NC and CC was caused by inconsistent labels in the training set. In other words, the difference in the sizes of the noisy and clean training sets is not a threat to the internal validity of RQ2~RQ4.

The third threat to the internal validity (for RQ2~RQ4) is the unknown impact of modeling/rebalancing/feature selection techniques on our conclusions. When investigating the influence of inconsistent labels on the defect prediction model (RQ2~RQ4), we adopted random forest (one of the best modeling techniques [69-73]), SMOTE (rebalancing technique), and CFS (feature selection technique) which are common in current practice [31, 32, 54, 65]. In this sense, our findings may be entirely bound to the techniques or combinations of techniques used. This means that different modeling/rebalancing/feature selection techniques may lead to different experimental conclusions. In order to mitigate this threat, we used more techniques and repeated our experiments in RQ2~RQ4 to investigate the influence of inconsistent labels on defect prediction. Specifically, we used three modeling techniques (Random Forest, Naive Bayes, and Logistic Regression), two rebalancing techniques (NON1[23] and SMOTE) and three feature selection techniques (NON2[24], CFS, and GainRatio [126]). In total, these technologies produce 18 (= 3*2*3) combinations. Note that, in the following, for simplicity of presentation, we only summarize our findings obtained from the overall experimental results of 18 combinations.

- Influence of technique combinations on RQ2. Under different combinations, the number of significant |*diff*| / |*pgr*| values may be different under different evaluation indicators. We count the number of significant performance changes. For a combination, if the |*diff*| / |*pgr*| values are significant under all evaluation indicators (F1, AUC, MAP, MRR, ER, RI, $P_{opt}$, and ACC), the number of significant performance changes is 8. Table 12 reports the distribution of 18 combinations on the number of significant performance changes. The "8", "7", "6", and "5" columns, respectively, represent the number of significant performance changes is 8, 7, 6, and 5. The number in a cell in the "8", "7", "6", and "5" columns is the cumulative result of 18 combinations. For a multi-version-project defect data set in the CVDP/CPDP context, the total number of each row is equal to 18. Table 12 shows that the influence of inconsistent labels on the performance of a defect prediction model is significant under most evaluation indicators (>5), regardless of which modeling/rebalancing/feature selection techniques are used. Therefore, the findings of RQ2 are robust to the combination of modeling-rebalancing-feature selection techniques.

- Influence of technique combinations on RQ3. Fig. 23 (a) and (b), respectively, show the distribution of the average DTPs and percentages of significant DTPs on 18 combinations in each multi-version-project defect data set. For each combination, the average DTPs refers to the average value of DTPs on all (NC, CC) pairs in a data set; the percentage of significant DTPs refers to the number of significant DTPs to the total number of DTPs in a data set. As shown in Fig. 23 (a), the median average DTPs on most combinations varies from 0.1~0.4, while the average DTPs of a single combination can be close to 0.8. From Fig. 23 (b), we can see that, the median percentage of significant DTPs on most combinations is more than 60% in CVDP and more than 40% in CPDP, while the percentage of significant DTPs of a single combination can be close to 1. Therefore, for most combinations, inconsistent labels lead to a different set of actually defects detected. In other words, overall, our findings of RQ3 are largely the same under the different combinations of modeling-rebalancing-feature selection techniques.

- Influence of technique combinations on RQ4. Fig. 24 reports the distribution of the proportions of "shift($k$) = 0" on 18 combinations in each multi-version-project defect data set. In Fig. 24, "shift(1) = 0", "shift(2) = 0", and "shift(3) = 0", respectively, represent the proportion of the features in the first, second, and third ranks in model CC that have not changed in model NC. As shown in Fig. 24, the proportions of "shift(1) = 0" / "shift(2) = 0" / "shift(3) = 0" vary over a wide range on 18 combinations. The change degree of proportion between most combinations is about 10%~30% in CVDP and 30%~50% in CPDP. This shows that the influence degree of inconsistent labels on the interpretation of the defect prediction model is different under different techniques. Nevertheless, we find that the three findings of RQ4 are still applicable to most combinations. Because for most combinations, the proportion of "shift(1) = 0" / "shift(2) = 0" / "shift(3) = 0" is far away from 100%, which means that a considerable proportion of ranks of the top three features in the model CC have changed.

---

[23] "NON1" means not using any rebalancing technique.
[24] "NON2" means not using any feature selection technique.

Furthermore, for most combinations, the features in the second and third ranks in CC are less stable to inconsistent labels compared with the features in the first rank. Moreover, overall, the ranks of the top three features in CC shift more dynamically in CVDP compared with in CPDP (the proportions of shift$(k) \neq 0$ in CVDP are higher than that in CPDP). Therefore, to a large extent, the findings of RQ4 still hold for most combinations.

In summary, our findings in RQ2~RQ4 are applicable to most technique combinations we investigate. Nonetheless, considering the difference of results between different combinations, we do not claim that our findings can be extended to other combinations. To mitigate the threat, it is necessary to replicate our study across a wider variety of technique combinations.

Table 12 Distribution of 18 combinations on number of significant performance changes

(a) Performance change measured by |diff|

| Dataset | CVDP | | | | CPDP | | | |
|---|---|---|---|---|---|---|---|---|
| | 8 | 7 | 6 | 5 | 8 | 7 | 6 | 5 |
| ECLIPSE-2007 | 2 | 11 | 4 | 1 | - | - | - | - |
| Metrics-Repo-2010 | 18 | 0 | 0 | 0 | 14 | 4 | 0 | 0 |
| JIRA-HA-2019 | 17 | 1 | 0 | 0 | 11 | 7 | 0 | 0 |
| JIRA-RA-2019 | 18 | 0 | 0 | 0 | 12 | 6 | 0 | 0 |
| MA-SZZ-2020 | 18 | 0 | 0 | 0 | 18 | 0 | 0 | 0 |

(b) Performance change measured by |pgr|

| Dataset | CVDP | | | | CPDP | | | |
|---|---|---|---|---|---|---|---|---|
| | 8 | 7 | 6 | 5 | 8 | 7 | 6 | 5 |
| ECLIPSE-2007 | 2 | 13 | 2 | 1 | - | - | - | - |
| Metrics-Repo-2010 | 18 | 0 | 0 | 0 | 14 | 4 | 0 | 0 |
| JIRA-HA-2019 | 17 | 1 | 0 | 0 | 11 | 7 | 0 | 0 |
| JIRA-RA-2019 | 18 | 0 | 0 | 0 | 12 | 6 | 0 | 0 |
| MA-SZZ-2020 | 18 | 0 | 0 | 0 | 18 | 0 | 0 | 0 |

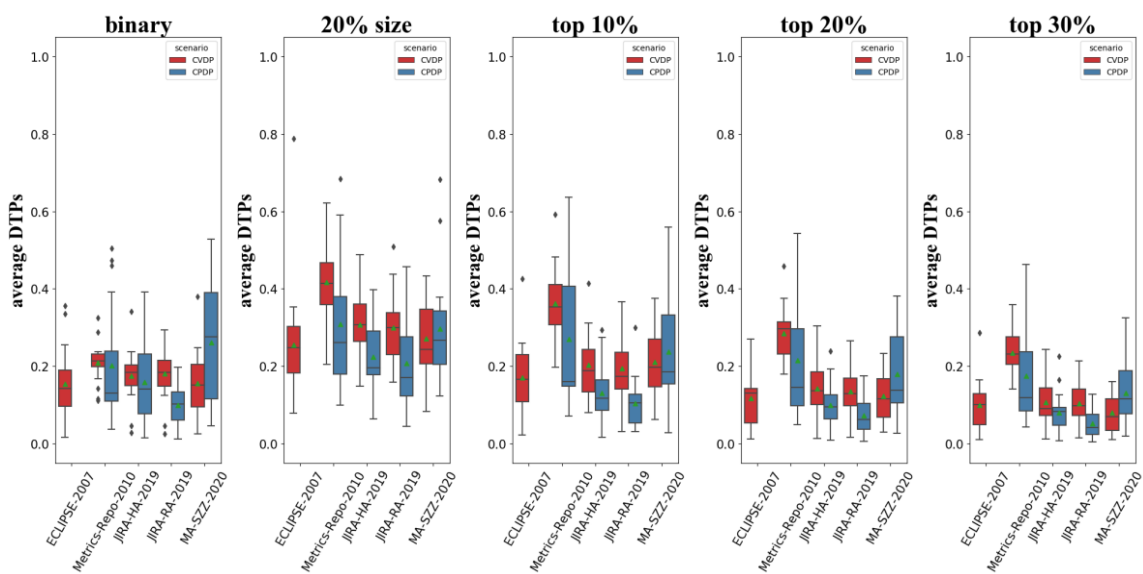

(a) Distribution of the average DTPs

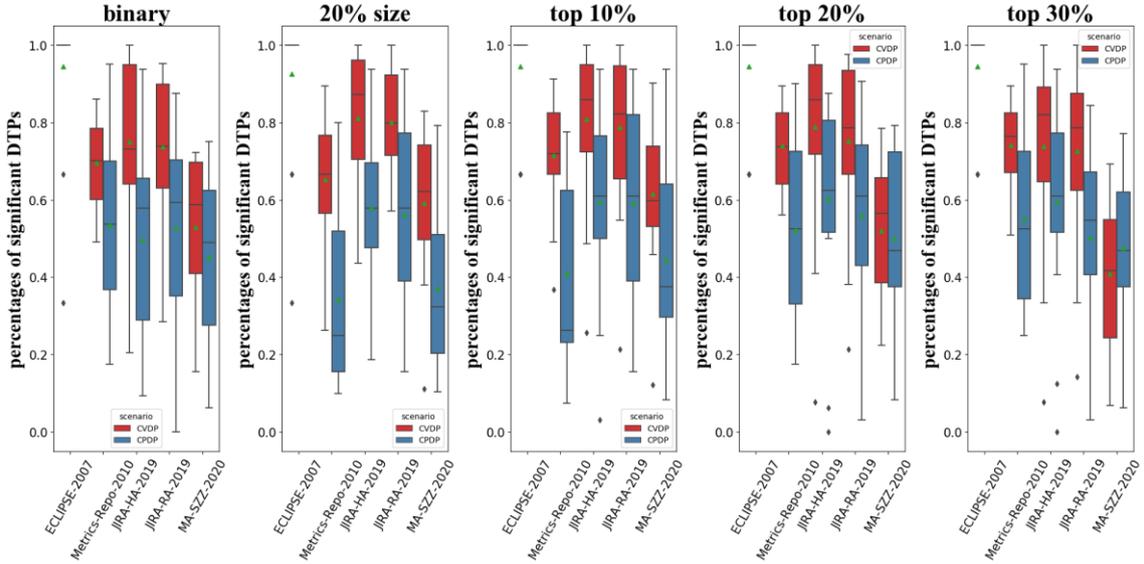

(b) Distribution of the percentages of significant DTPs

Fig. 23. Distribution of DTP (difference in true positives) on 18 combinations

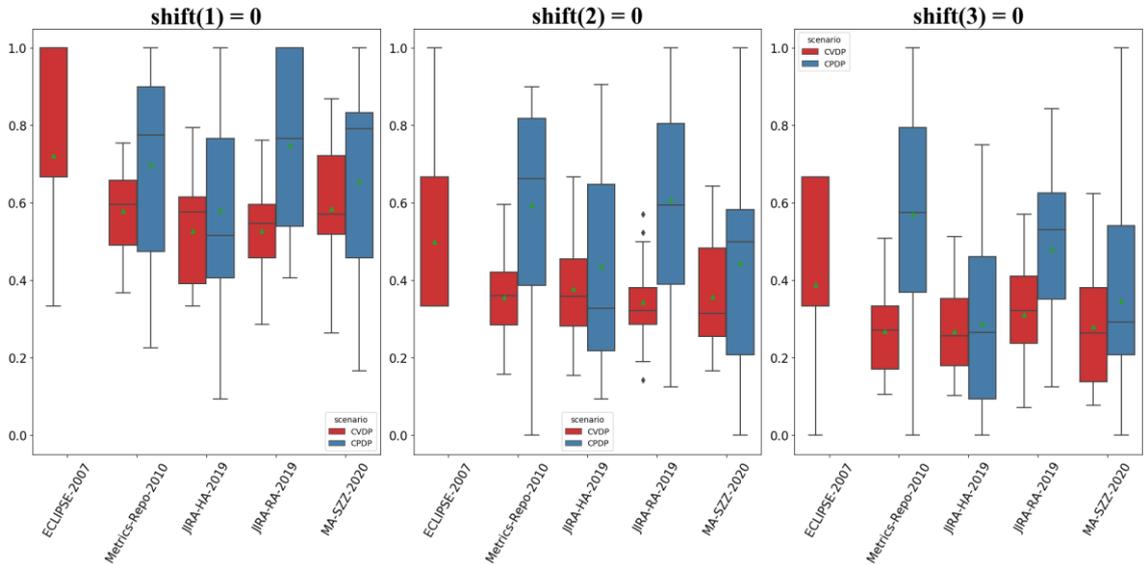

Fig. 24. Distribution of shift(*k*)=0 for the top 3 important features on 18 combinations

**11.3 External validity**

In this paper, we used six multi-version-project defect data sets from different sources as the subject data sets, including two data sets that have been widely used as benchmarks in previous studies, three recently published data sets, and one newly collected data set by ourselves. Consequently, we found that inconsistent labels exist in all the investigated data sets. Given this situation, it is reasonable to believe that inconsistent labels are likely to exist in other defect data sets collected by the same or similar approaches. Nonetheless, our subject defect data sets have their own particularities and peculiarities and might not be representative of defect data sets in general. Therefore, we do not claim that our findings can be generalized to all defect data sets. Indeed, this is an inherent problem to most (if not all) empirical studies, not unique to us. To mitigate this threat, there is a need to replicate our study across a wider variety of defect data sets in the future work.

# 12. Conclusion and future work

In this paper, we find that there is an inconsistent label phenomenon in multi-version-project data sets generated by different defect label collection approaches. By combining the latest idea of intrinsic and extrinsic bugs with the mechanism of diverse defect data collection approaches, we uncover the underlying rationale behind the generation of inconsistent labels and their real meaning: mislabels and labels corresponding to extrinsic bugs are the two causes of inconsistent labels; inconsistent labels generally indicate noise for defect prediction models.

We propose a Three Stage Inconsistent Label Identification (TSILI) approach to automatically detect inconsistent labels. In order to understand the degree of existence of inconsistent labels and evaluate the influence of inconsistent labels on defect prediction models, we conducted a large-scale empirical study on existing multi-version-project defect data sets covering common defect collection approaches. The experimental results show that:

(1) Most versions (>90%) of multi-version-project defect data sets collected by current common approaches contain varying degrees of inconsistent labels.

(2) Even the affected version approach, which is considered to be accurate by previous studies [13, 23], and the state-of-the-art defect data collection approach IND-JLMIV+R [90], lead to a considerable proportion of inconsistent labels.

(3) In both CVDP and CPDP contexts, the influence of inconsistent labels on a prediction model is all-round: inconsistent labels can considerably influence the prediction performance, the identified actual defects, and the model interpretation.

The key message of our study is to shed light that inconsistent labels have a significant impact on the performance of defect prediction models built with widely used modeling techniques (such as random forest, Naive Bayes, and logistic regression). Furthermore, our work reveals that existing defect label collection approaches are vulnerable to inconsistent labels and hence needed to be improved. Based on our findings, we strongly suggest that practitioners should detect inconsistent labels in defect data sets to understand their quality before building defect prediction models. If inconsistent labels are found, there is a need to correct or preprocess them to avoid the potential negative impact on defect prediction. In addition, there is a need to re-examine the experimental conclusions of previous studies using multi-version-project defect data sets with inconsistent labels.

What we need to emphasize is that our study is in no way a criticism to existing multi-version-project defect data sets investigated in our study. On the contrary, we provide a new perspective to understand the quality of defect data sets and contribute an automatic approach TSILI to detect label noise (i.e. inconsistent labels) in defect data sets. Indeed, providing high-quality defect data sets is the common goal of all researchers in our community. In the future, we plan to systematically evaluate the differences in inconsistent labels generated by different defect collection approaches and evaluate the impact of inconsistent labels on more classifiers.

# Acknowledgements

We are very grateful to Thomas Zimmermann, Rahul Premraj and Andreas Zeller for sharing their ECLIPSE-2007 data sets; Marian Jureczko and Diomidis D. Spinellis for sharing their Metrics-Repo-2010 data sets; Suraj Yatish, Jirayus Jiarpakdee, Patanamon Thongtanunam, and Chakkrit Tantithamthavorn for sharing their JIRA-HA-2019 and JIRA-RA-2019 data sets; Steffen Herbold, Alexander Trautsch, and Fabian Trautsch for sharing their IND-JLMIV+R-2020 data sets. Their data sets enable us to conduct the current study.

## Appendix A. The running time of our TSILI algorithm

The method we adopted to generate the source code databases required by the TSILI algorithm is to download the codes corresponding to each version from the official website of each target project, and then use the Understand[25] tool to parse the code to generate source code databases (.udb file). We write a Python[26] script to implement the TSILI algorithm. In the second stage of TSILI, the source code of each module is parsed and filtered (b9 in Fig. 17) based on the Python API (application programming interface) provided by the Understand tool.

In order to observe the time required for the TSILI algorithm to detect inconsistent labels on projects of different orders of magnitude, we selected Log4j, Hive, and Eclipse projects from multi-version-project defect data sets Metrics-Repo-2010 [1], JIRA-RA-2019 [2], and ECLIPSE-2007 [3], respectively, and then ran TSILI and recorded the time spent. Table 1 lists the details of these three projects and the single thread running time of TSILI. The 2nd column lists the versions included in each project. The 3rd column lists the order of magnitude of the project size, where $n$, *totalIns*, and *sumSLOC* represent the number of versions, the total number of instances of all versions, and the total number of code lines of all versions, respectively. The 4th column reports the running time of TSILI. These three projects (Log4j, Hive, and Eclipse) were selected because they represented orders of magnitude of the size of the projects in their respective data sets (the ECLIPSE-2007 data set only has the Eclipse project). In addition, the size of the total number of instances of these three projects varies in turn by one order of magnitude, which is conducive to observe the running time of TSILI under different orders of magnitude data sets.

Table 1 shows that the running time of TSILI is positively correlated with the size of the data set. For the project with hundreds of instances (i.e., Log4j project), the running time is at the second level. For the projects with ten thousands of instances (i.e., Eclipse project), the running time is at the hour level. Because TSILI is an offline algorithm, the hour-level (even minute-level or second-level) running time is acceptable in practice.

Table 1. Time consuming of the TSILI algorithm

| Dataset | Project (versions) | n, totalIns, sumSLOC | Running Time | Experimental environment |
|---|---|---|---|---|
| Metrics-Repo-2010 | Log4j (1.0, 1.1, 1.2) | n=3, totalIns=411, sumSLOC=74857 | ≈ 25 seconds | Inter(R) Core(TM) i7-7700 CPU @ 3.6GHz and 16G RAM |
| JIRA-RA-2019 | Hive (0.9.0, 0.10.0, 0.12.0) | n=3, totalIns=5285, sumSLOC=974774 | ≈ 11 minutes | |
| ECLIPSE-2007 | Eclipse (2.0, 2.1, 3.0) | n=3, totalIns=25203, sumSLOC=3089619 | ≈ 3 hours | |

## Appendix B. Metrics in the MA-SZZ-2020 data set

Table 2 describes the size, complexity, coupling, and inheritance metrics in the MA-SZZ-2020 data set we collected. In Table 2, column "Type" represents the type to which each metric belongs to, column "Name" gives the acronym of each metric, column "Definition" provides an informal description of the corresponding metric, and column "Tool for measuring metrics" gives the source of the tool that we measure metrics from. Note that inheritance metrics are indeed a form of coupling metrics. In practice, however, many researchers distinguish inheritance metrics from coupling metrics. Our study follows a metric classification framework similar to that in Briand et al. [4].

---

[25] https://scitools.com
[26] https://www.python.org

Table 2. List of metrics in the MA-SZZ-2020 data set

| Type | Name | Definition | Tool for measuring metrics |
|---|---|---|---|
| Size Metrics | SLOC (loc in data set) | the non-commentary source lines of code in a class | We used the Perl script developed in previous studies [5, 6] to collect metrics based on the udb database, where the udb database is generated by the commercial software Understand. |
| | NMIMP | the number of methods implemented in a class | |
| | NumPara | sum of the number of parameters of the methods implemented in a class | |
| | NM | the number of methods in a class, both inherited and non-inherited | |
| | NAIMP | the number of attributes in a class excluding inherited ones | |
| | NA | the number of attributes in a class including both inherited and non-inherited | |
| | Stms | the number of declaration and executable statements in the methods of a class | |
| | Nmpub | number of public methods implemented in a class | |
| | NMNpub | number of non-public methods implemented in a class | |
| | NIM | Number of Instance Methods | |
| | NCM | Number of Class Methods | |
| | NLM | Number of Local Methods | |
| | AvgSLOC | Average Source Lines of Code | |
| Complexity Metrics | CDE | Class Definition Entropy | |
| | CIE | Class Implementation Entropy | |
| | WMC | Weighted Method Per Class | |
| | SDMC | Standard Deviation Method Complexity | |
| | AvgWMC | Average Weight Method Complexity | |
| | CCMax | Maximum cyclomatic complexity of a single method of a class | |
| | NTM | Number of Trivial Methods | |
| Coupling Metrics | CBO | Coupling Between Object | |
| | DAC | Data Abstraction Coupling: Type is the number of attributes of other classes. | |
| | DACquote | Data Abstraction Coupling: Type is the number of other classes. | |
| | ICP | Information-flow-based Coupling | |
| | IHICP | Information-flow-based inheritance Coupling | |
| | NIHICP | Information-flow-based non-inheritance Coupling | |
| Inheritance Metrics | NOC | Number Of Child Classes | |
| | NOP | Number Of Parent Classes | |
| | DIT | Depth of Inheritance Tree | |
| | AID | Average Inheritance Depth of a class | |
| | CLD | Class-to-Leaf Depth | |
| | NOD | Number Of Descendants | |
| | NOA | Number Of Ancestors | |
| | NMO | Number of Methods Overridden | |
| | NMI | Number of Methods Inherited | |
| | NMA | Number Of Methods Added | |
| | SIX | Specialization IndeX = NMO * DIT / (NMO + NMA + NMI) | |
| | PII | Pure Inheritance Index. | |
| | SPA | static polymorphism in ancestors | |
| | SPD | static polymorphism in descendants | |
| | DPA | dynamic polymorphism in ancestors | |
| | DPD | dynamic polymorphism in descendants | |
| | SP | static polymorphism in inheritance relations | |
| | DP | dynamic polymorphism in inheritance relations | |

## Appendix C. Reasoning process of AP and RR formulas of a random model

Let *perf*(NC) be the performance of NC and *perf*(CC) be the performance of CC. It is more important for practitioners to evaluate *perfGain*, the relative performance of a model with respect to *random* [7]. In our context, *perfGain(NC)* = *perf(NC)* − *perf(random)* and *perfGain(CC)* = *perf(CC)* − *perf(random)*. Therefore, in this study, we also employ the absolute value of *pgr* (performance gain ratio) to evaluate the influence of inconsistent labels in a data set on prediction performance:

$$pgr = \frac{perfGain(NC) - perfGain(CC)}{perfGain(CC)} \times 100\%$$

Assume that a test set *T* consists of *N* instances, in which $n_1$ are defective. For a random model *random*, we calculate AP and RR as:

- $AP(random) = \frac{1}{N}\sum_{i=1}^{N}\left(\sum_{k=1}^{i}\frac{C_{i-1}^{k-1} \times C_{N-i}^{n_1-k}}{C_N^{n_1}} \times \frac{k}{i}\right)$

- $RR(random) = \sum_{i=1}^{N-n_1+1}\left(\frac{1}{i} \times \frac{C_{N-i}^{n_1-1}}{C_N^{n_1}}\right)$

First, the reasoning process of AP formula is as follows. We consider the calculation of AP as the sum of contribution of each position *i* (1≤ *i*≤ *N*) with defective modules. For each *i*, assuming that there are defective modules at the current position *i*, the remaining $n_1$-1 defective modules need to be placed on both sides of position *i*, that is, one part of the defective modules should be placed on 1 to *i*-1, and the other part should be placed on *i*+1 to *N*. At this time, (1) the probability that there is a defective module at the current position *i* is $\frac{C_{N-1}^{n_1-1}}{C_N^{n_1}}$; (2) the probability that the defective module at the current position *i* is the *k*th defective module (i.e., *k*-1 (*k* ≤ *i*) defective modules are placed on 1 to *i*-1, and $n_1$-*k* defective modules are placed on *i*+1 to *N*) is $\frac{C_{i-1}^{k-1} \times C_{N-i}^{n_1-k}}{C_{N-1}^{n_1-1}}$; (3) the contribution value of position *i* with defective modules is *k*/*i*. Therefore, the total AP contribution value of upstream defective module on current position *i* is calculated as:

$$AP(i) = \sum_{k=1}^{i}\left(\frac{C_{i-1}^{k-1} \times C_{N-i}^{n_1-k}}{C_{N-1}^{n_1-1}} \times \frac{C_{N-1}^{n_1-1}}{C_N^{n_1}} \times \frac{k}{i}\right) = \sum_{k=1}^{i}\left(\frac{C_{i-1}^{k-1} \times C_{N-i}^{n_1-k}}{C_N^{n_1}} \times \frac{k}{i}\right) \quad (k \leq i \leq N - n_1 + k)$$

Further, we can derive the formula of AP as follows:

$$AP = \frac{1}{N}\sum_{i=1}^{N}(AP(i)) = \frac{1}{N}\sum_{i=1}^{N}\sum_{k=1}^{i}\left(\frac{C_{i-1}^{k-1} \times C_{N-i}^{n_1-k}}{C_N^{n_1}} \times \frac{k}{i}\right)$$

Second, the reasoning process of RR formula is as follows. RR can be expressed as:

$$RR = \frac{1}{1}\frac{C_{N-1}^{n_1-1}}{C_N^{n_1}} + \frac{1}{2}\frac{C_{N-2}^{n_1-1}}{C_N^{n_1}} + \cdots + \frac{1}{N-n_1+1}\frac{C_{N-(N-n_1+1)}^{n_1-1}}{C_N^{n_1}}$$

The first summation term $\frac{1}{1}\frac{C_{N-1}^{n_1-1}}{C_N^{n_1}}$ indicates that if the first defective module appears in the first position, then the RR value is 1, and the probability of occurrence of this event is $\frac{C_{N-1}^{n_1-1}}{C_N^{n_1}}$. Because the total possibility is $C_N^{n_1}$ (regarding the defective modules as repeated events), the possibility of the current situation is to select $n_1$-1 positions from the subsequent *N*-1 positions to place the defective modules, that is, the total possibility is $C_{N-1}^{n_1-1}$. By analogy, the first defective module can appear at most in the *N*-$n_1$+1 position. After sorting out the above formulas, the results are as follows:

$$RR = \sum_{i=1}^{N-n_1+1}\left(\frac{1}{i} \times \frac{C_{N-i}^{n_1-1}}{C_N^{n_1}}\right)$$

The advantage of using PGR indicator is that the influence of the difficulty of the problem itself is considered. If the performance of a model *m* is significantly increased compared with the random model, it should be considered that model *m* has a good effect. By eliminating the impact of the difficulty of the problem itself, it will be more fair and meaningful to observe or compare the change magnitude of evaluation indicators on different data points (pairs of training set and test set built with different versions of different projects).